\documentclass[twocolumn,twocolappendix]{aastex701}
\usepackage{CJK}
\usepackage{xspace}

\usepackage{graphicx}
\usepackage{subcaption}
\usepackage{amsmath}

\newcommand{\groupnumber}{16\xspace}
\newcommand{\starnumber}{68\xspace}

\begin{document}
\begin{CJK*}{UTF8}{gbsn}
  \title{Gravity-mode main-sequence asteroseismology of \groupnumber open clusters and associations}%\footnote{}}

  \author[orcid=0000-0001-9313-251X,gname=Gang,sname=Li]{Gang Li (李刚)}
  \affiliation{Centre for Astrophysics, University of Southern Queensland, Toowoomba, QLD 4350, Australia}
  \affiliation{Institute of Astronomy, KU Leuven, Celestijnenlaan 200D, 3001 Leuven, Belgium}
  \email[show]{gang.li@unisq.edu.au}
  \correspondingauthor{Gang Li}

  \author[orcid=0009-0004-1774-7167, gname=Haotian,sname=Wang]{Haotian Wang (王昊天)}
  \affiliation{Institute of Astronomy, KU Leuven, Celestijnenlaan 200D, 3001 Leuven, Belgium}
  \email[]{haotian.wang@kuleuven.be}

  \author[orcid=0000-0003-0619-0207, gname=Anuj,sname=Gautam]{Anuj Gautam}
  \affiliation{Centre for Astrophysics, University of Southern Queensland, Toowoomba, QLD 4350, Australia}
  \email[]{anuj.gautam@unisq.edu.au}

  \author[orcid=0000-0002-9901-3113, gname=J. S. G., sname=Mombarg]{J. S. G. Mombarg}
  \affiliation{Universit\'{e} Paris-Saclay, Universit\'{e} de Paris, Sorbonne Paris Cit\'{e}, CEA, CNRS, AIM, 91191 Gif-sur-Yvette, France}
  \email[]{joey.mombarg@cea.fr}

  \author[orcid=0000-0002-5648-3107, gname=Simon J.,sname=Murphy]{Simon J. Murphy}
  \affiliation{Centre for Astrophysics, University of Southern Queensland, Toowoomba, QLD 4350, Australia}
  \email[]{simon.murphy@usq.edu.au}

  \author[orcid=0009-0005-7508-1591, gname=Mara,sname=Bernizzoni]{Mara Bernizzoni}
  \affiliation{Sydney Institute for Astronomy (SIfA), School of Physics, University of Sydney, NSW 2006, Australia}
  \email[]{mber0784@uni.sydney.edu.au}

  \author[orcid=0000-0002-5649-7461, gname=Mingjie,sname=Jian]{Mingjie Jian (简明杰)}
  \affiliation{Institute of Astronomy, University of Cambridge, Madingley Road, Cambridge CB3 0HA, UK}
  \email[]{jian-mingjie@outlook.com}

  \author[orcid=0009-0000-4652-2566,gname=Saakshi,sname=Wadhwa]{Saakshi Wadhwa}
  \affiliation{Centre for Astrophysics, University of Southern Queensland, Toowoomba, QLD 4350, Australia}
  \email[]{Saakshi.Wadhwa@unisq.edu.au}

  \author[orcid=0000-0002-2275-3877, gname=Dario J.,sname=Fritzewski]{Dario J. Fritzewski}
  \affiliation{Institute of Astronomy, KU Leuven, Celestijnenlaan 200D, 3001 Leuven, Belgium}
  \email[]{dario.fritzewski@kuleuven.be}

  \author[orcid=0000-0002-9633-4093, gname=Facundo D., sname=Moyano]{Facundo D. Moyano}
  \affiliation{International Centre of Supernovae (ICESUN), Yunnan Key Laboratory of Supernova Research, Yunnan Observatories, Chinese Academy of Sciences (CAS), Kunming 650216, China}
  \email[]{facundo.moyano@ynao.ac.cn}

  \author[orcid=0000-0002-8707-201X,gname=Prasad,sname=Mani]{Prasad Mani}
  \affiliation{Sydney Institute for Astronomy (SIfA), School of Physics, University of Sydney, NSW 2006, Australia}
  \email[]{prasadmani94@gmail.com}

  \author[orcid=0000-0002-3084-5157, gname=Chengyuan,sname=Li]{Chenyuan Li (李程远)}
  \affiliation{School of Physics and Astronomy, Sun Yat-sen University, Zhuhai 519082, China}
  \email[]{lichengy5@mail.sysu.edu.cn}

  \author[orcid=0000-0001-5222-4661, gname=Timothy R.,sname=Bedding]{Timothy R. Bedding}
  \affiliation{Sydney Institute for Astronomy (SIfA), School of Physics, University of Sydney, NSW 2006, Australia}
  \email[]{tim.bedding@sydney.edu.au}

  \author[orcid=0000-0003-1822-7126,gname=Conny,sname=Aerts]{Conny Aerts}
  \affiliation{Institute of Astronomy, KU Leuven, Celestijnenlaan 200D, 3001 Leuven, Belgium}
  \affiliation{Department of Astrophysics, IMAPP, Radboud University Nijmegen, PO Box 9010, 6500 GL Nijmegen, The Netherlands}
  \affiliation{Max-Planck-Institut f\"ur Astronomie, K\"onigstuhl 17, D-69117 Heidelberg, Germany}
  \email[]{conny.aerts@kuleuven.be}

  %\collaboration{all}{The Terra Mater collaboration}

  \begin{abstract}

    Coeval stellar populations offer a unique advantage for studying stellar physics because their members share a common age and broadly similar initial chemical composition, allowing the free parameters in stellar models to be constrained more efficiently.
    Meanwhile, asteroseismology---through the study of pulsation modes that propagate through stellar interiors---provides powerful constraints on internal structure and the physical processes operating beneath stellar atmospheres. The combination of coeval-population studies and asteroseismology therefore provides an unparalleled laboratory for testing and improving stellar evolutionary and structural models.
    Using light curves from the Kepler and TESS space missions, together with membership constraints based on Gaia astrometry, we identify g-mode period-spacing patterns in dozens of early-type ($T_{\rm eff} \gtrsim 7000\,\mathrm{K}$) main-sequence stars across \groupnumber stellar groups.
    From these patterns, we infer near-core rotation rates $f_\mathrm{rot}$ and measure the asymptotic spacing $\Pi_0$ of g modes, tracing both angular-momentum transport and the evolution of the convective stellar cores. Using our homogeneous set of g-mode stellar-group members spanning a wide range of ages, we show that near-core rotation in early-type stars decreases with stellar evolution. 
    We compare our results with existing model predictions. Our $\Pi_0$ measurements still show discrepancies with current stellar models for reasons that remain unclear, as reported in several previous studies. \added{We also report two g-mode pre-main-sequence pulsators and three pulsating blue stragglers}. Our sample provides a useful basis for calibrating stellar models in future ensemble asteroseismic studies of coeval stellar populations.

  \end{abstract}



  \section{Introduction}
  Asteroseismology is the study of stellar oscillations and what they reveal about stellar interiors \citep[see the reviews by][]{Aerts2010book, Bowman2026}. Over the past decade, space-based photometric missions such as CoRoT \citep{Auvergne2009CoRoT}, Kepler \citep{Borucki2010}, and TESS \citep{Ricker2015} have driven major advances in the field and deepened our understanding of stellar physics \citep[e.g.][]{Aerts2021RvMP}.
  In particular, gravity (g) modes are highly sensitive to the physical conditions in and near stellar cores, and therefore provide direct probes of near-core regions. 
  In early-type main-sequence stars ($T_{\mathrm{eff}} \gtrsim 7000\,\mathrm{K}$), high-radial-order ($20 \lesssim n \lesssim 100$), low-degree ($\ell \lesssim 2$) g modes can reach observable amplitudes at the surface because their envelopes contain only shallow (or no) convective layers \citep[e.g.][]{VanReeth2015ApJS, Li2020MNRAS_611}. 
  Stars exhibiting g-mode pulsations are either $\gamma$\,Doradus stars \citep[A- to F-type;][]{Balona1994, Kaye1999} or Slowly Pulsating B-type (SPB) stars \citep{Waelkens1991, DeCat2002A&A}. Here we use the umbrella term ``g-mode pulsators'' for both classes. These g-mode pulsators have proved to be powerful probes of multiple physical processes, including near-core rotation \citep{VanReeth2016_TAR, Li2020MNRAS_611, Garcia2022_60_gdor}, convective-core overshooting and mixing in the radiative envelope \citep{Miglio2008MNRAS, Pedersen2021, Mombarg2021}, angular-momentum (AM) transport \citep{Ouazzani2019A&A, Moyano2023, Mombarg2023calibrating_AM, Mombarg2026_calibrating_AM}, and even deep-seated magnetic fields \citep{Takata2026_gdor_magnetic_fields, Ihallaine_2026}.

Despite these successes, comprehensive modelling of gravity-mode main-sequence pulsators remains computationally demanding. Detailed modelling is at least a seven-dimensional problem \citep{Aerts2018ApJS}, and several physical descriptions cannot yet be derived from first principles and lack robust calibration, such as mixing length and convective overshooting \citep[e.g.][]{Joyce2023}. 
In addition, the most basic parameters of stellar evolution, mass, age, and metallicity, often lack accurate measurements, which leads to severe parameter degeneracies. This forces us to explore a wide parameter space and makes the computation of evolutionary grids time-consuming. 

Furthermore, the underlying physics of AM transport and envelope mixing also remains uncertain: multiple processes may operate simultaneously. 
For example, AM transport may be driven by meridional circulation \citep{Zahn1992, Maeder1998}, 
shear instabilities \citep{Heger2000}, 
magnetic dynamos \citep{Spruit2002, Fuller2019}, 
or internal gravity waves \citep{Rogers2013, Ratnasingam2019, Rogers2025, Mathis2025} (see \citealt{Aerts2019ARAA} for a review). 
Chemical mixing in radiative envelopes may arise from shear mixing \citep{Maeder1997}, internal gravity waves \citep{Rogers2017, Varghese2023, Varghese2025, Mombarg2025}, or surface fossil magnetic fields \citep{Keszthelyi2020} (see \citealt{Maeder2009} and \citealt{Salaris2017} for reviews). Yet their relative contributions and calibrations are still poorly constrained. 

Seeking additional constraints on stars greatly helps to expand the scope of asteroseismic modelling. 
Binaries with two pulsating components are already highly valuable \citep[e.g.][]{Johnston2019, Murphy2021_dsct_solar_like_binary, Hon2022, Beck2024}, provided that the two stars are coeval and chemically homogeneous \citep[e.g.][]{Kraus2009, Valle2016}. 
Previous synergistic binary-asteroseismology studies have strongly tested current stellar evolutionary theory, in both main-sequence and post-main-sequence phases \citep[e.g.][]{Schmid2016_10080943_modelling, Guo2017, LiYaGuang2018_solarlike_binary,Sekaran2020, Murphy2021_dsct_solar_like_binary, Schimak2026}. 

Even more powerfully, clusters with many coeval pulsating stars provide a much stronger set of constraints and tests of stellar models \citep[e.g.][]{Piau2005,Brogaard2015}. 
Recent studies highlight both the promise and current limitations: 
\citet{Tayar2025} found large star-to-star scatter in red-giant asteroseismic ages within the same cluster, sometimes in apparent tension with isochrone ages.
On the other hand, \citet{Fritzewski2024UBC1} showed that joint modelling of $g$-mode asteroseismology, gyrochronology, and isochrone fitting yields a more robust age for the young open cluster UBC\,1. 
In addition to a common age, global stellar parameters such as temperature, luminosity, and mass can be inferred from the stars' locations on the colour-magnitude diagram (CMD). 
\citet{LiGang_2025_NGC2516_modelling} modelled several g-mode pulsators in NGC\,2516 and inferred a seismic cluster age of $132\pm8\,\mathrm{Myr}$, but reported discrepancies between seismic and isochrone masses. These discrepancies may arise from inconsistencies between surface and interior properties, or may indicate that a detailed treatment of the effects of rotation on the CMD is needed. 
A series of recent studies has demonstrated the potential of cluster--asteroseismology synergistic research.

Asteroseismology in clusters has been burgeoning over the past decade. 
Early efforts used CoRoT data \citep[e.g.][]{Zwintz201_NGC2264}. 
The Kepler field contained four well-known clusters, NGC\,6791, NGC\,6819, NGC\,6811, and NGC\,6866 \citep[e.g.][]{Stello2011, Hekker2011, Basu2011, Corsaro_2012, Balona2013, Balona2013_NGC6819, Brogaard2021, Colman2022, Brogaard2023}, and asteroseismology helped constrain the properties of these clusters in the pre-Gaia era. 
The K2 mission scanned along the ecliptic plane and covered many additional clusters \citep[e.g.][]{Lund2016, Stello2016_M67, Sandquist2020, Murphy2022Pleiades}, enabling the detection of a seismic signature of deepening convective envelopes in post-main-sequence solar-like oscillators in M\,67 \citep{Reyes2025Natur}. 

Data from the ongoing TESS mission is also useful for asteroseismology in clusters, despite the shorter observing baselines and crowding in its large pixels. In particular, pressure-mode (p-mode) pulsators in open clusters---whose amplitude spectra can often be resolved with shorter light curves---have been identified with TESS data, including in the Pleiades \citep{Bedding2023ApJ}, NGC\,3532 \citep{Berry2025_dsct_in_NGC3532}, the Cep--Her complex \citep{Murphy2024Cep_Her_complex}, NGC\,1901 \citep{Bernizzoni2025_NGC1901}, and NGC\,2126 \citep{Dileep2025_NGC2126}. 
Gravity-mode period-spacing patterns have also been discovered using year-long light curves in the TESS continuous viewing zone in UBC\,1 and NGC\,2516 \citep{Fritzewski2024UBC1, LiGang_2024_NGC2516}. Such patterns have also been observed in shorter TESS light curves in the Pleiades and NGC\,3532 \citep{Fritzewski-Pleiades}. 

In this paper, we present g-mode period-spacing patterns for \starnumber member stars across \groupnumber stellar groups. Throughout this work, we use ``stellar groups'' as an umbrella term encompassing open clusters, young stellar associations, and substructures within larger stellar complexes. Our homogeneous sample of early-type main-sequence stars with measured near-core rotation rates and asymptotic spacings establishes a benchmark data set for calibrating AM transport and internal mixing in future stellar-evolution models. 
The paper is organised as follows. Section~\ref{sec:the_cluster_and_LC} describes the properties of the stellar groups studied in this work. Section~\ref{sec:signal_extraction} details the asteroseismic signal extraction, including frequency pre-whitening, period-spacing pattern identification, and fitting under the traditional approximation of rotation (TAR).
We present an overview of our seismic measurements in Section~\ref{sec:overall_look}, including the distribution of near-core rotation rates in Section~\ref{subsec:frot}, their mass-dependent evolution in Section~\ref{subsec:near_core_rotation_rate_with_mass}, and the corresponding results for asymptotic spacings in Sections~\ref{subsec:Pi0} and \ref{subsec:Pi0_with_mass}. We present a newly discovered pre-main-sequence pulsator in Section~\ref{sec:PMS_pulsator}, and discuss the pulsating blue stragglers in our sample in Section~\ref{sec:BSs}.
Finally, Section~\ref{sec:conclusions} summarises our main conclusions.

\section{The Stellar Groups in This Work}\label{sec:the_cluster_and_LC}

\subsection{The Collected Stellar Groups}
Gaia has enabled the large-scale identification of open clusters, leading to the release of various open-cluster catalogues \citep[e.g.][]{Cantat-Gaudin2020, Castro-Ginard2022, Qin2023, Perren2023, Hunt2024}. 
In this work, we relied on the all-sky cluster catalog and membership lists derived from Gaia DR3 by \citet{Hunt2024}.

Some clusters in this work were compiled before the publication of \citet{Hunt2024}, while some others are not included in their catalogue. For these systems, we adopted membership lists from elsewhere in the literature. In particular, we used the Gaussian-mixture-model memberships for the four well-known open clusters in the Kepler field \citep{Wang2026_NGC6866}, the NGC\,2516 memberships from \citet{Meingast2021}, the NGC\,3532 memberships from \citet{Pang2022}, the Pleiades memberships combined from \citet{Gagne2018} and \citet{Meingast2021} (see \citealt{Fritzewski-Pleiades}), the Huluwa~1--5 subgroup definitions and memberships for Vela~OB2 from \citet{Pang2021_huluwa}, and the Cep--Her complex memberships from \citet{Kerr2023SPYGLASSIV}, as adopted by \citet{Murphy2024Cep_Her_complex}.

For the Kepler field, by crossmatching with the catalogue by \citet{Hunt2024}, we found that, in addition to the four well-known clusters, three further open clusters were identified: CWNU\,1183 ($\sim50\,\mathrm{Myr}$), ASCC\,108 ($\sim190\,\mathrm{Myr}$), and Theia\,4877 ($\sim720\,\mathrm{Myr}$). However, these clusters were sparsely sampled by Kepler, with light curves available for only a few member stars. 

For stellar groups observed by TESS, we did not perform photometric reductions for every group across the sky. Instead, we prioritised targets based on asteroseismic requests from our collaborators (e.g., for spectroscopic follow-up, rotation measurements, or complementary g-mode analyses). In practice, whenever such a group had at least two consecutive TESS sectors of coverage, we reduced and analysed the corresponding light curves to search for g-mode period-spacing patterns. 

In total, we collected \groupnumber stellar groups for g-mode asteroseismic analysis, listed in the first column of Table~\ref{tab:cluster_gdor_sample}.

% Auto-generated AASTeX multi-page landscape table (one row per star)
\begin{longrotatetable}
\begin{deluxetable*}{l l r r l r r r l l l}
\tablecaption{Gravity-mode pulsators in \groupnumber stellar groups.\label{tab:cluster_gdor_sample}}
\tablehead{
\colhead{Stellar group} & \colhead{Age (Myr)} & \colhead{$A_V$ (mag)} & \colhead{Distance (pc)} &
\colhead{TIC or KIC} & \colhead{Gaia DR3 source\_id} &
\colhead{RA (deg)} & \colhead{Dec (deg)} &
\colhead{$f_{\rm rot}$ (d$^{-1}$)} & \colhead{$\Pi_0$ (s)} & \colhead{Mass ($\mathrm{M_\odot}$)}
}
\tabletypesize{\scriptsize}
\startdata
 Perseus Alcaeus & $6.5^{+2.2}_{-2.2}$ & $1.2$ & $286^{+5}_{-5}$ & TIC\,26532226$^{*}$ & 216759169703741952 & 57.308996 & 31.734505 & $1.23^{+0.09}_{-0.09}$ & $4130^{+400}_{-400}$ & $1.52^{+0.07}_{-0.10}$ \\
 Vela OB2 Huluwa & $14^{+8}_{-8}$ & $0.35$ & $352.0^{+2.9}_{-2.9}$ & TIC\,56640851$^{*}$ & 5543536661448630528 & 127.241234 & -34.493455 & $2.634^{+0.007}_{-0.007}$ & $4302^{+70}_{-60}$ & $2.19^{+0.05}_{-0.05}$ \\
 Cep-Her ORPH & $32^{+8}_{-8}$ & $0.27$ & $341.2^{+1.5}_{-1.6}$ & TIC\,349011395$^{*}$ & 4531730361801104512 & 282.504541 & 22.909099 & $2.128^{+0.013}_{-0.014}$ & $4350^{+70}_{-70}$ & $1.55^{+0.05}_{-0.05}$ \\
 Cep-Her ORPH & $32^{+8}_{-8}$ & $0.24$ & $341.2^{+1.5}_{-1.6}$ & TIC\,346151391$^{*}$ & 4531762659948423936 & 280.041233 & 21.097964 & $0.965^{+0.01}_{-0.009}$ & $4794^{+90}_{-80}$ & $1.58^{+0.05}_{-0.05}$ \\
 Cep-Her CUPV & $67^{+13}_{-13}$ & $0.083$ & $341.2^{+1.5}_{-1.6}$ & TIC\,303754397$^{*}$ & 2181689321609724544 & 309.023173 & 53.203000 & $1.956^{+0.026}_{-0.028}$ & $3429^{+190}_{-190}$ & $1.59^{+0.05}_{-0.05}$ \\
 Cep-Her CINR & $36^{+8}_{-8}$ & $0.087$ & $341.2^{+1.5}_{-1.6}$ & TIC\,277011700$^{*}$ & 2181783295493195264 & 306.492739 & 52.483826 & $1.57^{+0.03}_{-0.03}$ & $4632^{+280}_{-260}$ & $1.60^{+0.05}_{-0.05}$ \\
 Cep-Her CINR & $36^{+8}_{-8}$ & $0.076$ & $341.2^{+1.5}_{-1.6}$ & TIC\,277570969$^{*}$ & 2184113366783588096 & 311.067054 & 56.173412 & $0.750^{+0.017}_{-0.018}$ & $9688^{+400}_{-400}$ & $3.42^{+0.10}_{-0.10}$ \\
 CWNU\,1183 & $46^{+40}_{-20}$ & $0.11$ & $471.7^{+0.6}_{-0.6}$ & KIC\,3868382 & 2073188036719192960 & 295.851255 & 38.996251 & $0.8441^{+0.0004}_{-0.0004}$ & $4453^{+9}_{-10}$ & $1.676^{+0.027}_{-0.05}$ \\
 Pleiades & $125^{+8}_{-8}$ & $0.10$ & $134.841^{+0.018}_{-0.018}$ & TIC\,385552372 & 65296357738731008 & 56.393686 & 24.463107 & $2.54^{+0.07}_{-0.07}$ & $4860^{+1800}_{-1800}$ & $1.568^{+0.017}_{-0.03}$ \\
  &  &  &  & TIC\,67828699 & 63502259702709888 & 56.383383 & 21.246480 & $2.990^{+0.020}_{-0.020}$ & $5150^{+260}_{-260}$ & $1.89^{+0.04}_{-0.04}$ \\
  &  &  &  & TIC\,5536869 & 62413988007668352 & 49.457440 & 22.831990 & $1.88^{+0.07}_{-0.07}$ & $5245^{+1100}_{-1100}$ & $1.97^{+0.05}_{-0.05}$ \\
 NGC\,2516 & $132^{+8}_{-8}$ & $0.20$ & $407.09^{+0.09}_{-0.09}$ & TIC\,341043961 & 5295292121571898880 & 118.323853 & -57.552839 & $3.011^{+0.011}_{-0.012}$ & $5035^{+90}_{-110}$ & $1.499^{+0.021}_{-0.05}$ \\
  &  &  &  & TIC\,308307454 & 5290771204639131648 & 120.228699 & -60.754088 & $0.97^{+0.03}_{-0.04}$ & $4340^{+150}_{-160}$ & $1.515^{+0.023}_{-0.05}$ \\
  &  &  &  & TIC\,358466729 & 5290739146997100160 & 119.262316 & -60.610183 & $2.1191^{+0.003}_{-0.0029}$ & $4726^{+80}_{-90}$ & $1.533^{+0.023}_{-0.05}$ \\
  &  &  &  & TIC\,410452218 & 5290650705036608256 & 119.914707 & -61.194376 & $2.803^{+0.010}_{-0.01}$ & $4799^{+90}_{-90}$ & $1.628^{+0.025}_{-0.06}$ \\
  &  &  &  & TIC\,281582674 & 5289355171098491520 & 115.990674 & -61.694069 & $2.953^{+0.006}_{-0.006}$ & $4841^{+60}_{-60}$ & $1.70^{+0.04}_{-0.04}$ \\
  &  &  &  & TIC\,364398040 & 5290820029826947968 & 119.628849 & -60.629526 & $2.947^{+0.014}_{-0.014}$ & $4865^{+120}_{-120}$ & $1.77^{+0.03}_{-0.07}$ \\
  &  &  &  & TIC\,372912679 & 5290642699218901376 & 119.483950 & -61.365514 & $3.005^{+0.011}_{-0.013}$ & $4821^{+100}_{-110}$ & $1.82^{+0.03}_{-0.08}$ \\
  &  &  &  & TIC\,308992761 & 5276258269626067840 & 123.282602 & -65.390744 & $2.965^{+0.006}_{-0.007}$ & $4901^{+60}_{-70}$ & $1.91^{+0.09}_{-0.09}$ \\
  &  &  &  & TIC\,410451583 & 5290770586163654656 & 119.934187 & -60.671891 & $2.953^{+0.008}_{-0.008}$ & $5024^{+80}_{-70}$ & $2.27^{+0.11}_{-0.08}$ \\
  &  &  &  & TIC\,372913043 & 5290669774690987776 & 119.394940 & -61.058467 & $2.974^{+0.007}_{-0.006}$ & $4892^{+60}_{-60}$ & $2.95^{+0.13}_{-0.22}$ \\
  &  &  &  & TIC\,358466708 & 5290739387520374912 & 119.227792 & -60.598423 & $1.265^{+0.027}_{-0.03}$ & $7530^{+800}_{-700}$ & $3.54^{+0.29}_{-0.3}$ \\
 UBC\,1 & $225^{+80}_{-80}$ & $0.062$ & $322.42^{+0.16}_{-0.16}$ & TIC\,421334449 & 2142861786451443840 & 288.972108 & 57.891319 & $0.544^{+0.009}_{-0.009}$ & $4549^{+24}_{-24}$ & $1.649^{+0.025}_{-0.03}$ \\
 ASCC\,108 & $265^{+5}_{-5}$ & $0.37$ & $1152.1^{+1.7}_{-1.7}$ & KIC\,4774208 & 2073465354172422144 & 298.194971 & 39.869692 & $1.8336^{+0.0012}_{-0.0011}$ & $4340^{+25}_{-23}$ & $1.58^{+0.04}_{-0.05}$ \\
  &  &  &  & KIC\,4951030 & 2073467724994578304 & 298.309182 & 40.010125 & $2.534^{+0.027}_{-0.03}$ & $5060^{+500}_{-500}$ & $1.63^{+0.04}_{-0.05}$ \\
  &  &  &  & KIC\,4774194 & 2073465354172420736 & 298.193391 & 39.869311 & $1.8348^{+0.0010}_{-0.0010}$ & $4340^{+19}_{-19}$ & $2.74^{+0.15}_{-0.21}$ \\
 NGC\,3532 & $340^{+10}_{-20}$ & $0.12$ & $471.88^{+0.11}_{-0.11}$ & TIC\,305347034 & 5338678949185892864 & 165.743467 & -58.956847 & $0.49^{+0.03}_{-0.04}$ & $4513^{+210}_{-220}$ & $1.55^{+0.03}_{-0.03}$ \\
  &  &  &  & TIC\,306045270 & 5338648854402511232 & 166.450422 & -59.005159 & $0.911^{+0.022}_{-0.024}$ & $4706^{+150}_{-150}$ & $1.56^{+0.03}_{-0.03}$ \\
  &  &  &  & TIC\,305906846 & 5340222530409352704 & 166.283793 & -58.318407 & $1.705^{+0.021}_{-0.020}$ & $5063^{+220}_{-190}$ & $1.59^{+0.04}_{-0.03}$ \\
  &  &  &  & TIC\,306503414 & 5340158586938781440 & 166.816190 & -58.474797 & $2.011^{+0.019}_{-0.018}$ & $4217^{+170}_{-150}$ & $1.60^{+0.04}_{-0.03}$ \\
  &  &  &  & TIC\,306503983 & 5340160511084370176 & 166.912764 & -58.397806 & $2.75^{+0.04}_{-0.05}$ & $5873^{+500}_{-500}$ & $1.78^{+0.05}_{-0.04}$ \\
  &  &  &  & TIC\,305909136 & 5338663830900967296 & 166.335782 & -58.642940 & $2.85^{+0.03}_{-0.03}$ & $5414^{+400}_{-300}$ & $1.78^{+0.05}_{-0.04}$ \\
  &  &  &  & TIC\,306384085 & 5340151337033203456 & 166.732204 & -58.623414 & $2.705^{+0.029}_{-0.03}$ & $4667^{+170}_{-170}$ & $1.81^{+0.05}_{-0.04}$ \\
  &  &  &  & TIC\,306385060 & 5338659162324993280 & 166.613855 & -58.762990 & $2.800^{+0.027}_{-0.028}$ & $5249^{+220}_{-200}$ & $2.80^{+0.13}_{-0.3}$ \\
 NGC\,6866 & $590^{+170}_{-170}$ & $0.31$ & $1402.0^{+2.2}_{-2.2}$ & KIC\,8330312 & 2076065664799868288 & 300.952501 & 44.219388 & $1.338^{+0.005}_{-0.005}$ & $4241^{+90}_{-90}$ & $1.51^{+0.05}_{-0.05}$ \\
  &  &  &  & KIC\,8264304 & 2076064874525839744 & 300.921448 & 44.111050 & $1.552^{+0.021}_{-0.020}$ & $3897^{+270}_{-240}$ & $1.51^{+0.05}_{-0.05}$ \\
  &  &  &  & KIC\,8197019 & 2076063122179130624 & 300.848740 & 44.077954 & $1.886^{+0.01}_{-0.009}$ & $3896^{+110}_{-100}$ & $1.52^{+0.05}_{-0.05}$ \\
  &  &  &  & KIC\,8264720 & 2081882459267654400 & 301.023483 & 44.173077 & $1.450^{+0.013}_{-0.014}$ & $4232^{+150}_{-150}$ & $1.55^{+0.04}_{-0.05}$ \\
  &  &  &  & KIC\,8198031 & 2075873422065500800 & 301.111555 & 44.063979 & $1.8549^{+0.0014}_{-0.0013}$ & $4144^{+21}_{-21}$ & $1.55^{+0.04}_{-0.05}$ \\
  &  &  &  & KIC\,8330056 & 2076066145836177408 & 300.887723 & 44.201764 & $1.9125^{+0.0012}_{-0.0012}$ & $4096^{+15}_{-15}$ & $1.55^{+0.04}_{-0.05}$ \\
  &  &  &  & KIC\,8263970 & 2076063568855737728 & 300.841745 & 44.106563 & $1.9462^{+0.0012}_{-0.0012}$ & $4183^{+18}_{-18}$ & $1.56^{+0.04}_{-0.05}$ \\
  &  &  &  & KIC\,8265126 & 2081881905204354048 & 301.124823 & 44.175800 & $0.891^{+0.004}_{-0.004}$ & $4198^{+50}_{-40}$ & $1.57^{+0.04}_{-0.05}$ \\
  &  &  &  & KIC\,8127448 & 2076047591576773120 & 300.406599 & 43.947894 & $1.145^{+0.013}_{-0.013}$ & $4256^{+110}_{-110}$ & $1.57^{+0.04}_{-0.05}$ \\
  &  &  &  & KIC\,8264708 & 2075877648313310848 & 301.020006 & 44.159308 & $1.6361^{+0.0012}_{-0.0011}$ & $4240^{+40}_{-40}$ & $1.66^{+0.04}_{-0.07}$ \\
  &  &  &  & KIC\,8330092 & 2076066725646925824 & 300.895527 & 44.247220 & $1.36^{+0.04}_{-0.04}$ & $3643^{+210}_{-200}$ & $1.67^{+0.04}_{-0.07}$ \\
  &  &  &  & KIC\,8197788$^{*}$ & 2075876583161403392 & 301.046622 & 44.092546 & $2.290^{+0.012}_{-0.012}$ & $4512^{+90}_{-90}$ & $1.86^{+0.05}_{-0.08}$ \\
  &  &  &  & KIC\,8264667 & 2075877609644932608 & 301.010774 & 44.157189 & $0.684^{+0.005}_{-0.005}$ & $4240^{+130}_{-130}$ & $2.03^{+0.05}_{-0.13}$ \\
  &  &  &  & KIC\,8264550 & 2075877884522838400 & 300.979905 & 44.178839 & $1.536^{+0.014}_{-0.014}$ & $4521^{+170}_{-160}$ & $2.25^{+0.11}_{-0.15}$ \\
  &  &  &  & KIC\,8264293$^{*}$ & 2076065699159591296 & 300.919389 & 44.195388 & $2.427^{+0.015}_{-0.016}$ & $7935^{+400}_{-400}$ & $3.18^{+0.17}_{-0.10}$ \\
 NGC\,2632 & $700^{+100}_{-100}$ & $0.16$ & $183.49^{+0.04}_{-0.04}$ & TIC\,175239414$^{*}$ & 664329190407201280 & 129.993050 & 20.158186 & $2.119^{+0.005}_{-0.004}$ & $3842^{+40}_{-40}$ & $1.69^{+0.05}_{-0.04}$ \\
  &  &  &  & TIC\,184873310$^{*}$ & 664845308036464256 & 130.720922 & 20.819176 & $1.083^{+0.028}_{-0.029}$ & $4113^{+190}_{-180}$ & $1.72^{+0.05}_{-0.04}$ \\
  &  &  &  & TIC\,14503989$^{*}$ & 663386977661550336 & 125.459595 & 19.489010 & $0.958^{+0.004}_{-0.004}$ & $4255^{+20}_{-21}$ & $1.74^{+0.04}_{-0.06}$ \\
 Theia\,4877 & $720^{+400}_{-270}$ & $0.22$ & $1268^{+4}_{-4}$ & KIC\,8240902 & 2078327841249527168 & 294.484252 & 44.155876 & $0.896^{+0.004}_{-0.004}$ & $4437^{+50}_{-50}$ & $1.55^{+0.04}_{-0.04}$ \\
 NGC\,1901 & $820^{+120}_{-120}$ & $0.12$ & $416.34^{+0.25}_{-0.25}$ & TIC\,30935460$^{*}$ & 4661283304330138112 & 76.225559 & -68.685193 & $1.11^{+0.07}_{-0.08}$ & $3565^{+1200}_{-900}$ & $1.540^{+0.04}_{-0.029}$ \\
  &  &  &  & TIC\,179437576$^{*}$ & 4658745635794924160 & 80.047359 & -68.040575 & $1.1519^{+0.0006}_{-0.0005}$ & $3970^{+14}_{-14}$ & $1.60^{+0.04}_{-0.03}$ \\
 NGC\,6811 & $1100^{+200}_{-200}$ & $0.088$ & $1110.1^{+1.3}_{-1.3}$ & KIC\,9716563 & 2128134824634071424 & 294.410699 & 46.486817 & $0.9081^{+0.0009}_{-0.0008}$ & $3861^{+10}_{-10}$ & $1.453^{+0.04}_{-0.018}$ \\
  &  &  &  & KIC\,9716350 & 2128145166915272832 & 294.324900 & 46.462534 & $0.8631^{+0.0026}_{-0.0026}$ & $4079^{+60}_{-60}$ & $1.458^{+0.04}_{-0.018}$ \\
  &  &  &  & KIC\,9654789$^{*}$ & 2128119568910037632 & 294.119940 & 46.310836 & $1.1423^{+0.0009}_{-0.0009}$ & $3856^{+18}_{-19}$ & $1.474^{+0.04}_{-0.020}$ \\
  &  &  &  & KIC\,9655399$^{*}$ & 2128109089189944448 & 294.341681 & 46.347348 & $1.0263^{+0.0008}_{-0.0008}$ & $3899^{+15}_{-15}$ & $1.479^{+0.04}_{-0.020}$ \\
  &  &  &  & KIC\,9594007 & 2128118362018931328 & 294.203584 & 46.241467 & $1.6714^{+0.0012}_{-0.0012}$ & $3975^{+22}_{-22}$ & $1.481^{+0.04}_{-0.020}$ \\
  &  &  &  & KIC\,9655151 & 2128121080738610176 & 294.254905 & 46.383155 & $1.5523^{+0.0008}_{-0.0007}$ & $3823^{+15}_{-14}$ & $1.485^{+0.04}_{-0.020}$ \\
  &  &  &  & KIC\,9896743$^{*}$ & 2128152000202526592 & 294.233321 & 46.704713 & $1.6060^{+0.0012}_{-0.0013}$ & $4023^{+20}_{-19}$ & $1.500^{+0.04}_{-0.020}$ \\
  &  &  &  & KIC\,9716358 & 2128144376641265152 & 294.328495 & 46.401731 & $0.3059^{+0.0013}_{-0.0013}$ & $3835^{+27}_{-26}$ & $1.519^{+0.04}_{-0.020}$ \\
  &  &  &  & KIC\,9716107 & 2128145510512609920 & 294.236498 & 46.449388 & $1.8019^{+0.0012}_{-0.0012}$ & $3998^{+19}_{-19}$ & $1.545^{+0.04}_{-0.022}$ \\
  &  &  &  & KIC\,9776816 & 2128146644384064512 & 294.373671 & 46.510974 & $1.392^{+0.006}_{-0.006}$ & $4092^{+60}_{-60}$ & $1.589^{+0.04}_{-0.028}$ \\
  &  &  &  & KIC\,9533489 & 2080064515564300160 & 294.673745 & 46.122633 & $0.85^{+0.04}_{-0.05}$ & $4267^{+300}_{-300}$ & $1.589^{+0.04}_{-0.028}$ \\
 NGC\,6819 & $2300^{+300}_{-300}$ & $0.32$ & $2510^{+5}_{-5}$ & KIC\,5024455$^{*}$ & 2076298417671932800 & 295.320873 & 40.101100 & $1.646^{+0.001}_{-0.001}$ & $5043^{+5}_{-5}$ & $1.48^{+0.10}_{-0.10}$ \\
 NGC\,6791 & $8200^{+400}_{-400}$ & $0.65$ & $4189^{+19}_{-19}$ & KIC\,2438249$^{*}$ & 2051105479535706752 & 290.294602 & 37.758718 & $0.6682^{+0.0025}_{-0.0026}$ & $5432^{+100}_{-90}$ & $1.73^{+0.10}_{-0.10}$ \\
\enddata
    \tablecomments{Stars marked with an asterisk are newly reported in this work. Extinctions and distances are the median posterior values from \citet{Hunt2024} where available; the distance uncertainties span the 16th--84th percentiles. The values for Vela OB2 Huluwa and Perseus Alcaeus are from \citet{Pang2021_huluwa} and \citet{Olivares2023}, respectively. The Huluwa distance uncertainty is propagated from its tabulated parallax uncertainty. The Cep-Her extinctions are individual stellar values, while the listed distance and its propagated uncertainty describe the mean of the five Cep-Her stars, using the geometric distances adopted by \citet{Kerr2024_Cep_Her}.}
    \end{deluxetable*}
    \end{longrotatetable}

Our collection spans a wide age range. We have four stellar groups younger than $\sim 50\,\mathrm{Myr}$ (Alcaeus in Perseus, Huluwa in Vela OB2, and the Orpheus and Cinyras associations in Cep--Her), providing the opportunity to discover pre-main-sequence (PMS) g-mode pulsators.
Beyond the PMS phase, our sample extends from an age of $\sim 67\,\mathrm{Myr}$ (the Cupavo association in the Cep--Her complex) to $\sim 1100\,\mathrm{Myr}$ with NGC\,6811. 
Unfortunately, we could not collect any normal $\gamma$\,Dor pulsators in stellar groups with ages between $\sim 1100\,\mathrm{Myr}$ and $\sim 2300\,\mathrm{Myr}$. 
This age interval roughly corresponds to the epoch when many open clusters dissolve into the Galactic field, and when stars with masses of $\sim1.5\,\mathrm{M_\odot}$, typical of $\gamma$\,Dor stars, leave the main sequence. 
Interestingly, NGC\,6819 ($\sim2300\,\mathrm{Myr}$) and NGC\,6791 ($\sim8200\,\mathrm{Myr}$) show rich populations of blue stragglers located in the g-mode instability strip. We indeed found one g-mode pulsator with clear period-spacing patterns in each of the two clusters. Detailed descriptions of our PMS pulsators and blue-straggler pulsators are given in Sections~\ref{sec:PMS_pulsator} and \ref{sec:BSs}. 

\subsection{Adopted Ages of the Stellar Groups}\label{subsec:age_discussion}

Table~\ref{tab:cluster_gdor_sample} lists all \groupnumber stellar groups. We found substantial \added{discrepancies} between the ages given by \cite{Hunt2024} and the well-established ages in the literature. We therefore reviewed the literature and adopted the stellar-group ages \added{listed} in the second column of Table~\ref{tab:cluster_gdor_sample}. We provide the details of the adopted ages, in chronological order, in the following paragraphs.

\citet{Pavlidou2021} identified Alcaeus as a subgroup of the Perseus star-forming region and estimated its age to be around 5\,Myr. \citet{Kounkel2022} reported an age of 4.5\,Myr for Alcaeus. \citet{Olivares2023} found an age of 10\,Myr, while \citet{Pavlidou2026} most recently reported 5--8\,Myr. We therefore adopted $6.5\pm2.2\,\mathrm{Myr}$ for Alcaeus, calculated as the mean and standard deviation of the previous measurements.

For Vela~OB2, we adopted the Huluwa membership and age scale of \citet{Pang2021_huluwa}, who divided the structure into five subgroups, Huluwa~1--5, with ages of 10--22\,Myr for the older Huluwa~1--3 and 7--20\,Myr for the younger Huluwa~4--5. Within this framework, we adopted $14\pm8\,\mathrm{Myr}$ for Huluwa, encompassing the full age range reported by \citet{Pang2021_huluwa}. An independent spectroscopic and three-dimensional kinematic analysis by \citet{Armstrong2022_VelaOB2} adopted a different decomposition of the region, distinguishing the extended Vela~OB2 association from the $\gamma$\,Vel and P\,Puppis clusters, while deriving ages broadly consistent with the $10$--$20\,\mathrm{Myr}$ range. Thus, the Huluwa subgroup labels and ages used here should be understood specifically within the framework of \citet{Pang2021_huluwa}.

We identified five g-mode pulsators in three young associations within the Cep--Her complex. We assigned their ages following \citet{Kerr2024_Cep_Her}: $32\pm8\,\mathrm{Myr}$ for Orpheus, $36\pm8\,\mathrm{Myr}$ for Cinyras, and $67\pm13\,\mathrm{Myr}$ for Cupavo.

For CWNU\,1183 in the Kepler field, \citet{Ironi2025} derived an age of $48\pm6\,\mathrm{Myr}$ but noted that the age uncertainty may be underestimated. We therefore adopted the age from \citet{Hunt2024}, who reported a consistent result with a larger uncertainty, $46^{+40}_{-20}\,\mathrm{Myr}$.

Ages for some other clusters were adopted as follows. For NGC\,2516, we used $132\pm8\,\mathrm{Myr}$, measured by four g-mode pulsators simultaneously \citep{LiGang_2025_NGC2516_modelling}. The age was also supported by the \added{gyrochronological} age of this cluster by \cite{Fritzewski2020-NGC2516}. 
For the Pleiades, the most widely accepted age is $125\pm8\,\mathrm{Myr}$ from the lithium-depletion boundary \citep{Stauffer1998}, although this age is sensitive to both the membership list used and the underlying model physics, particularly the treatment of magnetic fields \citep{Dahm2015}. 
\cite{Fritzewski-Pleiades} pointed out that the g-mode properties of Pleiades member stars are similar to those of NGC\,2516, supporting a similar age. 
We adopted $225\pm75\,\mathrm{Myr}$ for UBC\,1, derived by asteroseismology and gyrochronology by \cite{Fritzewski2024UBC1}. 
\cite{Yazici2025} reported an isochrone-fitting age of $265\pm5\,\mathrm{Myr}$ for ASCC\,108. 
\cite{HeChenyu2025_NGC332} performed isochrone-fitting for NGC\,3532 and derived an age of $340^{+10}_{-20}\,\mathrm{Myr}$. 

Age measurements for NGC\,1901 are quite scattered. \citet{Hunt2024} reported an age of $283^{+220}_{-120}\,\mathrm{Myr}$, and \citet{Carraro2007_NGC1901} reported $400\pm100\,\mathrm{Myr}$. Older ages have also been suggested: \citet{Kos2018_NGC1901_age} reported $\log(\mathrm{age/yr})=8.84$ ($\sim692\,\mathrm{Myr}$) from Padova isochrone fitting, \cite{Bernizzoni2025_NGC1901} derived an age from gyrochronology of $820\pm120\,\mathrm{Myr}$, \citet{Cantat-Gaudin2020} reported $891\,\mathrm{Myr}$, and \cite{Malhotra2026_NGC1901} obtained $\log(\mathrm{age/yr})=8.987$ ($\sim971\,\mathrm{Myr}$). We adopted the \added{gyrochronological} age because our g-mode asteroseismic results favoured an older age (see the discussion in Section~\ref{subsec:Pi0_with_mass}). 

Age measurements for NGC\,2632 and NGC\,6866 remain highly model dependent. For NGC\,2632, \citet{Brandt2015} reported an age of $810\pm70\,\mathrm{Myr}$ after including rotation, compared with the previously well-accepted age of around 600\,Myr. In contrast, \citet{Gossage2018} reproduced an age of $610\pm25\,\mathrm{Myr}$ based on models with different initial rotation assumptions, as summarized in their Table~3. We therefore adopted an average age of $700\pm100\,\mathrm{Myr}$. For NGC\,6866, \citet{Wang2026_NGC6866} reported a PARSEC-isochrone age of $690^{+140}_{-30}\,\mathrm{Myr}$, a MESA-isochrone age of $467^{+70}_{-50}\,\mathrm{Myr}$, and a g-mode asteroseismic age of $759^{+54}_{-82}\,\mathrm{Myr}$. 
In addition, \citet{Brogaard2023} reported an age of $430\pm50\,\mathrm{Myr}$ based on red-giant oscillations. We thus adopted an average age of $590\pm170\,\mathrm{Myr}$ for NGC\,6866.

We did not find a dedicated age study for Theia\,4877, so we adopted the age from \citet{Hunt2024}, which is $720^{+400}_{-270}\,\mathrm{Myr}$. For NGC\,6811, we adopted an age of $1.1\pm0.2\,\mathrm{Gyr}$, measured using an eclipsing binary in the cluster by \citet{Sandquist2016} and consistent with \citet{Meibom2011}.

Several age estimates are available for NGC\,6819. \citet{Basu2011} reported an age of $2.4\pm0.4\,\mathrm{Gyr}$ based on 12 red-giant stars in NGC\,6819 and summarized that the seismic age is consistent with ages determined from several isochrone models. We also included the age measured from eclipsing binaries \citep{Brewer2016}, and the white-dwarf cooling-sequence age by \cite{Bedin2015}. Therefore, we adopted a mean age of $2.3\pm0.3\,\mathrm{Gyr}$ for NGC\,6819.

For NGC\,6791, solar-like asteroseismic age estimates include $7.9\pm1.0\,\mathrm{Gyr}$ from \citet{Basu2011} and $8.2\pm0.3\,\mathrm{Gyr}$ from \citet{McKeever2019}. Other estimates include an age of $8.3\pm0.3\,\mathrm{Gyr}$ from binary systems \citep{Brogaard2021_binary_age_NGC6791}, a white-dwarf cooling-sequence age of $8.0\pm0.2\,\mathrm{Gyr}$ \citep{Garcia_Berro_2010}, and a CMD age of $8.5\pm0.7\,\mathrm{Gyr}$ \citep{Dufresne2026}. We adopted the average value of $8.2\pm0.4\,\mathrm{Gyr}$ for NGC\,6791.

We also note that \citet{Bouma2023_gyrointerp} recovered \added{gyrochronological} ages of $117\pm6\,\mathrm{Myr}$ for the Pleiades, $306^{+11}_{-9}\,\mathrm{Myr}$ for NGC\,3532, $738^{+22}_{-20}\,\mathrm{Myr}$ for NGC\,2632/Praesepe, $946^{+16}_{-15}\,\mathrm{Myr}$ for NGC\,6811, and $2518^{+32}_{-33}\,\mathrm{Myr}$ for NGC\,6819 using the \texttt{gyro-interp} framework; these values are broadly consistent with the ages adopted in this work. 

\section{Asteroseismic Signal Extraction}\label{sec:signal_extraction}

In this section, we briefly outline our analysis of the g-mode asteroseismic signals, following procedures that have been applied to hundreds of g-mode pulsators in the Kepler and TESS data \citep[e.g.][]{VanReeth2015-method, VanReeth2016_TAR, Li2020MNRAS_611, Garcia2022}.

\subsection{Target Selection and CMD Construction}

To construct CMDs for each stellar group and select its early-type members, we downloaded Gaia $G$-band magnitudes and the colour index $G_{\rm BP}-G_{\rm RP}$ from the Gaia Archive\footnote{\url{https://gea.esac.esa.int/archive/}}, using the membership tables described above. We selected only early-type main-sequence stars for the subsequent asteroseismic analysis, because stars cooler than the red edge of the classical instability strip (roughly 7000\,K) do not exhibit classical pulsations, and any solar-like oscillations would remain undetectable given the long cadence and low amplitudes. This temperature criterion approximately corresponds to the well-known Kraft break \citep{Kraft1967, BeyerWhite2024}, where stars transition from having substantial convective envelopes to much thinner ones. We identified this criterion visually, as the slope of the main sequence changes slightly near this temperature.

To place all stellar-group members and $\gamma$\,Dor stars on a homogeneous intrinsic Gaia CMD, we corrected the Gaia photometry for interstellar extinction and reddening. For the 13 groups included in the catalogue of \citet{Hunt2024}, namely CWNU\,1183, the Pleiades, NGC\,2516, UBC\,1, ASCC\,108, NGC\,3532, NGC\,6866, NGC\,2632, Theia\,4877, NGC\,1901, NGC\,6811, NGC\,6819, and NGC\,6791, we adopted the median posterior extinction, $A_{V,50}$, and distance modulus, $\mu_{50}$, reported in that catalogue. We converted $A_V$ to the Gaia passbands using
\begin{align}
    A_G &= 0.857 A_V, \\
    E(G_{\rm BP}-G_{\rm RP}) &= 0.518 A_V,
\end{align}
following \citet{BauerFasching2024}. The extinction-corrected colour and absolute magnitude were then calculated as
\begin{align}
    (G_{\rm BP}-G_{\rm RP})_0
        &= (G_{\rm BP}-G_{\rm RP})
           - E(G_{\rm BP}-G_{\rm RP}), \\
    M_{G,0} &= G-A_G-\mu,
\end{align}
where $\mu=5\log_{10}(d/10\,{\rm pc})$. The same corrections were applied to the stellar-group members and the identified $\gamma$\,Dor stars.

For Huluwa, we adopted $A_V=0.35$~mag from \citet{Pang2021_huluwa} and used the individual stellar distances provided in its membership catalogue. For Alcaeus, we adopted $A_V=1.19$~mag, obtained as the median of the individual maximum-a-posteriori extinction estimates reported by \citet{Olivares2023}, and used their mean distance of 286~pc. For the Cep--Her complex, we used the per-star dereddened Gaia colour, extinction-corrected absolute $G$ magnitude, and distance estimates provided by \citet{Kerr2024_Cep_Her}; therefore, no additional cluster-wide extinction correction was applied.

\subsection{Light-curve Reduction}

For the Kepler data, we first queried the official mission light curves from the Mikulski Archive for Space Telescopes (MAST), and downloaded the simple-aperture-photometry flux (SAP\_FLUX) time series for our analysis. For stars in NGC\,6819 and NGC\,6791, we also used the image-subtraction light curves from the Kepler Increased Resolution Image Subtraction (IRIS) catalogue of \citet{Colman2022}, which were extracted from the Kepler superstamp images.

The light curve reduction for the TESS data was more complex. We extracted custom light curves from the full-frame images following the procedure described by \citet{Garcia2022}. Briefly, we downloaded $20\times20$ pixel TESS cutouts, defined target and background apertures on a median image, performed aperture photometry to obtain the simple-aperture-photometry (SAP) flux, corrected instrumental systematics using principal-component-analysis (PCA) cotrending, removed quality-flagged cadences and outliers, and finally stitched and normalised the multi-sector light curves. We also computed the contamination rate, defined as the fraction of flux from nearby stars entering the target aperture. For some targets, we manually adjusted the aperture to reduce the contamination as much as possible. We selected the background aperture to be at least six pixels away from the target centre, as we found that some large-amplitude variables could induce significant variations even in the background. 

The Kepler long-cadence light curves and TESS calibrated full-frame images used in this work were obtained from the Mikulski Archive for Space Telescopes (MAST).
The corresponding dataset DOIs are
\dataset[10.17909/T9488N]{https://doi.org/10.17909/T9488N}
for Kepler and
\dataset[10.17909/0cp4-2j79]{https://doi.org/10.17909/0cp4-2j79}
for TESS. The compiled stellar data used in this work are available at \dataset[10.17909/sq4y-sy72]{https://doi.org/10.17909/sq4y-sy72}.

\subsection{Frequency Extraction}

For the Kepler data, we used the full four-year light curve as input to our frequency-extraction code. For the TESS data, to mitigate systematic uncertainties and window-function effects, we selected the light curves with the longest continuous TESS coverage, avoiding sparse temporal coverage (e.g., short segments separated by multi-year gaps). The TESS sectors used for the frequency extraction are listed in Table~\ref{tab:tess_sectors}.

\begin{table}
\centering
\caption{TESS sectors used for the frequency extraction.}
\label{tab:tess_sectors}
\begin{tabular}{lll}
\hline
Cluster & TIC & TESS sector(s) \\
\hline
Cep-Her & TIC303754397 & 15, 16 \\
Cep-Her & TIC346151391 & 53, 54 \\
Cep-Her & TIC277011700 & 14, 15, 16 \\
Cep-Her & TIC277570969 & 82, 83, 84 \\
Cep-Her & TIC349011395 & 53, 54 \\
NGC2632 & TIC175239414 & 44, 45, 46 \\
NGC2632 & TIC184873310 & 44, 45, 46 \\
NGC2632 & TIC14503989 & 44, 45, 46 \\
NGC1901 & TIC179437576 & Years 1, 3, and 5 \\
NGC1901 & TIC30935460 & Years 1, 3, and 5 \\
Huluwa & TIC56640851 & 88, 89 \\
Alcaeus & TIC26532226 & 42, 43, 44 \\
\hline
\end{tabular}
\vspace{0.5em}
\begin{minipage}{0.95\linewidth}
\footnotesize
\textit{Note.} For published cluster studies, details of the TESS sectors used and the frequency extraction are provided in \cite{LiGang_2024_NGC2516} for NGC\,2516, \cite{Ligang_2026_NGC3532} for NGC\,3532, \cite{Fritzewski-Pleiades} for Pleiades, and \cite{Fritzewski2024UBC1} for UBC\,1.
\end{minipage}
\end{table}

We adopted the iterative pre-whitening algorithm described by \citet{Li2019_splitting_gdor}. In each iteration, we calculated the amplitude spectrum, identified the highest peak, and fitted a sinusoidal function to the light curve within a narrow range around that peak in frequency and amplitude. The residuals between the best-fitting sinusoid and the original light curve were then used as input for the next iteration. 
We stopped when no peak exceeded a signal-to-noise ratio (SNR) of four, with the noise level calculated within a $\pm2.5\,\mathrm{d^{-1}}$ window.

\subsection{Traditional Approximation of Rotation (TAR) Fit}

After extracting all frequencies from the light curves, we focused on the low-frequency region ($\lesssim 5\,\mathrm{d^{-1}}$) to identify g modes. Period spacings ($\Delta P$) are defined as the period differences between adjacent g-mode overtones with the same angular degree $\ell$ and azimuthal order $m$ (i.e., $\Delta n = 1$). In the asymptotic limit, the period spacing is constant with radial order for a non-rotating and chemically homogeneous star \citep{Shibahashi1979, Tassoul1980}. However, rotation causes the period spacing to vary quasi-linearly with period \citep{Bouabid2013, VanReeth2016_TAR, Ouazzani2017}. We therefore searched for period-spacing patterns showing approximately linear variations of $\Delta P$ with period, following the approach of \citet{Li2019_splitting_gdor}.

We next assigned $(\ell, m)$ to the identified period-spacing patterns. \citet{Saio2018} showed that the most common g-mode patterns are sectoral modes, which means those with $\ell=m$ (e.g., $\ell=m=1$ and $\ell=m=2$). Here we use the convention that prograde modes have positive $m$ values. In addition, we measured the slopes of the period-spacing patterns and compared them with the results of \citet{Li2020MNRAS_611}, since different modes form distinct groups in the diagram of pattern slope versus mean period. 
Both $\ell=1$ and $\ell=2$ sectoral modes show decreasing period spacing with increasing period. 
The most common modes are expected to be $\ell=1$ and $m=1$. When we found two period-spacing patterns where one had roughly half the mean period of the other, we identified the shorter-period pattern as $\ell=2$, $m=2$ g modes. 
In some stars, we also detected Rossby modes. These are pure inertial modes with the Coriolis force as the restoring force. 
In the observer's reference frame, they appear at about 1.5 times the $\ell=1$, $m=1$ g-mode period, showing increasing period spacings and a distinctive amplitude profile \citep[e.g.][]{VanReeth2015ApJS, Saio2018_r_modes, Li2019_r_mode}.

After identifying each observed period-spacing pattern, we applied the TAR to fit the data. We refer to \citet{VanReeth2016_TAR}, \citet{Saio2018}, and \citet{Rui2024} for summaries of the theory and its practical applications in asteroseismology. In short, the period spacing is calculated as
\begin{equation}
  P_{n} = \frac{\Pi_0}{\sqrt{\lambda_{\ell,m,s}}}\left(n + \alpha_\mathrm{g}\right),
\end{equation}
where $\lambda_{\ell,m,s}$ is the eigenvalue of Laplace's tidal equation, and $s = 2f_\mathrm{rot}/f_\mathrm{co}$ is the spin parameter (twice the ratio of the near-core rotation rate to the co-rotating pulsation frequency). We fixed the g-mode phase $\alpha_\mathrm{g}$ to 0.5. 
The asymptotic spacing $\Pi_0$ is defined as $\Pi_0 = 2\pi^2\left[\int_\mathrm{gc} (N\,\mathrm{d}r/r)\right]^{-1}$, 
where $N$ is the Brunt--V\"ais\"al\"a frequency and the integral is taken over the g-mode cavity (gc). 
Since $\Pi_0$ generally decreases as the central hydrogen abundance decreases, it provides an asteroseismic constraint on the evolutionary stage; in combination with spectroscopic constraints, it can be used to estimate stellar ages \citep{Mombarg2019}. 
For a more detailed description, we refer to a series of related papers, such as \cite{VanReeth2015-method}, \cite{Li2020MNRAS_611}, and \cite{Ligang_2026_NGC3532}. 

Using this TAR framework, we fitted the observed period-spacing patterns to infer the near-core rotation rate $f_\mathrm{rot}$ and asymptotic spacing $\Pi_0$ for each target; the resulting measurements are summarised in Section~\ref{sec:overall_look}.

\section{Overview of the Seismic Measurements}\label{sec:overall_look}

\begin{figure*}[t]
  \centering
  \includegraphics[width=1\textwidth]{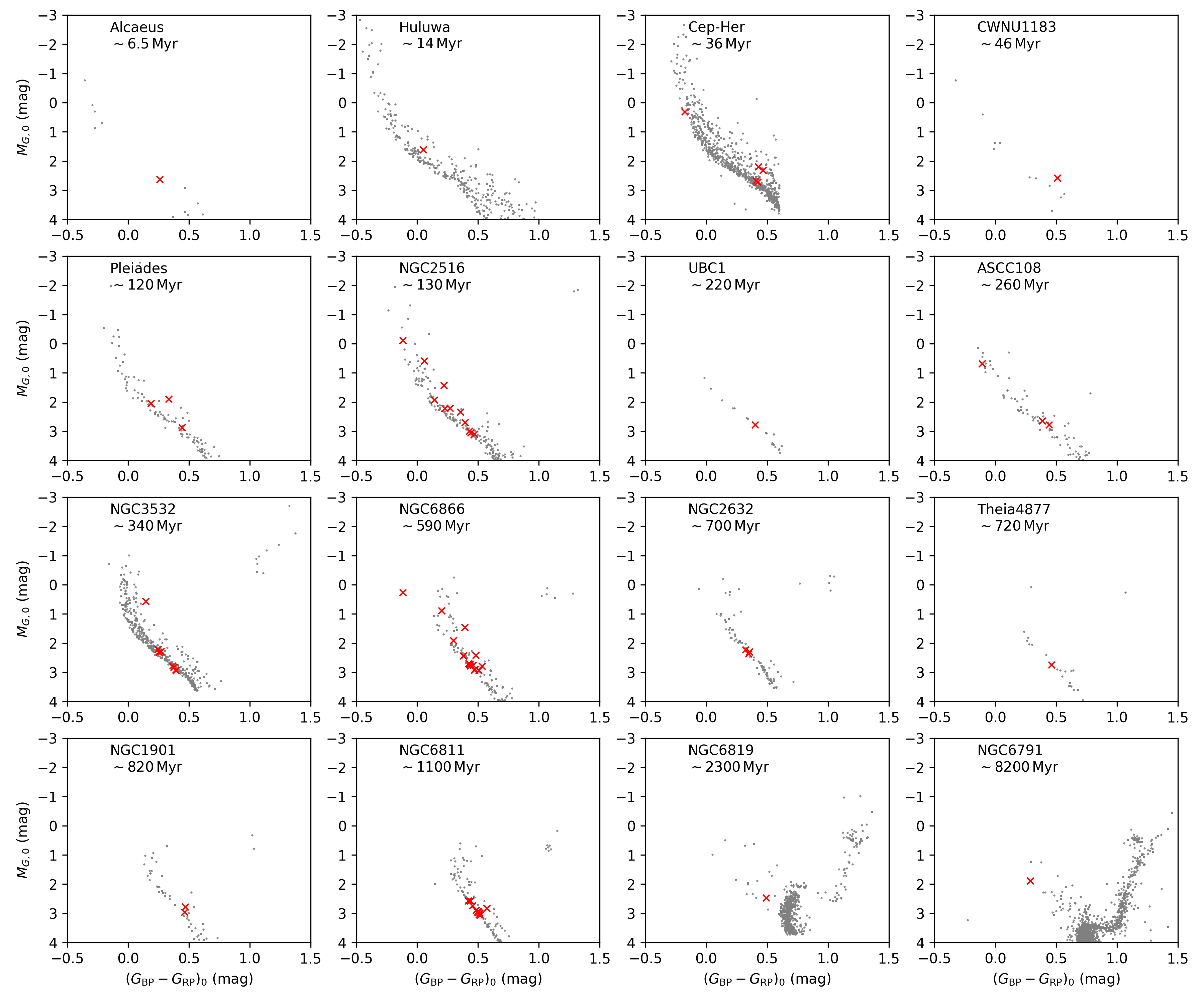}
  \caption{Colour--magnitude diagrams of the \groupnumber stellar groups in this work. The axes show the extinction-corrected intrinsic Gaia colour $(G_\mathrm{BP}-G_\mathrm{RP})_0$ and absolute magnitude $M_{G,0}$. The grey dots indicate stellar-group members. The red crosses mark the g-mode pulsators with clear period-spacing patterns.}
  \label{fig:all_cluster}
\end{figure*}

Table~\ref{tab:cluster_gdor_sample} lists all g-mode pulsators with clear period-spacing patterns in our sample of \groupnumber stellar groups. Figure~\ref{fig:all_cluster} shows the CMDs of all \groupnumber stellar groups in this work, highlighting the locations of the g-mode pulsators. 
In total, we identified \starnumber such pulsators. For the Kepler clusters, most targets were already reported in the 611-$\gamma$\,Dor sample of \citet{Li2020MNRAS_611}, demonstrating the completeness of that catalogue. 
\added{By cross-matching the Kepler target list with the Gaia DR3 membership catalogues of newly identified clusters in the Kepler field, we added seven additional Kepler targets to our sample. }
For clusters observed by TESS, we incorporated our previous studies of NGC\,2516 \citep{LiGang_2024_NGC2516}, the Pleiades \citep{Fritzewski-Pleiades}, UBC\,1 \citep{Fritzewski2024UBC1}, and NGC\,3532 \citep{Ligang_2026_NGC3532}. 
In addition, we added g-mode pulsators in the Huluwa subgroups of Vela~OB2 and the associations within Cep--Her, as well as in the open clusters NGC\,1901 and NGC\,2632 (the Beehive cluster).

\subsection[Near-core Rotation Rates]{Near-core Rotation Rates $f_\mathrm{rot}$}\label{subsec:frot}

\begin{figure*}
  \centering
  \includegraphics[width=1\linewidth]{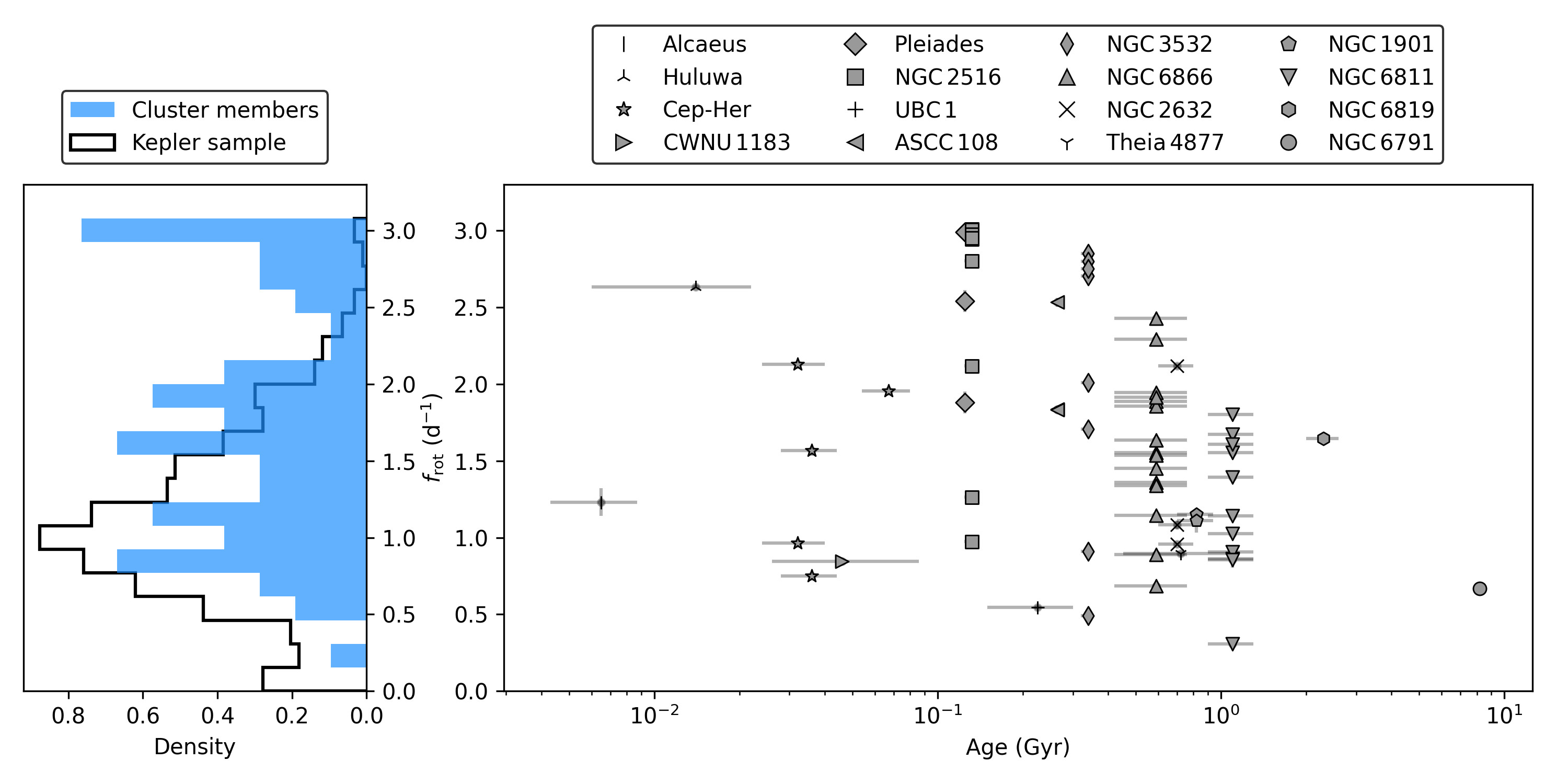}
  \caption{Left: distributions of near-core rotation rates for stellar-group members in this work and for Kepler field stars from \citet{Li2020MNRAS_611}. Right: near-core rotation rate as a function of stellar age for the stellar-group sample. Different stellar groups are indicated by the symbols shown in the legend at the top.}
  \label{fig:frot_age}
\end{figure*}

The left panel of Fig.~\ref{fig:frot_age} shows the distribution of near-core rotation rates for the stellar-group members in this work. \added{This distribution is distinct} from that of the Kepler field stars reported by \citet{Li2020MNRAS_611}. The stellar-group members show three peaks: a high-rotation peak above 2.5\,$\mathrm{d^{-1}}$, dominated by the Pleiades, NGC\,2516, and NGC\,3532; an intermediate peak between $1.5$ and $2.0$\,$\mathrm{d^{-1}}$; and a peak around $1.0$\,$\mathrm{d^{-1}}$, mainly populated by older stellar groups with ages comparable to or greater than that of NGC\,6866. 
By comparison, the Kepler field stars show a dominant peak around $1.0$\,$\mathrm{d^{-1}}$ and a slow-rotation plateau below $0.5$\,$\mathrm{d^{-1}}$.

The fact that the Kepler field stars show slower near-core rotation rates than the stellar-group members in this work suggests that Kepler g-mode pulsators may, on average, be older than NGC\,6866, implying that the excitation mechanism may preferentially operate in older main-sequence stars.
However, we emphasise that this comparison is subject to observational biases.
One source of observational bias arises from the TESS light curves. Due to their short time baselines, the frequency resolution $\delta f$ is limited; consequently, the observations are less sensitive to the long-period modes exhibited by slow rotators because the period resolution decreases as $\delta P = P^{2}\,\delta f$.
Another observational bias is the incompleteness of the stellar-group sample. Our sample contains few stellar groups older than $1\,\mathrm{Gyr}$, partly because of sample incompleteness and partly because unbound associations disperse while dynamical evolution preferentially dissolves low-mass open clusters over time. Open-cluster dissolution timescales depend strongly on cluster mass and environment, ranging from less than $1\,\mathrm{Gyr}$ for low-mass systems to several Gyr for more massive clusters \citep{Lamers2006, Krumholz2019}.

We do not see a slow-rotation plateau in our stellar-group sample comparable to that found in the Kepler data \citep[$f_\mathrm{rot}<0.5\,\mathrm{d^{-1}}$,][see also Fig.~\ref{fig:frot_age}]{Li2020MNRAS_611}. Binary interactions are a possible pathway for producing slowly rotating g-mode pulsators, through mechanisms such as inverse tides \citep{Li2020_gdor_in_EB, Fuller2021, Fellay2026} or Cassini-state trapping \citep{Fuller2024}. In bound-cluster environments, dynamical encounters preferentially disrupt wide binaries, whereas tightly bound (or ``hard'') binaries tend to become even more tightly bound \citep{Heggie1975, Hills1975}. This process could in principle enhance the role of close binaries in producing slow rotators. The paucity of slow rotators in our sample is therefore more plausibly explained by observational selection effects: the shorter TESS time baselines provide poorer frequency and period resolution than the four-year Kepler data, making slowly rotating g-mode patterns harder to identify.

The right panel of Fig.~\ref{fig:frot_age} shows the evolution of near-core rotation rates with host-group age. Although the rotation rates are broadly distributed, the upper envelope shows an increasing trend up to $\sim100\,\mathrm{Myr}$, followed by a gradual decline at older ages. The young stars in Huluwa, CWNU\,1183, and Cep--Her have rotation rates ranging from $\sim0.5\,\mathrm{d^{-1}}$ to $\sim2.5\,\mathrm{d^{-1}}$. Around $\sim100\,\mathrm{Myr}$, the stars show the fastest rotation rates, reaching $\sim3.0\,\mathrm{d^{-1}}$ in NGC\,2516 and the Pleiades. Thereafter, the maximum rotation rate decreases with age, dropping to $\sim2.0\,\mathrm{d^{-1}}$ at $1\,\mathrm{Gyr}$ in NGC\,6811. As stated above, our observations are more sensitive to fast rotators; therefore, the upper envelope should be observationally complete. Observed surface-rotation distributions of solar-like stars in young clusters also show a qualitatively similar spin-up, which \citet{Gallet2013} attributed to PMS contraction following disk dispersal (see their Fig.~1).

\subsection[Mass-dependent Evolution of Near-core Rotation Rates]{Mass-dependent Evolution of Near-core Rotation Rates $f_\mathrm{rot}$}\label{subsec:near_core_rotation_rate_with_mass}

\begin{figure*}[!t]
  \centering
  \begin{minipage}[t]{0.7\textwidth}
    \vspace{0pt}
    \includegraphics[width=\linewidth]{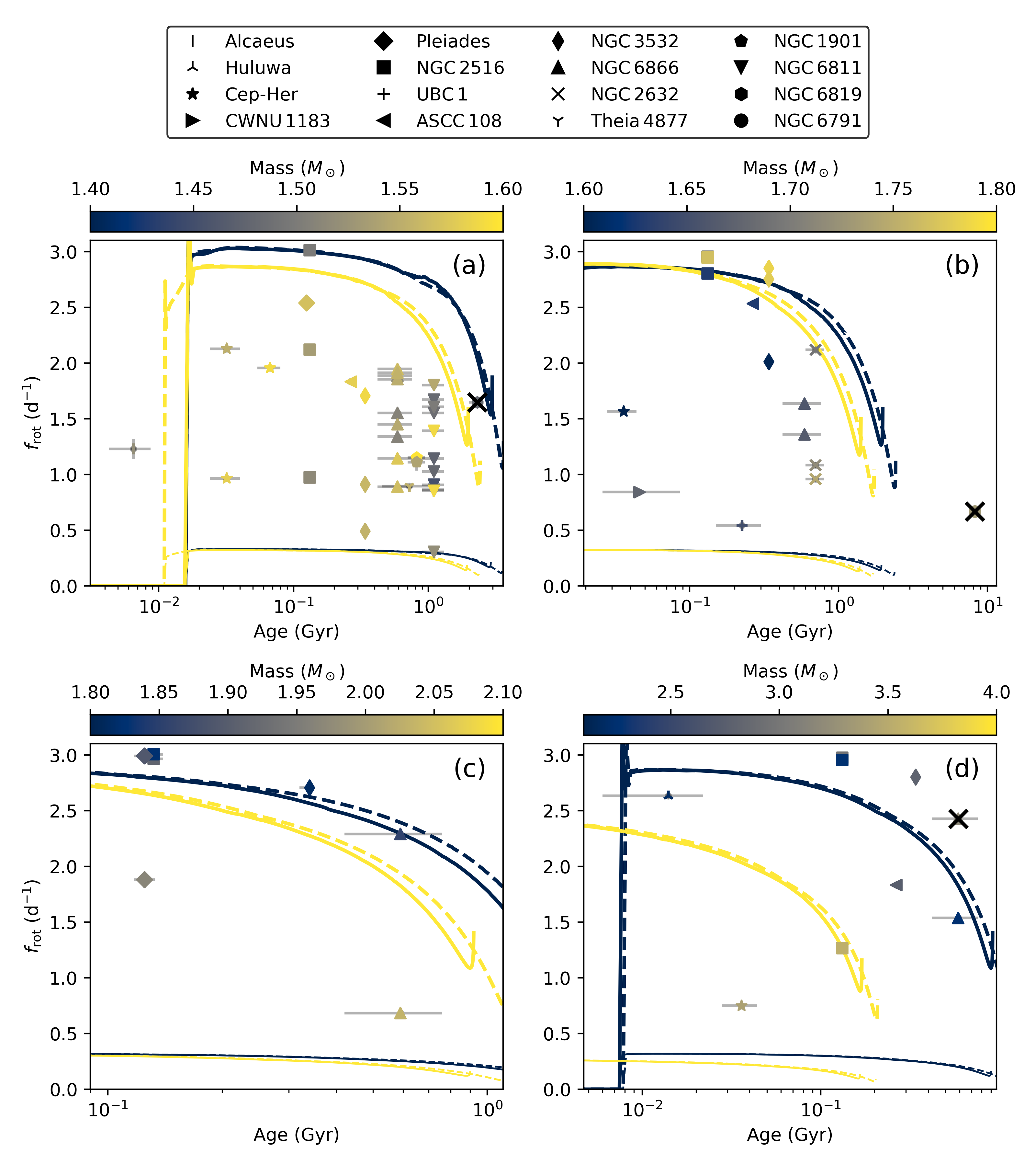}
  \end{minipage}
  \hfill
  \begin{minipage}[t]{0.25\textwidth}
    \vspace{0pt}
    \caption{Near-core rotation rates $f_\mathrm{rot}$ of our g-mode stellar-group members in different mass ranges. The \added{symbols are} the observed stars, with shapes indicating the stellar-group names, as in Fig.~\ref{fig:frot_age}. \added{Panels (a)--(d) correspond, respectively, to mass ranges of $[1.4, 1.6]$, $[1.6, 1.8]$, $[1.8, 2.1]$, and $[2.1, 4.0]$ $\mathrm{M_\odot}$.} The black crosses mark the three blue stragglers in NGC\,6866, NGC\,6819, and NGC\,6791. The lines show theoretical models. The line thickness, colour, and linestyle encode the input parameters: initial rotation, mass, and overshooting, respectively. The thin and thick lines correspond to models with initial rotation rates of $0.05\,\Omega_\mathrm{crit}$ and $0.55\,\Omega_\mathrm{crit}$, respectively. The black lines show models at the lower boundary of the mass range in each bin, while the blue lines show models at the upper mass boundary. The solid lines represent models with an overshooting parameter of $f_\mathrm{CBM}=0.005$, while the dashed lines have $f_\mathrm{CBM}=0.025$. }
    \label{fig:frot_age_mass_bin}
  \end{minipage}
\end{figure*}

In this section, we compare our observed near-core rotation rates, $f_\mathrm{rot}$, with stellar evolution models. Because these stars are members of coeval stellar groups, their ages are known and their masses can be estimated from their locations in the CMDs, enabling a direct comparison with theoretical models. We used the g-mode asteroseismic models of \citet{Mombarg2024AA_14000Gaia}. We refer to previous papers for a more detailed description of the input physics of these models, including the treatments of helium enrichment \citep{Verma2019}, metallicity \citep{Nieva2012}, overshooting, rotational effects on stellar structure \citep{Paxton2013ApJS, Paxton2019ApJS}, angular-momentum transport \citep{Mombarg2023calibrating_AM}, and element mixing in radiative layers.

We briefly summarize the rotation treatment used by \citet{Mombarg2024AA_14000Gaia}. Rotation is turned off during the pre-main-sequence phase and is switched on when the model reaches the zero-age main sequence, defined by the criterion that the nuclear luminosity exceeds 90\% of the total luminosity. At this stage, the model is assigned an initial rotation rate (we use $5\%$ or $55\%$ of the critical rotation rate $\Omega_\mathrm{crit}$ in Fig.~\ref{fig:frot_age_mass_bin}) and is allowed to relax to equilibrium. This explains the sudden rise in the model rotation rate in \added{panels (a) and (d)} of Fig.~\ref{fig:frot_age_mass_bin}. It also explains why, in Section~\ref{sec:PMS_pulsator} and Fig.~\ref{fig:pre-main-sequence-model}, we needed to recalculate some models when discussing rotational effects during the pre-main-sequence phase. The centrifugal correction to the stellar structure is applied using the shellular approximation \citep[e.g.][]{Endal1976, Meynet1997, Paxton2013ApJS}. AM transport is implemented in a fully diffusive framework by assuming a constant viscosity of $\nu = 10^{7}\,\mathrm{cm^2\,s^{-1}}$, calibrated using several slowly rotating $\gamma$\,Dor stars \citep{Mombarg2023calibrating_AM}. This prescription does not enforce rigid rotation, but the relatively high viscosity leads to efficient AM transport and suppresses strong radial differential rotation. \cite{Moyano2023} suggested that combining near-core rotation rates with reliable projected surface rotation velocities ($v\sin i$) would provide a stronger test of the internal AM-transport efficiency.

Figure~\ref{fig:frot_age_mass_bin} displays the near-core rotation rates, $f_\mathrm{rot}$, as a function of stellar age in different mass bins. We also overplot the theoretical rotation curves for multiple choices of stellar-model input parameters, as described in the caption. From Fig.~\ref{fig:frot_age_mass_bin}, we find that the observed rotation-rate distribution depends strongly on stellar mass. For stars with masses below $1.6\,\mathrm{M_\odot}$ (\added{panel (a)} of Fig.~\ref{fig:frot_age_mass_bin}), most targets lie well below the rotation rates predicted by the models with $55\%\,\Omega_\mathrm{crit}$, indicating that they are moderate rather than rapid rotators. For stars between $1.6$ and $1.8\,\mathrm{M_\odot}$ (\added{panel (b)} of Fig.~\ref{fig:frot_age_mass_bin}), the fraction of slow rotators decreases. At masses above $1.8\,\mathrm{M_\odot}$, most stars rotate rapidly, with $f_\mathrm{rot}$ following the $55\%\,\Omega_\mathrm{crit}$ curve or lying even higher.

This mass-dependent rotation distribution has also been reported in NGC\,2516 and NGC\,3532 \citep{LiGang_2024_NGC2516, Ligang_2026_NGC3532}. In these two clusters, with ages of $\sim100$ and $\sim300$\,Myr, respectively, g-mode pulsators with masses below $1.6\,\mathrm{M_\odot}$ also rotate more slowly than the more massive pulsators in the same clusters.
The mass-dependent rotation rate matches the surface rotation rates reported previously for stars in the same mass range \citep{Royer2007}.
Our larger sample also suggests that an additional mechanism may contribute to the spin-down of stars in this mass range.
One possible explanation is residual magnetic braking near the Kraft-break transition, where the presence of surface convective envelopes may allow magnetised winds to remove AM, although the efficiency is expected to decrease as these envelopes become thinner \citep[e.g.][]{Kraft1967, BeyerWhite2024, Wang_Wang_Ong_2026ApJ}. Stars around $1.6\,\mathrm{M_\odot}$ are expected to have shallow convective envelopes, which are also important for the excitation of $\gamma$\,Dor g modes \citep{Dupret2005}, so some level of magnetic braking may still operate. %Check Conny's comments, hard to identify the handwriting
We note, however, that \citet{BeyerWhite2024} placed the transition between slow and fast rotators at $1.32$--$1.41\,\mathrm{M_\odot}$. The lower-mass pulsators discussed in this work are therefore not slow rotators in an absolute sense; rather, they rotate more slowly relative to the more massive stars above $1.6\,\mathrm{M_\odot}$ in our sample.

\subsection[Asymptotic Spacing]{Asymptotic Spacing $\Pi_0$}\label{subsec:Pi0}
\begin{figure*}[!t]
  \centering
  \includegraphics[width=1\linewidth]{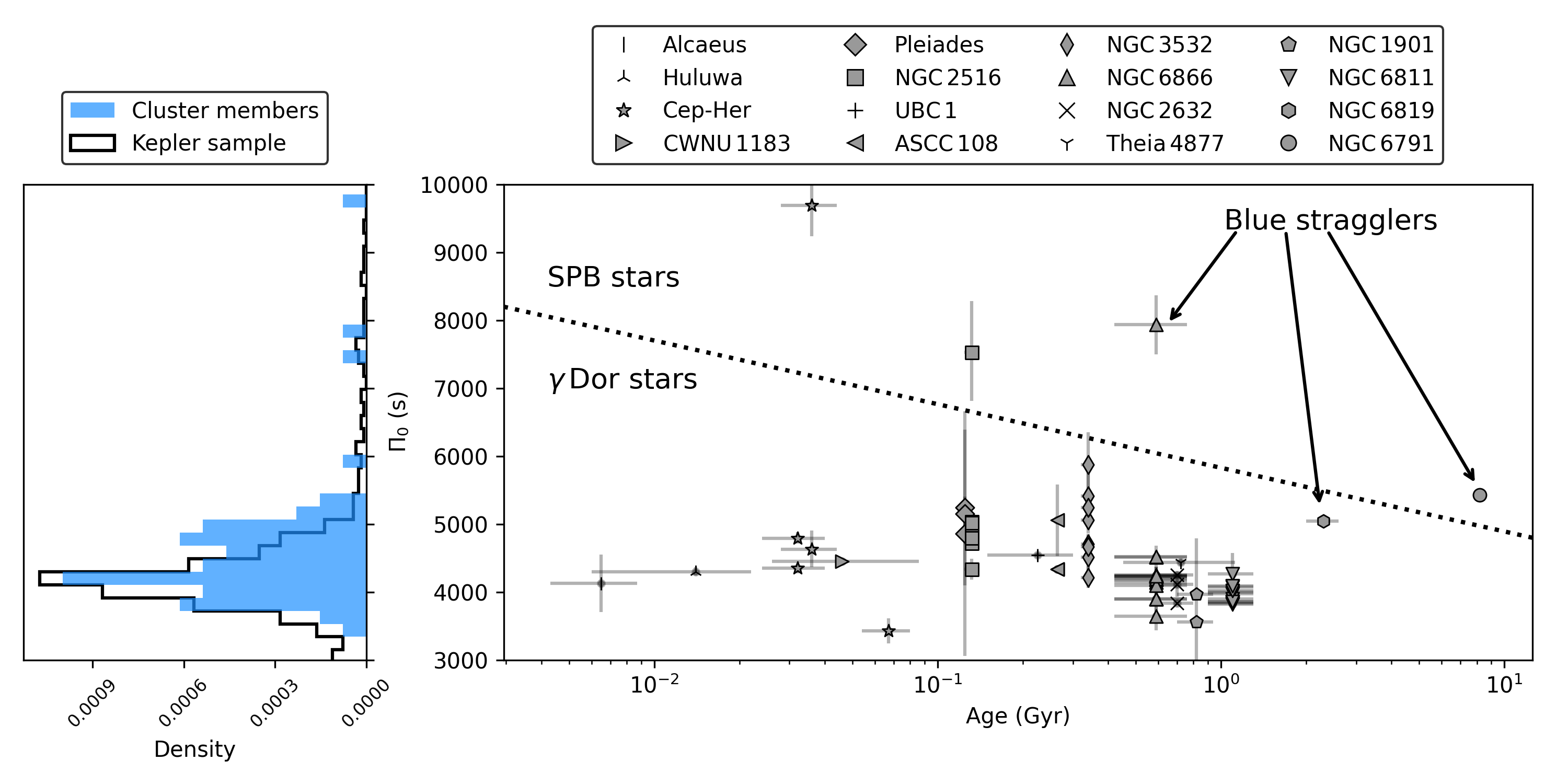}
  \caption{Same as Figure~\ref{fig:frot_age}, but showing the distributions of the asymptotic spacing $\Pi_0$ in this work and in \citet{Li2020MNRAS_611}, as well as $\Pi_0$ as a function of stellar age. The dotted line is an approximate guide drawn by eye to separate the SPB and $\gamma$\,Dor regimes; it is not intended to represent a precisely defined boundary.}
  \label{fig:Pi0_age}
\end{figure*}

The left panel of Fig.~\ref{fig:Pi0_age} shows the distribution of the asymptotic spacing $\Pi_0$ for the stellar-group members. We also compare our results with those for Kepler field stars from \citet{Li2020MNRAS_611}. We find that the stellar-group members have larger $\Pi_0$ values than the field stars. 
Compared with the Kepler sample, our stellar-group sample shows an excess of pulsators with $\Pi_0$ values from $\sim 4500\,\mathrm{s}$ to $\sim 5500\,\mathrm{s}$, mainly due to stellar groups younger than $300\,\mathrm{Myr}$, and a paucity of pulsators with $\Pi_0$ values below $\sim 3800\,\mathrm{s}$. As $\Pi_0$ decreases slightly with evolution, it is often used as an age indicator, implying that our stellar-group sample is younger than the field-star sample, probably because older stellar groups are underrepresented in our sample.

The right panel of Fig.~\ref{fig:Pi0_age} shows the evolution of $\Pi_0$ with host-group age. First, we find that most stars have $\Pi_0 \lesssim 6000\,\mathrm{s}$. Based on their locations on the CMDs, these stars typically have masses below $2.0\,\mathrm{M_\odot}$ and therefore fall within the typical parameter range of $\gamma$\,Dor stars. We also find that $\Pi_0$ increases slightly near the beginning of the main sequence up to an age of $\sim300\,\mathrm{Myr}$ and then declines slightly at older ages. The models of \citet{LiGang_2025_NGC2516_modelling} indicate that $\Pi_0$ can also increase during the early main-sequence evolution of stars with masses above approximately $2\,\mathrm{M_\odot}$. However, we note that differences in metallicity and initial stellar mass can also affect $\Pi_0$ \citep{Ouazzani2019A&A}. 

Pre-main-sequence stars or very young main-sequence stars may still undergo rapid structural changes, which could lead to a fluctuation in $\Pi_0$. We provide a brief discussion of pre-main-sequence g-mode pulsators in Section~\ref{sec:PMS_pulsator}. In addition, we find five stars with much larger $\Pi_0$ values. Two of them, located in Cep--Her and NGC\,2516, have larger masses inferred from their CMD positions and are therefore classified as SPB stars. We identify the remaining three stars as blue stragglers in NGC\,6866, NGC\,6819, and NGC\,6791, and discuss them in Section~\ref{sec:BSs}.
%Check Conny's comments between page 10 and 11, ask Conny about that

\subsection[Mass-dependent Evolution of Asymptotic Spacing]{Mass-dependent Evolution of Asymptotic Spacing $\Pi_0$}\label{subsec:Pi0_with_mass}

\begin{figure*}[!t]
  \centering
  \begin{minipage}[t]{0.7\textwidth}
    \vspace{0pt}
    \includegraphics[width=\linewidth]{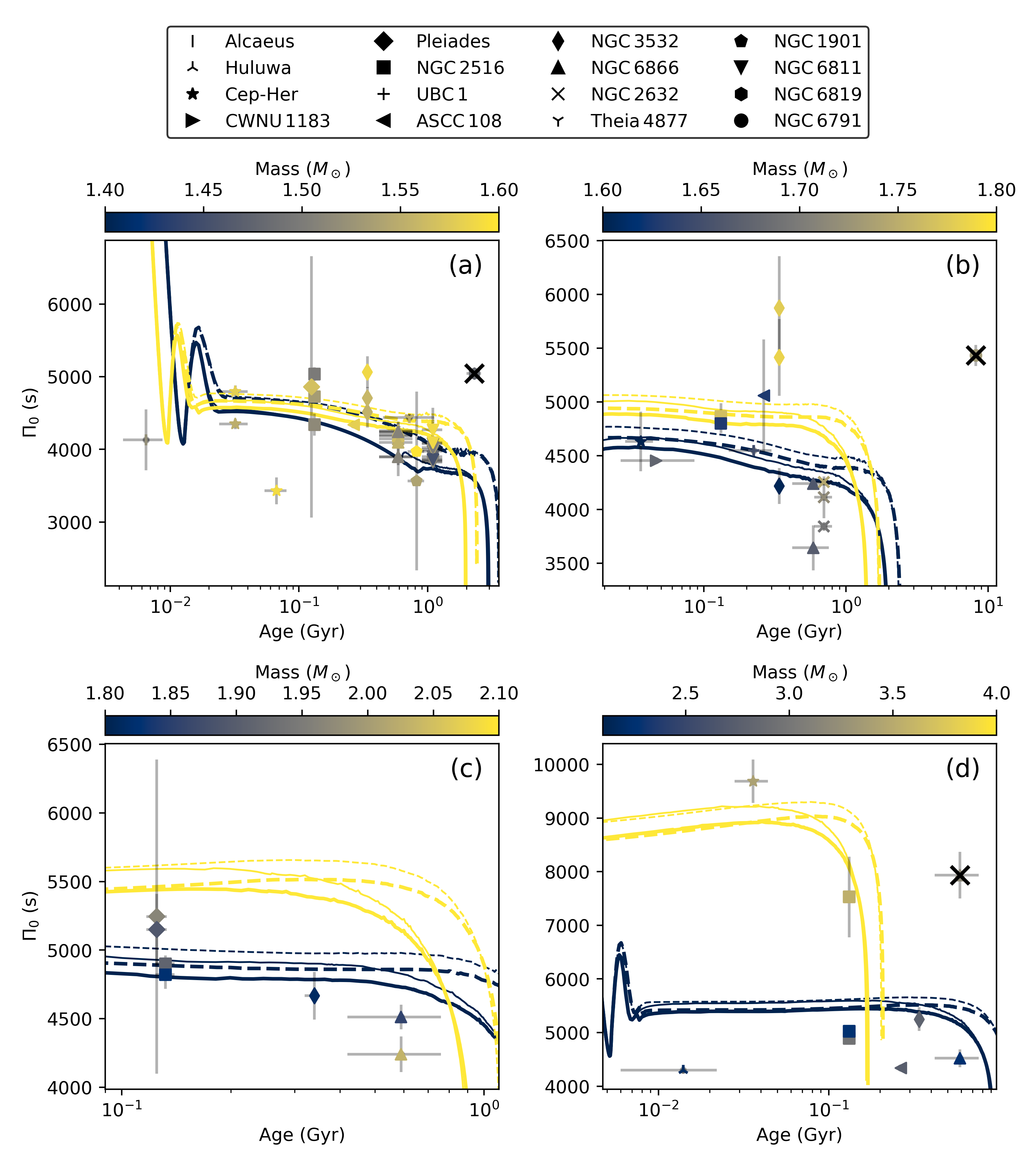}
  \end{minipage}
  \hfill
  \begin{minipage}[t]{0.25\textwidth}
    \vspace{0pt}
    \caption{Same as Fig.~\ref{fig:frot_age_mass_bin} but with asymptotic spacing $\Pi_0$ as the $y$-axis. \added{Panels (a)--(d) correspond, respectively, to mass ranges of $[1.4, 1.6]$, $[1.6, 1.8]$, $[1.8, 2.1]$, and $[2.1, 4.0]$ $\mathrm{M_\odot}$.}}
    \label{fig:Pi0_age_mass_bin}
  \end{minipage}
\end{figure*}

Analogous to Section~\ref{subsec:near_core_rotation_rate_with_mass} and Fig.~\ref{fig:frot_age_mass_bin}, we examine the evolution of $\Pi_0$ in different mass bins. Figure~\ref{fig:Pi0_age_mass_bin} shows the observed $\Pi_0$ as a function of age for four mass ranges: $1.4$--$1.6$, $1.6$--$1.8$, $1.8$--$2.1$, and $2.1$--\added{$4.0\,\mathrm{M_\odot}$}. We also overplot the theoretical $\Pi_0$ values computed from the same model grid of \citet{Mombarg2024AA_14000Gaia}.

In each panel of Fig.~\ref{fig:Pi0_age_mass_bin}, we show two sets of theoretical models: one at the lower boundary of the mass range (black curves) and one at the upper boundary (blue curves). Because $\Pi_0$ generally increases with mass for stars that are not too close to the terminal-age main sequence, the observed $\Pi_0$ values are expected to lie between these two sets of theoretical curves.

In \added{panel (a)} of Fig.~\ref{fig:Pi0_age_mass_bin} (the $1.4$--$1.6\,\mathrm{M_\odot}$ mass range), the theoretical curves reproduce the observed trend well. In this mass range, excluding the pre-main-sequence stars and the blue straggler, $\Pi_0$ decreases from $\sim4500\,\mathrm{s}$ near the beginning of the main sequence to $\sim4000\,\mathrm{s}$ near the terminal-age main sequence. Therefore, $\Pi_0$ can be considered as a good age indicator in this mass range. As mentioned in Section~\ref{subsec:age_discussion}, the two g-mode pulsators in NGC\,1901, shown as hexagons, have masses around $1.6\,\mathrm{M_\odot}$ and $\Pi_0$ values below $4000\,\mathrm{s}$. If we adopt an age of $820\,\mathrm{Myr}$ for this cluster, as suggested by the \added{gyrochronological} result of \cite{Bernizzoni2025_NGC1901}, the observed $\Pi_0$ values agree with the theoretical predictions. By contrast, adopting an age of $\sim400\,\mathrm{Myr}$ or younger \citep{Hunt2024, Carraro2007_NGC1901} would place the two stars in clear disagreement with the models, which predict $\Pi_0$ values larger than those observed at such ages.

The situation becomes more complex in the higher-mass bins. Starting from $1.6\,\mathrm{M_\odot}$ (\added{panels (b), (c), and (d)} of Fig.~\ref{fig:Pi0_age_mass_bin}), we find more outliers: several stars have $\Pi_0$ values that lie outside of the theoretical predictions. A straightforward explanation for this $\Pi_0$ discrepancy is that the mass estimates for higher-mass stars are too crude. These stars rotate rapidly, so gravity darkening is non-negligible and can shift their positions significantly in the CMD \citep[e.g.][]{Espinosa_Lara2011, Bouchaud2020, Murphy_2026}. However, \cite{Murphy_2026} suggests a mass uncertainty of only $\sim0.1\,\mathrm{M_\odot}$, which is insufficient to explain the observed $\Pi_0$ discrepancy. Differences between the metallicities adopted in the models and those of the observed stars could also contribute to this discrepancy. Nevertheless, most clusters in our sample with available spectroscopic metallicity measurements have approximately solar metallicities. Alternatively, the discrepancy may point to missing physics in the models, such that the surface parameters ($T_\mathrm{eff}$ and $L$) cannot be fully reconciled with the internal seismic diagnostic $\Pi_0$.

We also find that all three blue stragglers appear as outliers in the $\Pi_0$--age diagrams in Fig.~\ref{fig:Pi0_age_mass_bin}. We discuss these pulsating blue stragglers in detail in Section~\ref{sec:BSs}.

\subsection{Inferring Initial Rotation Distributions from Evolved Populations}
The known ages of the stellar groups allow us to infer the initial rotation distributions through forward population modelling. Following \cite{Mombarg2026_calibrating_AM}, we grouped the stellar groups listed in Table~\ref{tab:cluster_gdor_sample} into four age bins, excluding the blue stragglers. The bins were chosen to include as many stars as possible while retaining a relatively narrow age range. For each bin, we generated synthetic populations of 1000 stars at both the youngest and oldest ages of the bin. We drew stellar masses from the observed mass distribution and adopted the probability density function (PDF) of the core-boundary mixing scale ($f_{\rm CBM}$) from \cite{MombargMathis2026}. The initial rotation rate at the zero-age main sequence, expressed as a fraction of the critical rotation frequency, was drawn from a normal distribution with mean $\mu_{\omega_0}$ and standard deviation $\sigma_{\omega_0}$. We then evolved the sampled models to the selected ages and constructed the predicted distributions of near-core rotation frequency and $\Pi_0$. Thus, the distributions shown below represent the evolved populations at the boundaries of each age bin, rather than the initial rotation distributions themselves.

For each age bin, we optimized the free parameters $\mu_{\omega_0}$ and $\sigma_{\omega_0}$ by maximizing the likelihood. To compute this likelihood, we first obtained a kernel density estimate (KDE) from the rotation frequencies of the simulated population. The log-likelihood, $\mathcal{L}$, was then given by
\begin{equation}
  \mathcal{L} = \sum_i \log({\rm KDE}(f_{{\rm rot},i})),
\end{equation}  
where $f_{{\rm rot},i}$ are the observed rotation frequencies. 

\begin{figure}
    \centering
    \centering
    \begin{subfigure}{0.49\textwidth}
        \centering
        \includegraphics[width=0.85\linewidth]{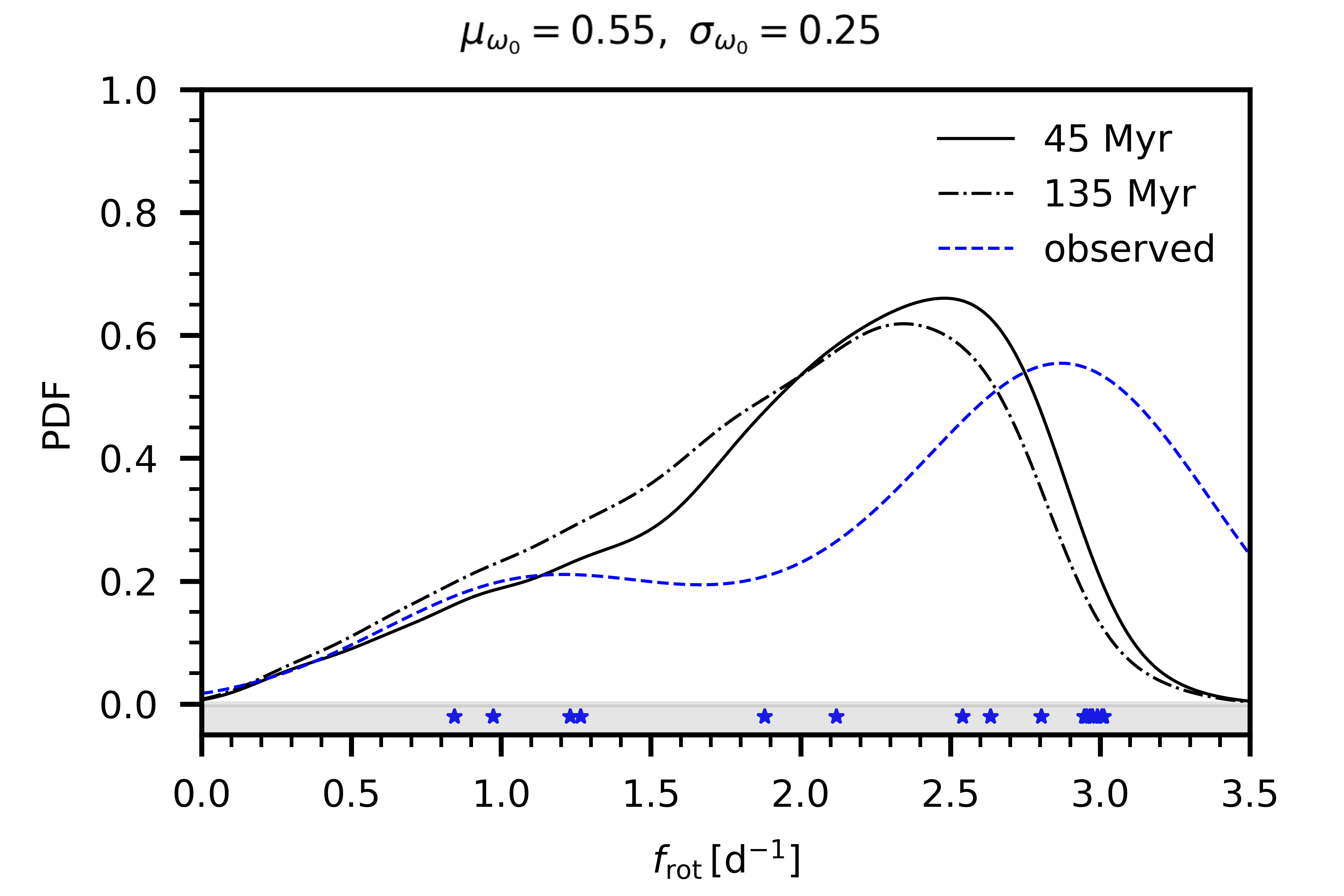}
        \caption{Alcaeus, Huluwa, CWNU\,1183, Pleiades, NGC\,2516}
    \end{subfigure}
    \hfill
        \centering
    \begin{subfigure}{0.49\textwidth}
        \centering
        \includegraphics[width=0.85\linewidth]{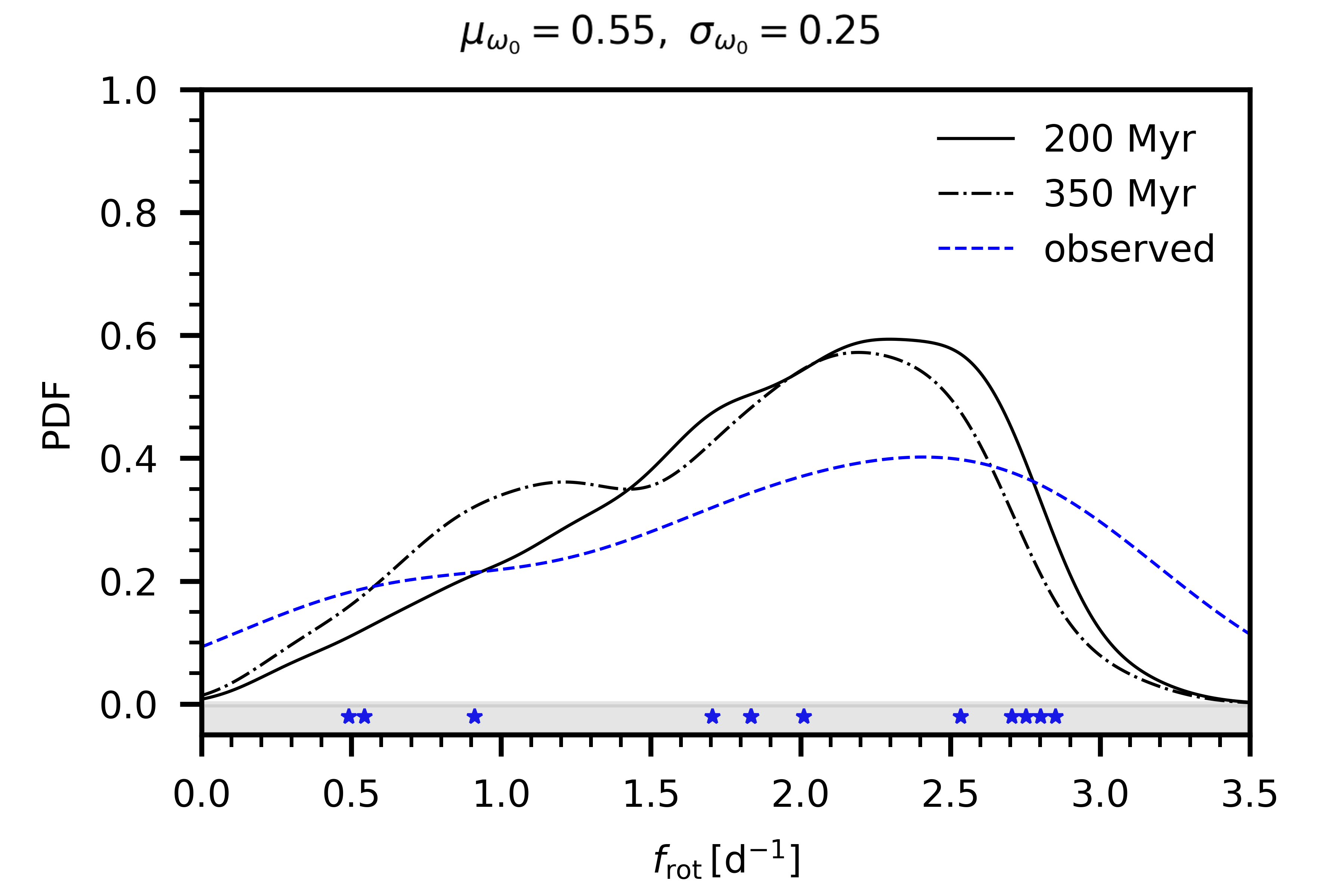}
        \caption{UBC\,1, ASCC\,108, NGC\,3532}
    \end{subfigure}
    \hfill
        \centering
    \begin{subfigure}{0.49\textwidth}
        \centering
        \includegraphics[width=0.85\linewidth]{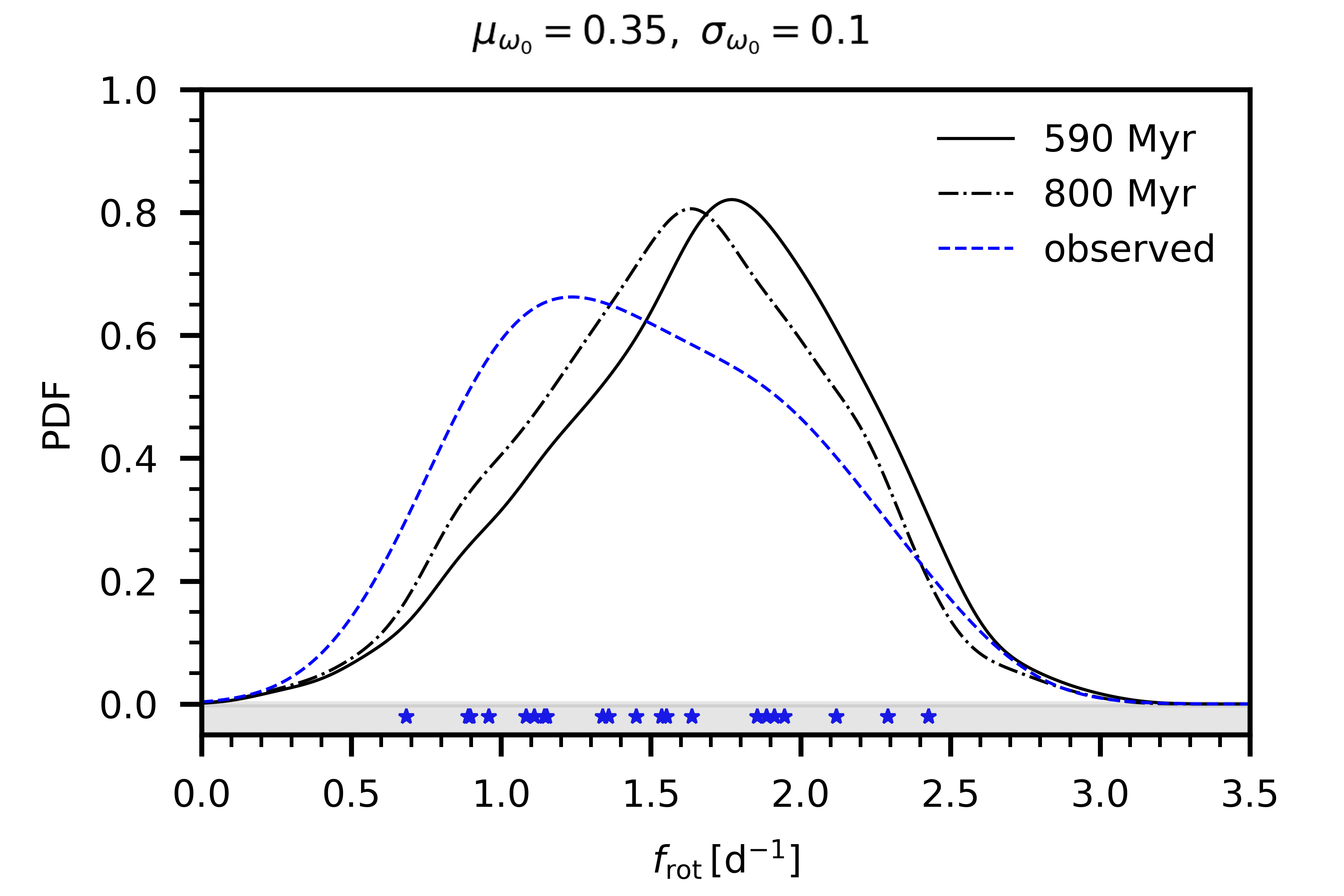}
        \caption{NGC\,1901, NGC\,2632, Theia\,4877, NGC\,6866}
    \end{subfigure}
    \hfill
    \begin{subfigure}{0.49\textwidth}
        \centering
        \includegraphics[width=0.85\linewidth]{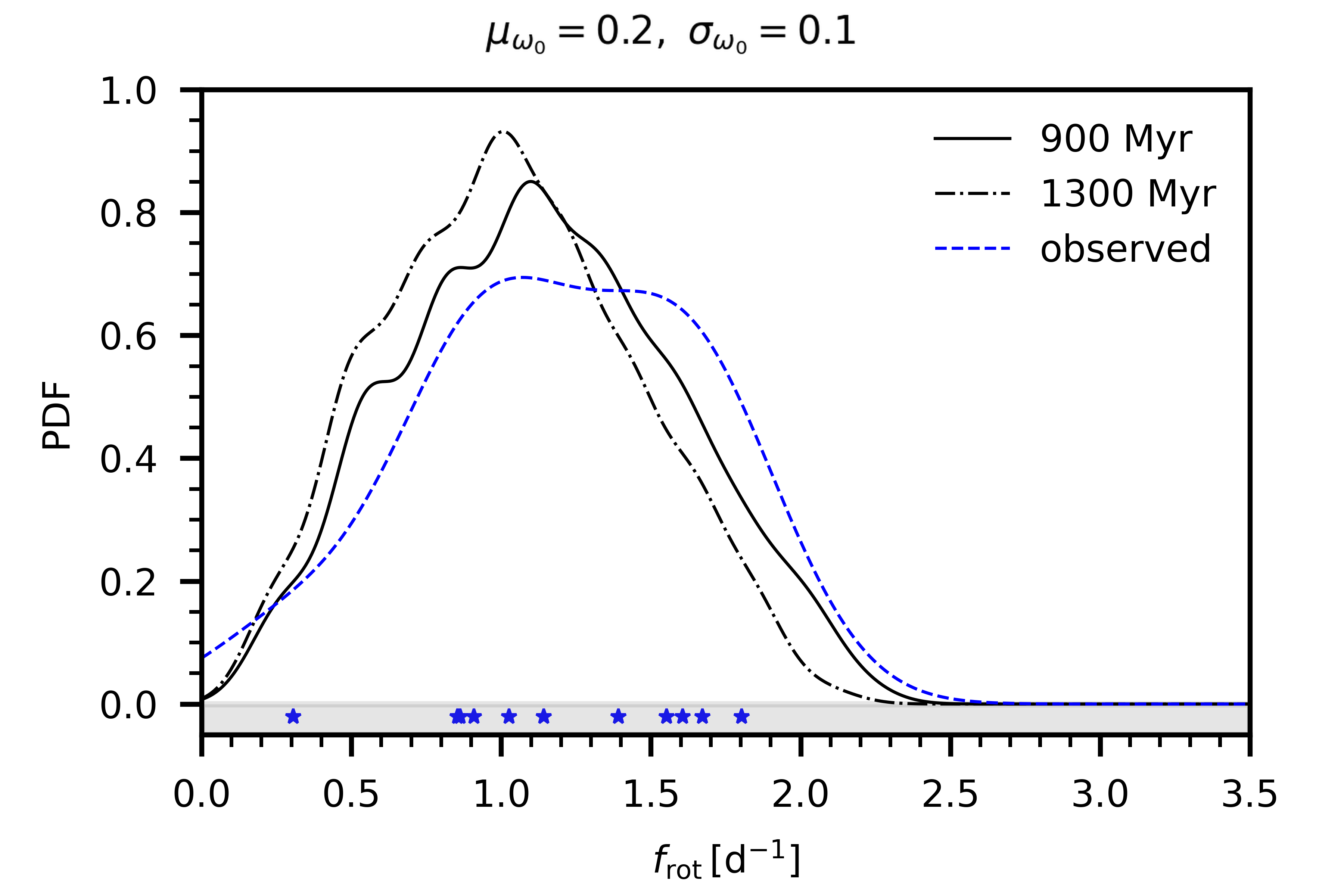}
        \caption{NGC\,6811}
    \end{subfigure}
    \caption{Predicted and observed rotation-frequency distributions in four stellar-group age bins. Although NGC\,1901 is slightly older than 800\,Myr, we included it in the third age bin. Star symbols indicate the individual measurements.}
    \label{fig:frot_pdfs}
\end{figure}

We divided the stellar groups into four age bins: 45--135\,Myr (22 stars), 200--350\,Myr (12 stars), 590--800\,Myr (21 stars), and 900--1300\,Myr (11 stars). \added{Figure~\ref{fig:frot_pdfs} compares the observed near-core rotation-frequency distributions with the simulated distributions corresponding to the maximum-likelihood values of $(\mu_{\omega_0}, \sigma_{\omega_0})$.} Because the model near-core rotation rates generally decrease over most of the main-sequence evolution (Fig.~\ref{fig:frot_age_mass_bin}), the predicted distributions shift towards lower rotation frequencies with increasing age, producing a larger fraction of slow rotators in the older populations. In general, the observed distributions agree with our predictions. However, for the two youngest bins, the large number of stars with rotation frequencies around $3\,\mathrm{d^{-1}}$ would require initial rotation rates above the upper limit of the grid, namely 55\% of the critical rotation rate. The synthetic populations therefore do not fully represent the rapidly rotating end of these young samples, and the inferred initial-rotation parameters are limited by the boundary of the model grid. With the treatment of rotation adopted in our models, most stars above this threshold reach critical rotation on the main sequence. The mismatch may therefore indicate either substantial differential rotation or limitations in the adopted rotational-evolution prescription. For the two older bins, we inferred mean initial rotation rates of 20--30\% of the critical rotation rate, in agreement with the results of \cite{Mombarg2024} and \cite{Mombarg2026_calibrating_AM}. 
\added{\cite{Mombarg2026_calibrating_AM} applied this population-modelling approach to the field-star sample analysed by \cite{Aerts2026}, using the normalized central hydrogen mass fraction, $X_{\rm c}/X_{\rm ini}$, as a proxy for age. Here, we group the cluster stars by the independently determined ages of their host clusters. Despite this difference in binning, the observed near-core rotation frequencies show a qualitatively similar decline with age in the field and cluster samples.}
We also note that the g-mode pulsators in the younger stellar groups have a higher average mass than those in the older stellar groups. This difference partly reflects the shorter main-sequence lifetimes of higher-mass stars, which have already evolved off the main sequence in the older groups.

\begin{figure}
    \centering
    \centering
    \begin{subfigure}{0.49\textwidth}
        \centering
        \includegraphics[width=0.85\linewidth]{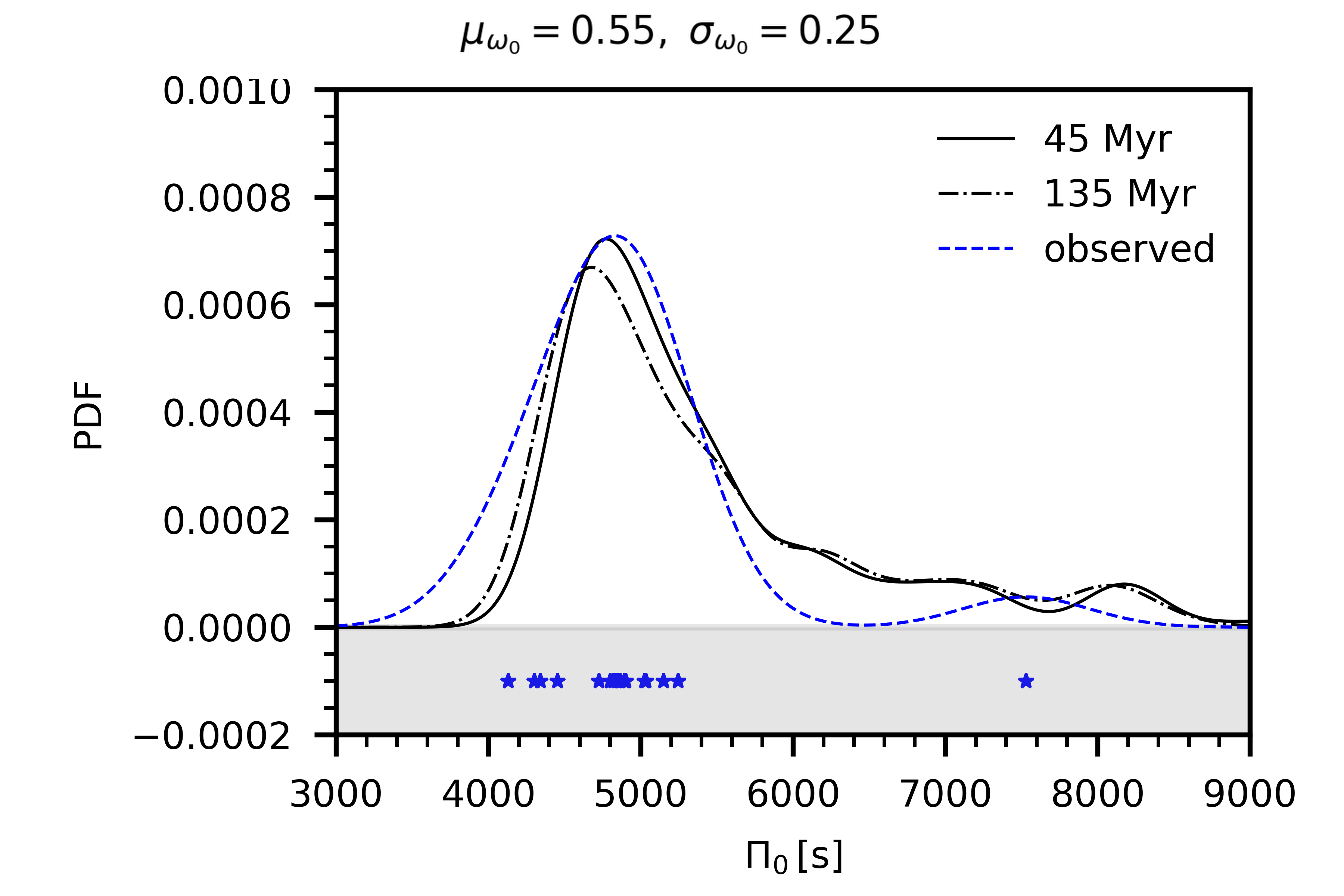}
        \caption{Alcaeus, Huluwa, CWNU\,1183, Pleiades, NGC\,2516}
    \end{subfigure}
    \hfill
        \centering
    \begin{subfigure}{0.49\textwidth}
        \centering
        \includegraphics[width=0.85\linewidth]{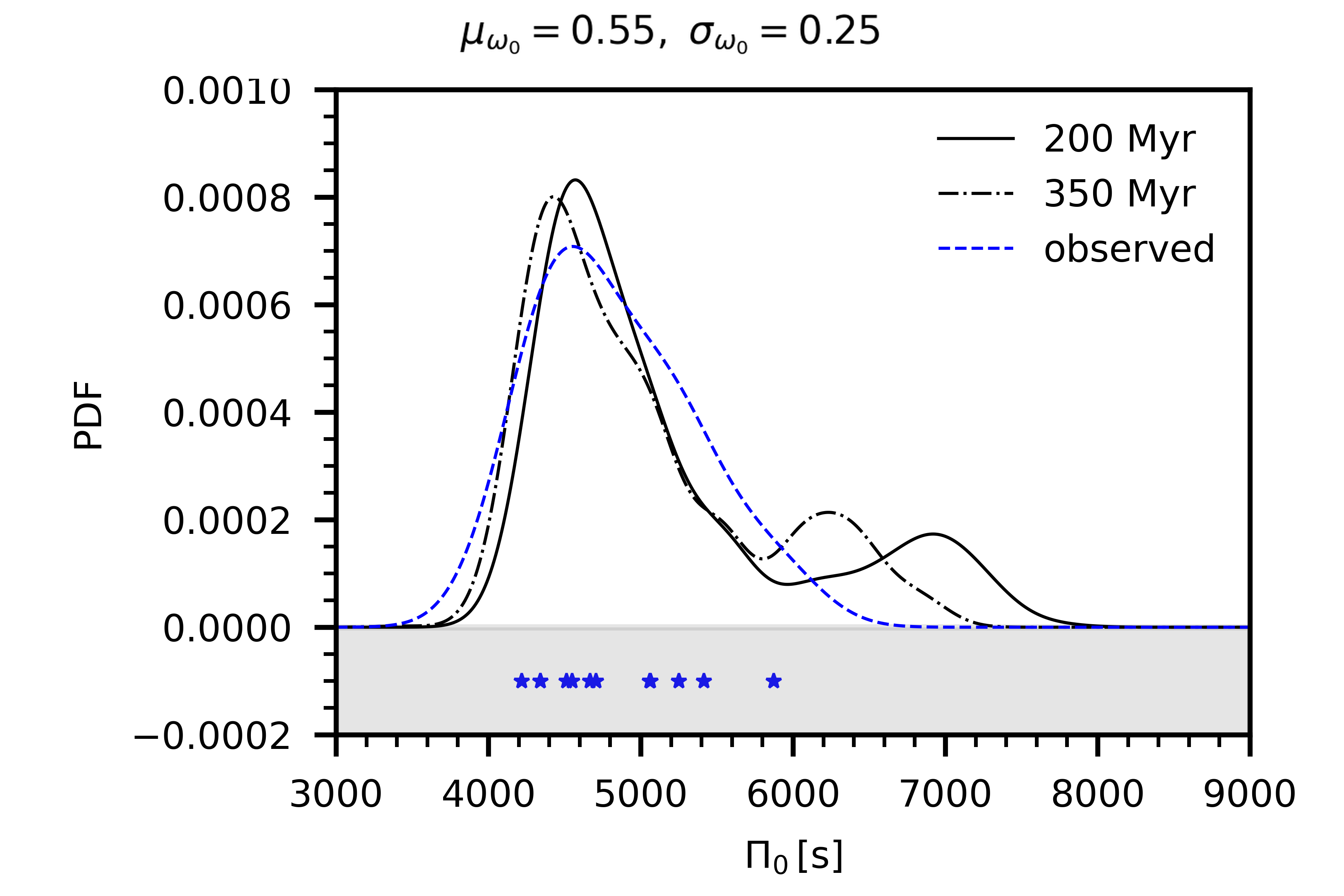}
        \caption{UBC\,1, ASCC\,108, NGC\,3532}
    \end{subfigure}
    \hfill
        \centering
     \begin{subfigure}{0.49\textwidth}
        \centering
        \includegraphics[width=0.85\linewidth]{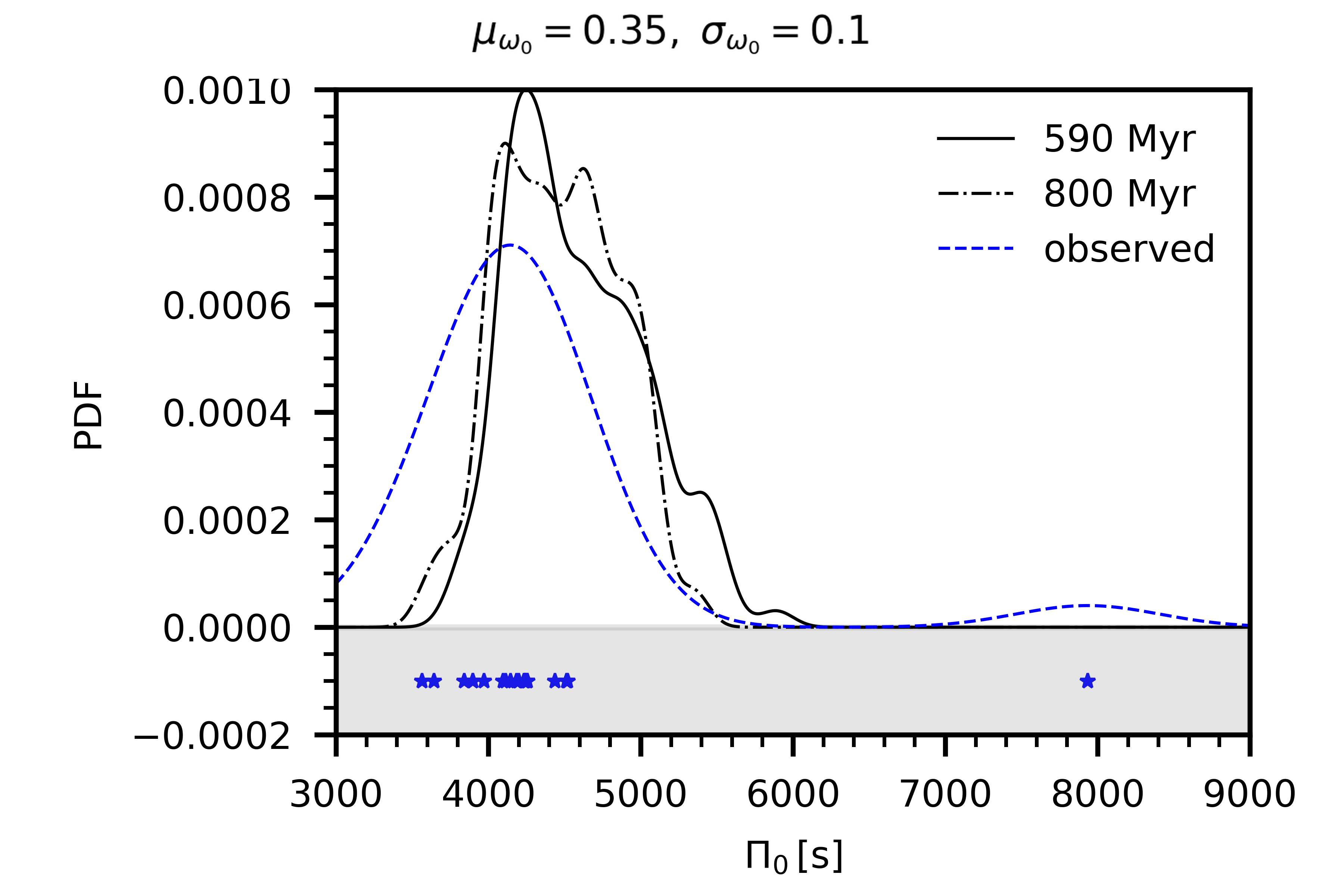}
        \caption{NGC\,1901, NGC\,2632, Theia\,4877, NGC\,6866}
    \end{subfigure}
    \hfill
    \begin{subfigure}{0.49\textwidth}
        \centering
        \includegraphics[width=0.85\linewidth]{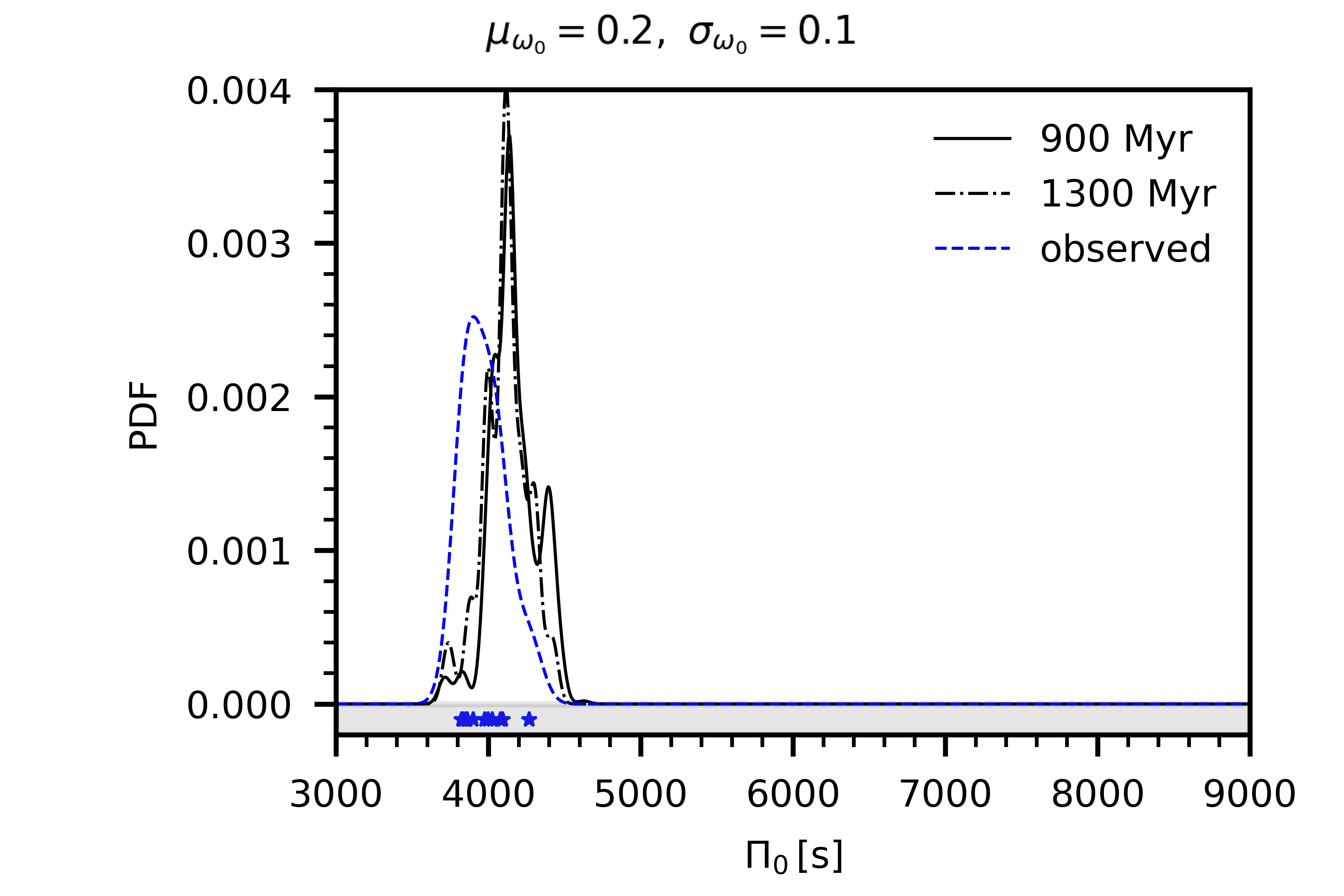}
        \caption{NGC\,6811}
    \end{subfigure}
    \caption{Predicted and observed $\Pi_0$ distributions in four stellar-group age bins. Star symbols indicate the individual measurements.}
    \label{fig:Pi0_pdfs}
\end{figure}

Similarly, Fig.~\ref{fig:Pi0_pdfs} shows the predicted and observed distributions of $\Pi_0$. We find good agreement for the two youngest bins, whereas for the two older bins, the observed $\Pi_0$ values are lower than predicted by approximately $200\,\mathrm{s}$ on average. Less efficient chemical mixing in the radiative envelope could reduce $\Pi_0$ at a given age. However, our models already assume low levels of mixing, making this explanation unlikely. The oldest bin contains only NGC\,6811, so the discrepancy may partly reflect an underestimated age for this cluster.

\section{Pre-main-sequence g-mode pulsators}\label{sec:PMS_pulsator}

Pre-main-sequence g-mode pulsators have long been predicted to exist. \citet{Bouabid2011_PMS} found that the instability strip of pre-main-sequence $\gamma$\,Dor stars spans nearly the same temperature range as that of main-sequence $\gamma$\,Dor stars. \citet{Steindl2021_PMS_IS} used an accretion model and found that an appropriate choice of the mass accretion rate and the fraction of injected accretion energy can reproduce the observed instability strip of pre-main-sequence pulsators. 

Pre-main-sequence g-mode pulsators have been discovered by the CoRoT and TESS missions, but period spacings have generally not been reported \citep[e.g.][]{Zwintz2013_PMS_gdor_CoRoT, Steindl2021_PMS_IS}. Owing to the different internal chemical stratification (e.g. the lack of a strong chemical gradient), the period-spacing patterns of pre-main-sequence g-mode stars are expected to be smoother than those of main-sequence g-mode pulsators. One PMS pulsator with a period-spacing pattern detected in K2 data was reported by \citet{Ketzer2020_PMS_gdor_K2}. Meanwhile, KIC\,9850387 was initially proposed to be a PMS g-mode pulsator with a clear period-spacing pattern \citep{Zhang2020_KIC9850387}, but subsequent studies disagreed \citep{Sekaran2020, Sekaran2021_9850387}. Recently, Wadhwa et al. (in prep.) discovered the g-mode pulsator TIC\,26532226, which shows a clear period-spacing pattern, in the Perseus complex; this star is included in Table~\ref{tab:cluster_gdor_sample}.

\begin{figure}[h]
  \centering
  \includegraphics[width=0.46\textwidth]{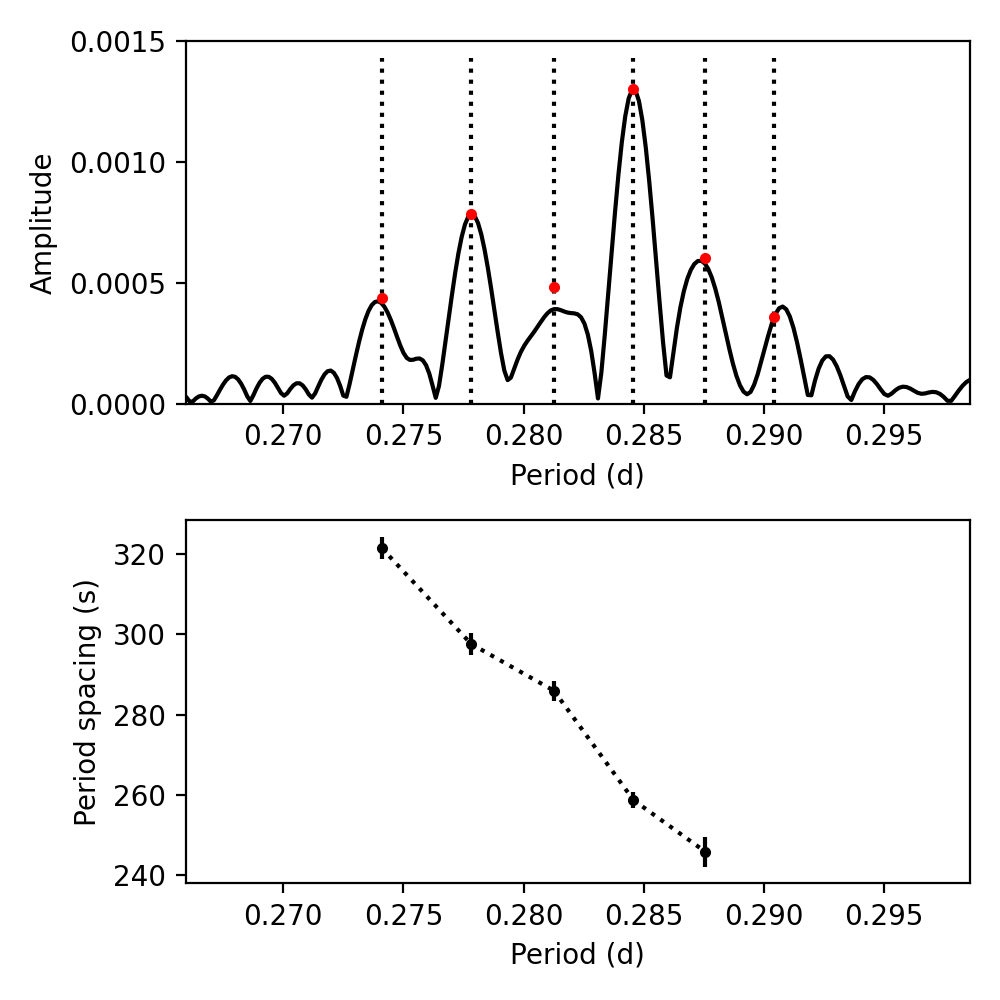}
  \caption{Top: amplitude spectrum of the pre-main-sequence pulsator TIC\,56640851, assigned to the Huluwa~5 subgroup of Vela~OB2 by \citet{Pang2021_huluwa}. The black curve traces the amplitude spectrum. The red dots mark peaks identified as part of a g-mode period\mbox{-}spacing pattern; the vertical dotted lines guide the eye. Bottom: period\mbox{-}spacing pattern of TIC\,56640851, identified as $\ell=1$, $m=1$ g modes.}
  \label{fig:TIC56640851_Huluwa}
  
\end{figure}

In this work, we investigated two additional young stellar complexes: the Vela~OB2 complex \citep{Pang2021_huluwa} and the Cep--Her complex \citep{Kerr2024_Cep_Her}. As listed in Table~\ref{tab:cluster_gdor_sample}, we found one g-mode pulsator with a clear period-spacing pattern in Vela~OB2 (TIC\,56640851, shown in Fig.~\ref{fig:TIC56640851_Huluwa}) and five in Cep--Her. In addition, CWNU\,1183, a newly identified young cluster in the Kepler field of view with an age of around 46\,Myr, hosts one clear g-mode pulsator.

For the period-spacing pattern of the Vela~OB2 complex member TIC\,56640851 shown in Fig.~\ref{fig:TIC56640851_Huluwa}, we find that the period spacings are small, approaching the period resolution of the two-consecutive-sector TESS light curve. 
Nevertheless, we are still able to identify the g-mode periods owing to the high amplitudes of the mode (top panel of Fig.~\ref{fig:TIC56640851_Huluwa}) and obtain a linearly decreasing period-spacing pattern (bottom panel of Fig.~\ref{fig:TIC56640851_Huluwa}). We infer a relatively fast near-core rotation rate of $2.634\pm0.007\,\mathrm{d^{-1}}$ and an asymptotic spacing $\Pi_0$ of $4302^{+70}_{-60}\,\mathrm{s}$. The mass given by \citet{Pang2021_huluwa} is $\sim2.19\,\mathrm{M_\odot}$; at this mass, a much higher $\Pi_0$ of around $6000\,\mathrm{s}$ would be expected for a main-sequence star compared with the observed value.

\begin{figure}[h]
  \centering
  \includegraphics[width=0.5\textwidth]{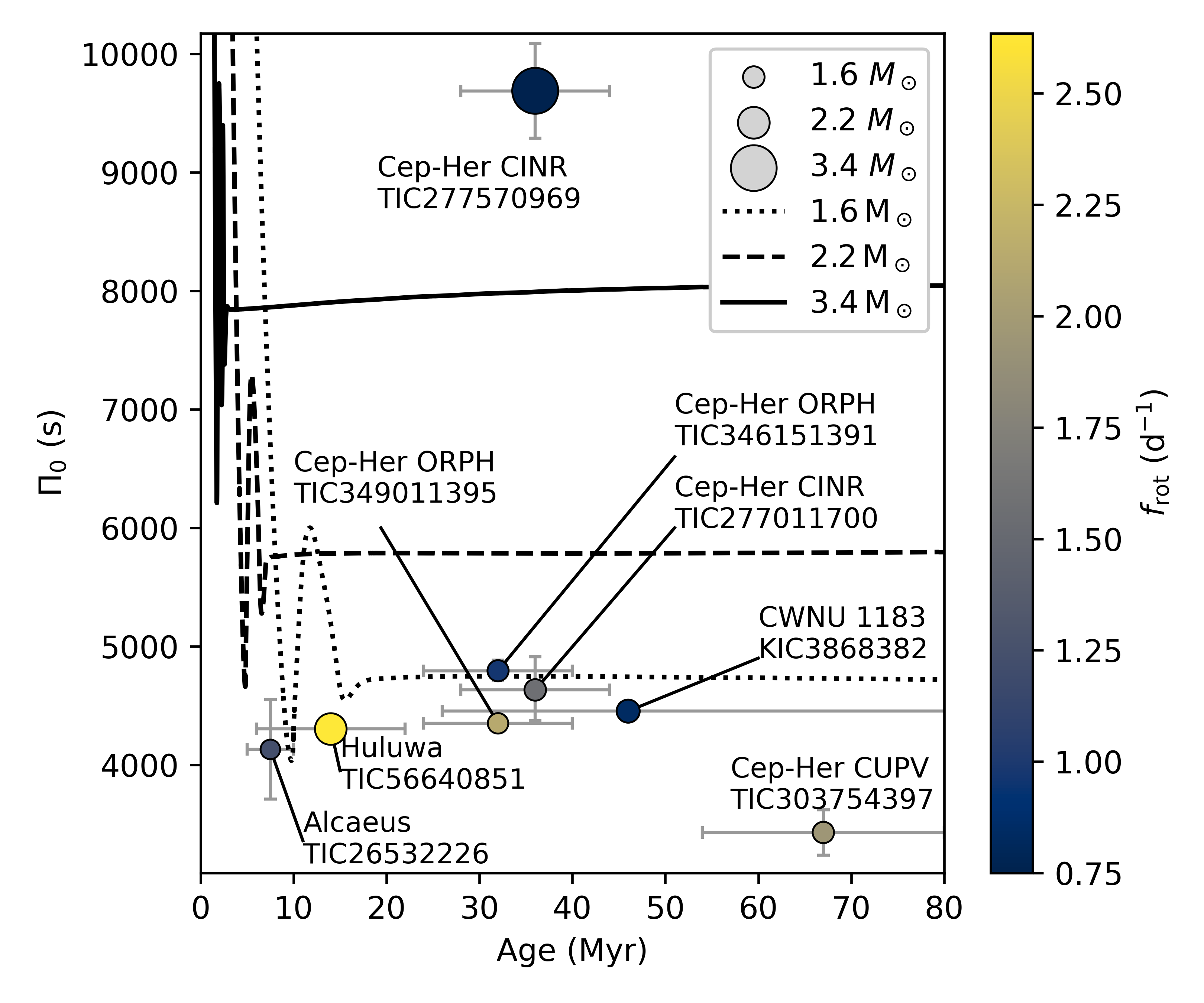}
  \caption{Asymptotic spacing $\Pi_0$ as a function of stellar age for young g-mode pulsators. The lines show theoretical models calculated by \citet{Gautam2026} with an initial equatorial velocity of $8\,\mathrm{km\,s^{-1}}$. The dotted, dashed, and solid lines correspond to models with masses of $1.6\,\mathrm{M_\odot}$, $2.2\,\mathrm{M_\odot}$, and $3.4\,\mathrm{M_\odot}$, respectively. These three masses are chosen to be consistent with the observed masses of the stars, which are indicated by the circle sizes. The colour indicates the near-core rotation rate. }
  \label{fig:pre-main-sequence-model}
\end{figure}

To investigate how $\Pi_0$ evolves with age, Fig.~\ref{fig:pre-main-sequence-model} displays the observed asymptotic spacing $\Pi_0$ versus stellar age for stars in young stellar groups. We also show representative theoretical evolutionary tracks computed from the pre-main-sequence models of \citet{Gautam2026}. The displayed tracks use models with initial metal mass fraction $Z=0.015$, corresponding to a near-solar composition relative to the solar bulk metallicity $Z_{\odot,\mathrm{bulk}}=0.0142$ adopted from \citet{Asplund_2009}. Rotation was included from the beginning of the pre-main-sequence evolution, with the tracks shown here initiated with solid-body equatorial rotation of $8\,\mathrm{km\,s^{-1}}$. We find that $\Pi_0$ changes rapidly during the pre-main-sequence phase, whereas it evolves slowly with age after the zero-age main sequence. For stars with $M \sim 1.6\,\mathrm{M_\odot}$, $\Pi_0$ remains around $4700\,\mathrm{s}$, and, in general, $\Pi_0$ increases with stellar mass. Similar rapid evolution of $\Pi_0$ during the pre-main-sequence phase was also reported by \citet{Bouabid2011_PMS}, who used the stellar-evolution code CLES \citep{Scuflaire2008_CLES} rather than MESA.

From Fig.~\ref{fig:pre-main-sequence-model}, we infer that the stars from Cep--Her and CWNU\,1183 should have already entered their main-sequence phases. For the stars with mass around $1.6\,\mathrm{M_\odot}$, we find that their $\Pi_0$ are slightly smaller than predicted (shown by the dotted line). This difference can be explained by adjusting stellar parameters, e.g., decreasing stellar mass or increasing initial velocity would decrease the theoretical $\Pi_0$ values. 

We find two stars for which the theoretical $\Pi_0$ values do not match the observations. First, TIC\,303754397 ($\sim1.59\,\mathrm{M_\odot}$), a member of the Cupavo association in the Cep--Her complex, has an anomalously small $\Pi_0$. This discrepancy may arise from an insecurely identified period-spacing pattern due to short TESS observation, or from uncertainties in the isochrone-derived stellar mass or evolutionary stage. Alternatively, if the star has undergone binary mass transfer, its internal chemical stratification and hence its g-mode period spacings may differ from those predicted by single-star models \citep{Guo2017, Wu2026}. 
Second, TIC\,277570969 ($\sim3.42\,\mathrm{M_\odot}$), a member of the Cinyras association in the Cep--Her complex, shows an extremely large $\Pi_0$. Previous studies have revealed similar $\Pi_0$ discrepancies, which are more significant in the high-mass range. The physics behind such discrepancies remains unclear. Rotation may play a role, because gravity darkening can shift the position of a star in the CMD and therefore bias the mass estimate from isochrone fitting. However, we obtain a relatively low rotation rate for this star ($0.750\,\mathrm{d^{-1}}$), so the purely observational effect of rotation should be marginal.

Based on the measured $\Pi_0$ values and isochrone-constrained ages, only two stars, TIC\,26532226 in the Alcaeus group of the Perseus star-forming complex and TIC\,56640851 in Vela~OB2, are clearly pre-main-sequence g-mode pulsators. Both stars show small $\Pi_0$ values that can only be explained by the oscillatory $\Pi_0$ track during the pre-main-sequence phase. Within the membership and age framework of \citet{Pang2021_huluwa}, TIC\,56640851 is assigned to Huluwa~5, the youngest subgroup of Vela~OB2, with an age of $7.5$--$14.5\,\mathrm{Myr}$. Our seismic model with $2.2\,\mathrm{M_\odot}$ (the dashed line in Fig.~\ref{fig:pre-main-sequence-model}) reaches a $\Pi_0$ minimum at an age of $5\,\mathrm{Myr}$, suggesting that this star is extremely young. We also find that, although the current minimum value of $\Pi_0$ is still larger than the observed value, increasing the initial velocity in the model would lower the minimum $\Pi_0$ and could explain the observed $\Pi_0$.

\section{Blue stragglers}\label{sec:BSs}

\begin{figure*}[t]
  \centering
  \begin{subfigure}[t]{0.49\textwidth}
    \centering
    \includegraphics[width=\linewidth]{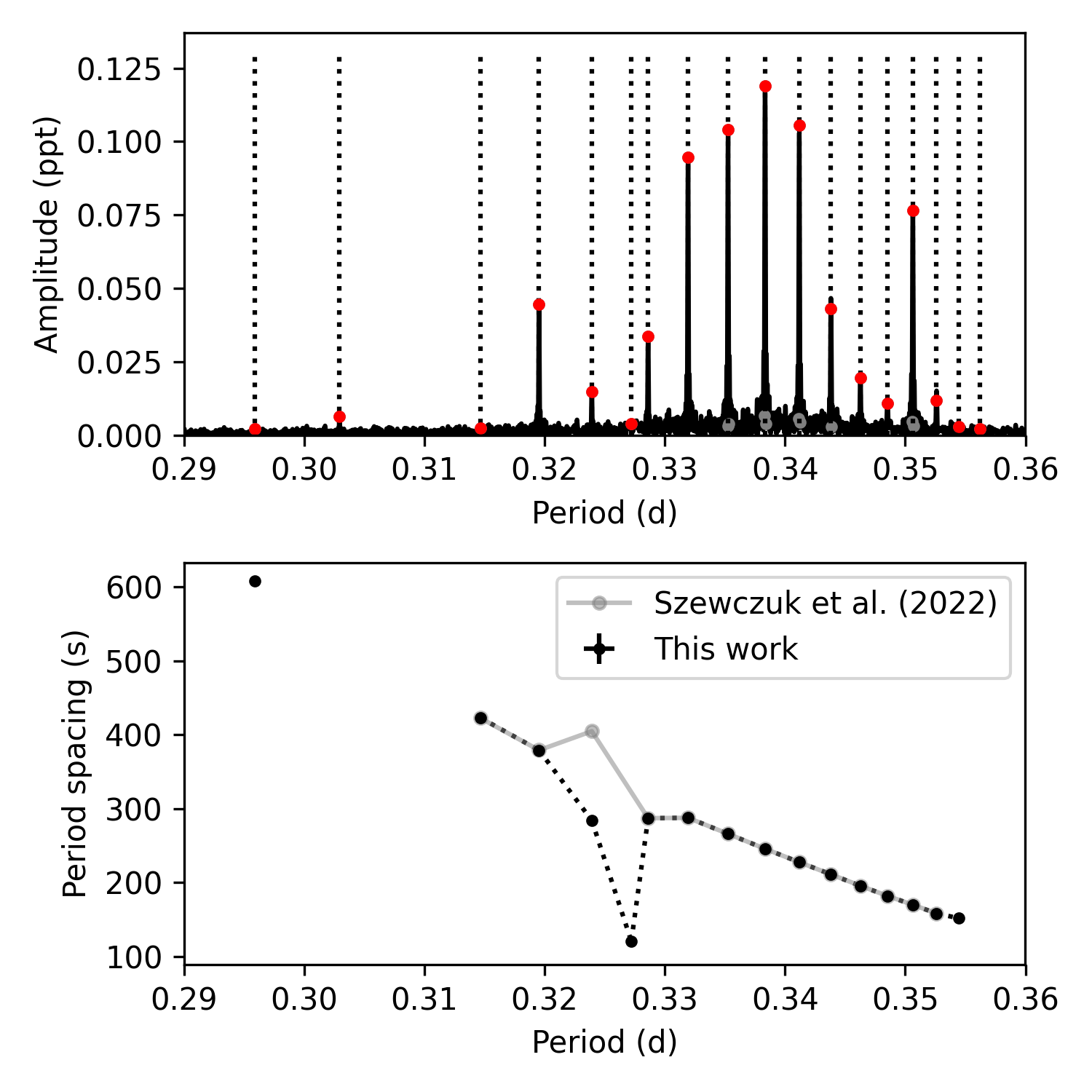}
    \caption{$\ell=1$, $m=1$ g modes.}
    \label{fig:KIC8264293_l1m1}
  \end{subfigure}
  \hfill
  \begin{subfigure}[t]{0.49\textwidth}
    \centering
    \includegraphics[width=\linewidth]{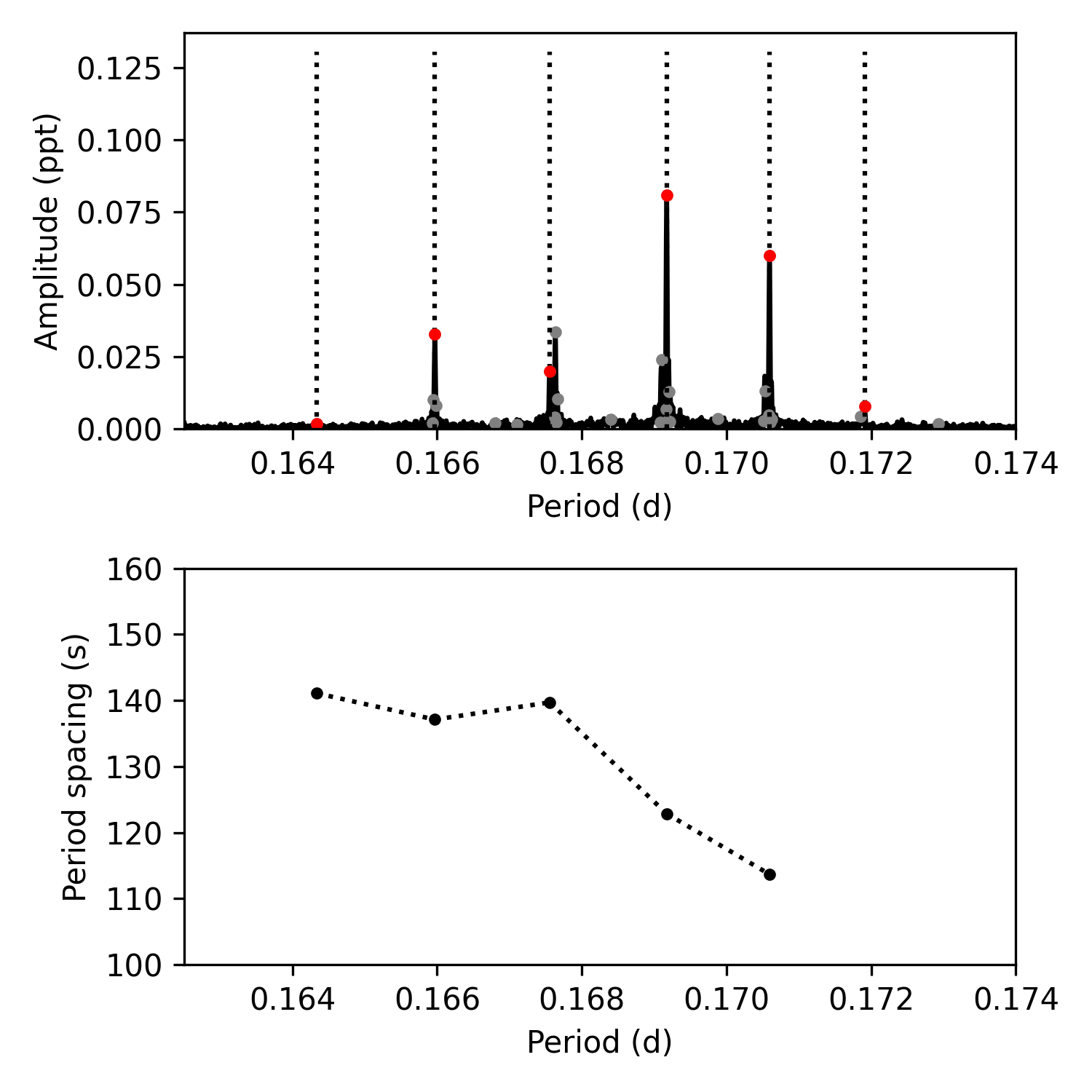}
    \caption{$\ell=2$, $m=2$ g modes.}
    \label{fig:KIC8264293_l2m2}
  \end{subfigure}
  \caption{\added{Period-spacing patterns of the blue straggler KIC\,8264293 in NGC\,6866. The left and right panels show the $\ell=1$, $m=1$ and $\ell=2$, $m=2$ patterns, respectively. The top panels show the amplitude spectra. Red dots mark peaks identified as g modes, while gray dots indicate peaks detected during pre-whitening but not adopted as part of a g-mode period-spacing pattern. The vertical lines mark the locations of the identified g-mode peaks, each corresponding to a red dot. The bottom panels show the corresponding period-spacing patterns. The error bars are smaller than the symbols.}}
  \label{fig:KIC8264293}
\end{figure*}

\begin{figure*}[t]
  \centering
  \begin{subfigure}[t]{0.49\textwidth}
    \centering
    \includegraphics[width=\linewidth]{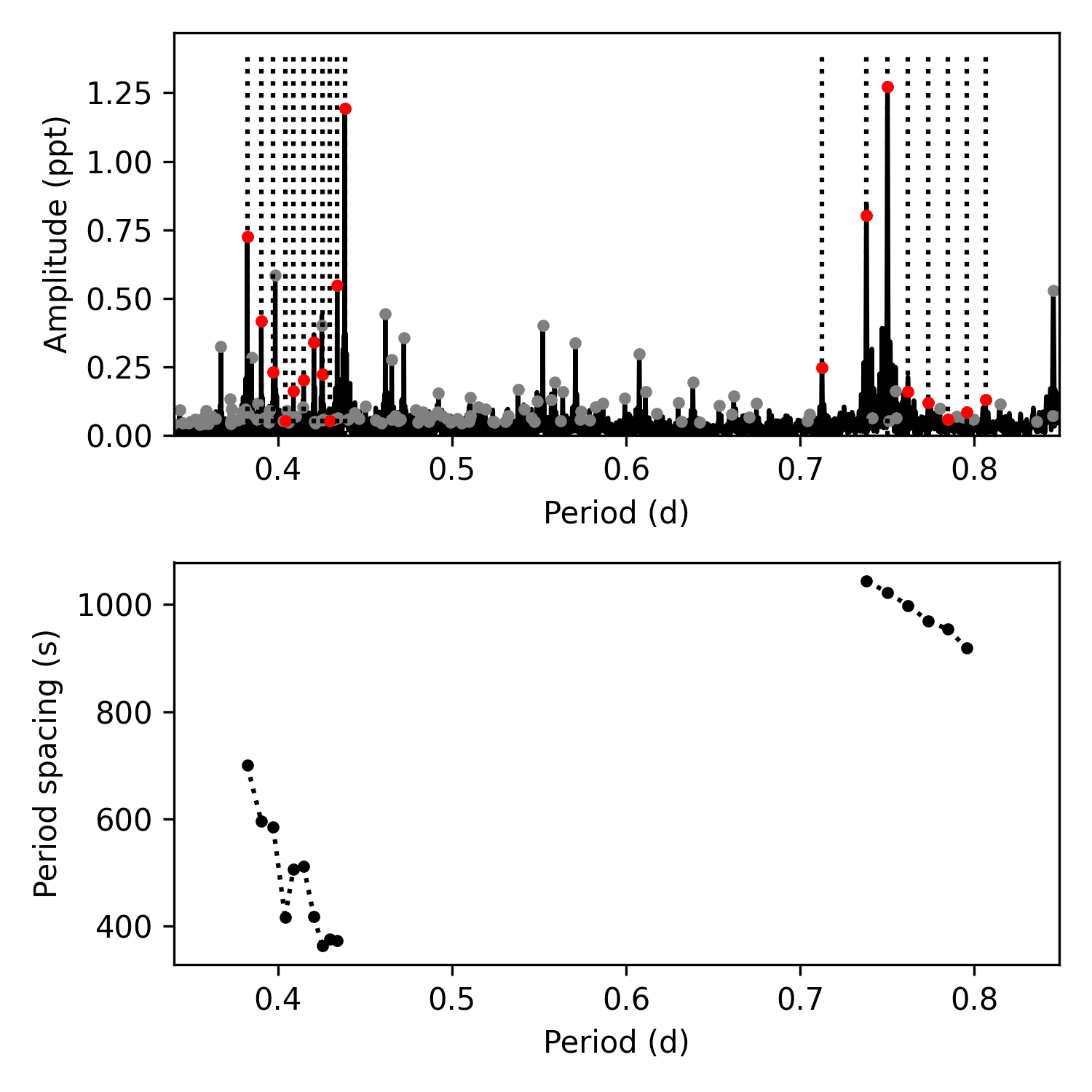}
    \caption{Same as Fig.~\ref{fig:KIC8264293}, but for the blue straggler KIC\,5024455 in NGC\,6819. In the bottom panel, the left period-spacing pattern is identified as $\ell=1$, $m=1$ g modes, while the right pattern is identified as $\ell=1$, $m=0$ g modes. }
    \label{fig:KIC5024455}
  \end{subfigure}
  \hfill
  \begin{subfigure}[t]{0.49\textwidth}
    \centering
    \includegraphics[width=\linewidth]{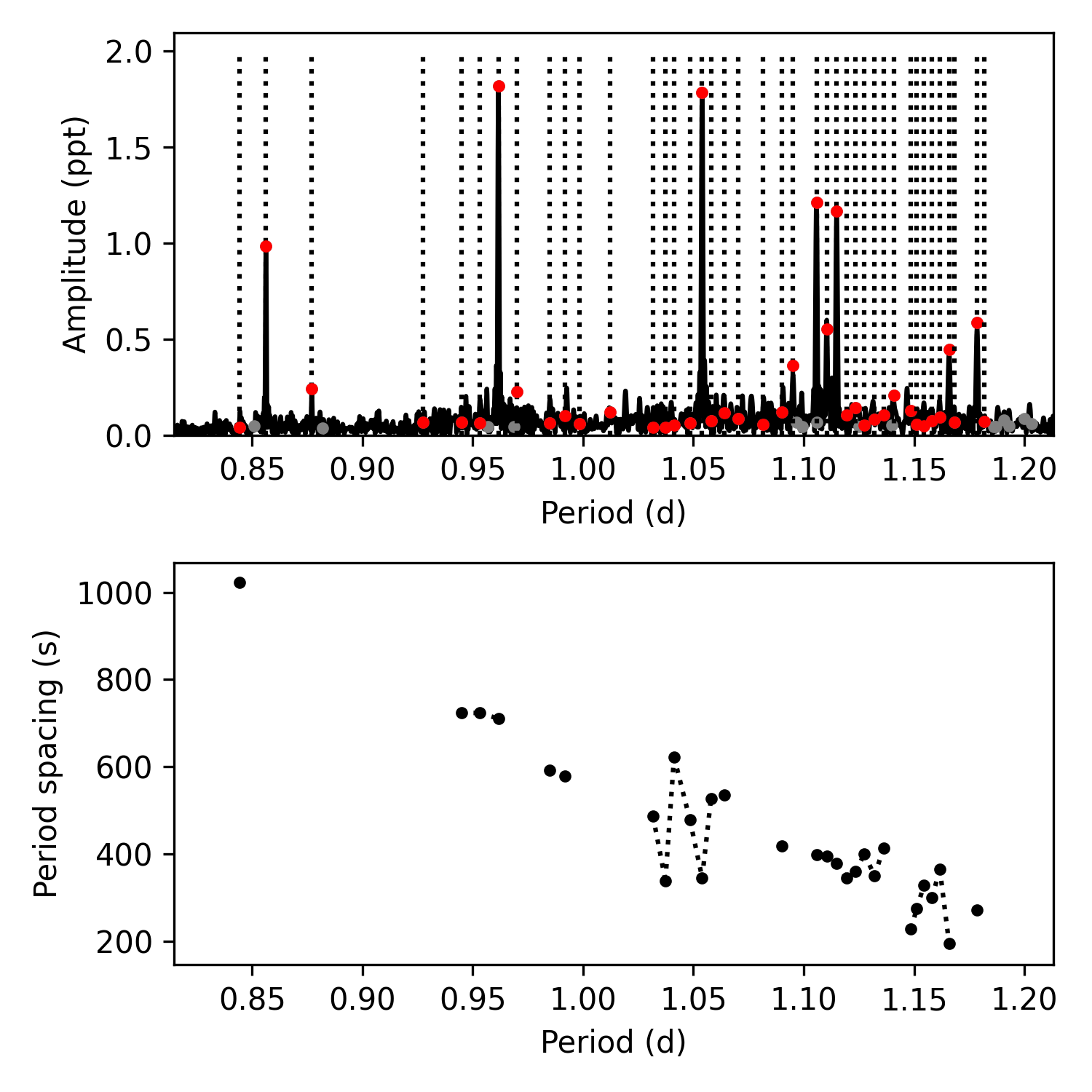}
    \caption{Same as Fig.~\ref{fig:KIC8264293}, but for the blue straggler KIC\,2438249 in NGC\,6791. }
    \label{fig:KIC2438249}
  \end{subfigure}
  \caption{Amplitude spectra and period-spacing patterns of two pulsating blue stragglers.}
  \label{fig:KIC5024455_KIC2438249}
\end{figure*}

We identify three pulsating blue stragglers with clear period-spacing patterns: KIC\,8264293 in NGC\,6866, KIC\,5024455 in NGC\,6819, and KIC\,2438249 in NGC\,6791. We exhibit their amplitude spectra and period-spacing patterns in Figs.~\ref{fig:KIC8264293}, \ref{fig:KIC5024455}, and \ref{fig:KIC2438249}. 

\added{Among the cluster members, we manually identified stars lying above the cluster main sequence in the CMDs as blue-straggler candidates. Cross-matching these candidates with the catalogue of blue stragglers in open clusters by \citet{Rain2021_BSS} shows that KIC\,8264293 and KIC\,2438249 are included, whereas KIC\,5024455 is not. The latter is nevertheless listed as a member of NGC\,6819 in the catalogue of \citet{Hunt2024}.}

We find that the mass of KIC\,2438249 given by \citet{Hunt2024} is too small, falling below the typical lower boundary of $\sim1.5\,\mathrm{M_\odot}$ for $\gamma$\,Dor stars. We therefore redetermined the masses of the three pulsating blue stragglers. 
For KIC\,5024455 and KIC\,2438249, we manually searched for the best-fitting MIST isochrones around the literature values, then shifted the ages younger until the main sequence extended to the locations of the blue stragglers. We adopted the mass of the isochrone point with the same $G$-band magnitude as the blue straggler as the mass listed in Table~\ref{tab:cluster_gdor_sample}. We assigned typical uncertainties of $\pm0.1\,\mathrm{M_\odot}$, slightly larger than the usual uncertainties reported by \citet{Hunt2024}. 
We obtained $1.48\pm0.10\,\mathrm{M_\odot}$ for KIC\,5024455 and $1.73\pm0.10\,\mathrm{M_\odot}$ for KIC\,2438249. 
We note that the mass estimates for the blue stragglers are preliminary; for example, they do not account for their merger or mass-transfer histories or adequately consider reddening. We therefore expect the estimated masses of the blue stragglers to have substantial systematic uncertainties.

For KIC\,8264293, we adopted a spectroscopic mass estimate of $3.18^{+0.17}_{-0.10}\,\mathrm{M_\odot}$ based on HERMES spectroscopy, Gaia astrometry, extinction corrections, and bolometric corrections. A more detailed description of the spectroscopic analysis will be given by Wang et al. (in prep.).

Figure~\ref{fig:Pi0_age} shows that all three pulsating blue stragglers have higher $\Pi_0$ values than normal $\gamma$\,Dor stars. This is expected for KIC\,8264293, as it lies in the mass range of SPB stars, but it is unusual for KIC\,5024455 and KIC\,2438249 because they have typical $\gamma$\,Dor-star masses ($<2.0\,\mathrm{M_\odot}$). Their unusually large $\Pi_0$ values may reflect an interesting mass-transfer or merger formation history.

Several recent studies have investigated how the formation processes of blue stragglers reshape their internal structures and how these changes are reflected in g-mode period-spacing patterns. \citet{Henneco2024} studied merger products involving massive stars near the Hertzsprung Gap and found that their g-mode period-spacing patterns can exhibit pronounced deep dips caused by strong chemical gradients. \citet{Henneco2025} further investigated the merger of two main-sequence stars using three-dimensional magnetohydrodynamic simulations. They found that the merger process can generate a double-peaked Brunt--V\"ais\"al\"a frequency profile owing to the formation and subsequent recession of a new convective core, leading to period-spacing patterns that differ significantly from those of single stars. \citet{Guo2026_FT_of_PS} showed that the periodicity of the period-spacing modulation, measured in radial order, is directly related to the location of sharp features in the Brunt--V\"ais\"al\"a frequency profile. This provides a new diagnostic of the internal chemical structure and evolutionary state of a star, and has recently been applied to a sample of 26 SPB stars \citep{Guo2025_FT_PS_26_SPB}. Blue stragglers are expected to retain these unusual chemical stratification profiles, producing additional modulations in their g-mode period-spacing patterns that preserve information about their formation history, as demonstrated by collision models \citep{Briganti2025} and mass-transfer models \citep{Wu2026}.

\cite{Szewczuk2022_8264293} conducted detailed seismic modelling of KIC\,8264293, treating it as a normal single-star evolutionary object and arguing that its parallax was inconsistent with membership in NGC\,6866. They derived a mass of $3.54\,\mathrm{M_\odot}$, an age of $6\,\mathrm{Myr}$, and a central hydrogen abundance of $X_\mathrm{c}\sim0.71$. However, the membership probability of this star in NGC\,6866 is 70\% according to \cite{Cantat-Gaudin2020}, 81\% according to \cite{Hunt2024}, and 99.8\% according to \cite{Wang2026_NGC6866}. We therefore consider this star to be a member of NGC\,6866.

\added{As shown in Fig.~\ref{fig:KIC8264293}, we identified two period-spacing patterns in KIC\,8264293: one corresponding to prograde sectoral modes with $(\ell,m)=(1,1)$ (Fig.~\ref{fig:KIC8264293_l1m1}), and the other to prograde sectoral modes with $(\ell,m)=(2,2)$ (Fig.~\ref{fig:KIC8264293_l2m2}).} For the $\ell=1$, $m=1$ pattern, we detected an overall smooth period-spacing pattern with a very deep dip at $P\sim0.326\,\mathrm{d}$. The dip is primarily caused by a mode with a small amplitude at $P\sim0.327219(4)\,\mathrm{d}$. Its $\mathrm{SNR}$ is 7.75, with an amplitude of $3.8\pm0.4\,\mathrm{ppm}$, higher than the well-accepted $\mathrm{SNR}>4$ criterion \citep{Breger1993}. This mode also passed the criterion implemented in the code \textsc{Star Shadow} \citep{IJspeert2024}. In the analysis of \cite{Szewczuk2022_8264293}, this low-amplitude mode was not reported; instead, they found a local fluctuation in the period-spacing pattern around $0.326\,\mathrm{d}$. However, such a strong fluctuation is not expected for an extremely young star with $X_\mathrm{c}\sim0.71$, because stars near the zero-age main sequence have not yet developed strong chemical gradients outside their convective cores. 
Additionally, the deep dip detected here does not resemble the dips caused by the coupling between g modes in the near-core regions and inertial modes in the convective cores predicted by \cite{Ouazzani2020} and \cite{Saio2021}, as those features are generally broader and shallower. Changing the parameters in the coupling process or including magnetic fields may produce a deeper dip \citep{Barrault2025_dip, Barrault2025_magnetic}, but this has not yet been investigated. 
Alternatively, the deep dip may be related to unusual internal chemical gradients associated with the formation of the blue straggler, as predicted by \cite{Henneco2024}. If so, this deep dip provides valuable constraints on the internal structure of the star. Combining this information with the CMD of NGC\,6866 may allow us to trace the formation history of the blue straggler.

For KIC\,5024455 (Fig.~\ref{fig:KIC5024455}), we detected two period-spacing patterns, both with $\ell=1$: one with $m=1$ and the other with $m=0$. The $m=0$ pattern is smooth, whereas the $m=1$ pattern exhibits some variations. The presence of $m=0$ modes in this rapid rotator indicates that the inclination of the star's rotation axis should be small \citep{Saio2018}. For KIC\,2438249 (Fig.~\ref{fig:KIC2438249}), we detected a long period-spacing pattern with numerous fluctuations.

\section{Conclusions}\label{sec:conclusions}

In this work, we performed a survey of g-mode pulsators with clear period-spacing patterns in coeval stellar populations using Gaia astrometry together with Kepler and TESS photometry. Reliable membership identification from Gaia DR3 allows us to estimate masses and ages that are difficult to determine for isolated field stars. The four-year high-precision Kepler photometry provides light curves with sufficient frequency resolution to resolve g-mode period-spacing patterns. Building on prior knowledge from hundreds of Kepler g-mode pulsators, we can also identify such patterns in TESS data, despite its lower frequency resolution.

We used the traditional approximation of rotation (TAR) to fit the g-mode period-spacing patterns and derived two asteroseismic quantities: the near-core rotation rate $f_\mathrm{rot}$ and the asymptotic spacing $\Pi_0$. These quantities trace AM transport and near-core evolution in early-type main-sequence stars. In total, we collected \starnumber g-mode pulsators with clear period-spacing patterns from \groupnumber stellar groups.

For the near-core rotation rates, we find that the stellar-group members in our sample generally rotate faster than the Kepler field stars studied by \citet{Li2020MNRAS_611}. This difference is probably caused by a combination of observational selection effects from the shorter TESS light curves and the relatively young age distribution of the stellar groups in our sample. We also find a paucity of slow rotators ($f_\mathrm{rot}<0.5\,\mathrm{d^{-1}}$) compared with the Kepler sample, which is probably due, at least in part, to the lower frequency resolution of the TESS light curves.

The near-core rotation rates show a clear evolutionary trend. The fastest rotators occur at ages around $\sim100\,\mathrm{Myr}$, beyond which the upper envelope of $f_\mathrm{rot}$ gradually decreases with age. By subdividing the stars into different mass ranges, we find that stars with masses between $1.4$ and $1.6\,\mathrm{M_\odot}$ rotate more slowly than higher-mass stars. This may indicate that stars below $1.6\,\mathrm{M_\odot}$ retain some residual magnetic braking associated with their surface convective envelopes, leading to angular-momentum loss through magnetised winds.

For the asymptotic spacing, $\Pi_0$, the stellar-group members show larger values than the Kepler field sample, indicating that they are generally younger. We find that $\Pi_0$ does not follow a strictly monotonic relation with age; instead, it increases slightly near the beginning of the main sequence and then decreases (Fig.~\ref{fig:Pi0_age}), consistent with stellar-model calculations.

The theoretical $\Pi_0$ values reproduce the observed trend for stars between $1.4$ and $1.6\,\mathrm{M_\odot}$, suggesting that $\Pi_0$ can provide useful asteroseismic age constraints for stars and clusters in this mass range. For example, the low $\Pi_0$ values ($<4000\,\mathrm{s}$) of the two g-mode pulsators in NGC\,1901 support the older age inferred from gyrochronology and disfavour the substantially younger ages inferred from isochrone fitting. However, above $1.6\,\mathrm{M_\odot}$, there is an increasing discrepancy between the observed and theoretical $\Pi_0$ values. This discrepancy may reflect uncertainties in CMD-based mass estimates or missing physics that prevents the stellar surface parameters from being fully reconciled with the internal seismic diagnostic. Future work should address this issue through spectroscopic follow-up of surface rotation ($v\sin i$), detailed corrections for gravity darkening, and seismic modelling beyond $\Pi_0$, such as fitting each identified mode, as in \cite{Mombarg2021} and \cite{LiGang_2025_NGC2516_modelling}.

Because these stars belong to coeval stellar groups, we can more reliably identify their evolutionary stages and distinguish pre-main-sequence pulsators from pulsating blue stragglers. 
We report a new pre-main-sequence g-mode pulsator with clear period-spacing patterns, TIC\,56640851, in Vela~OB2. Adopting the membership and age scale of \citet{Pang2021_huluwa}, the star is assigned to the Huluwa~5 subgroup. Our new rotating pre-main-sequence evolutionary model at this star's mass shows that the star happens to lie near the minimum of the oscillating $\Pi_0$ curve, making it an excellent age diagnostic. The asteroseismic age inferred from $\Pi_0$ is consistent with Huluwa~5 being the youngest subgroup.

We report three pulsating blue stragglers with clear period-spacing patterns: KIC\,8264293 in NGC\,6866, KIC\,5024455 in NGC\,6819, and KIC\,2438249 in NGC\,6791. Their $\Pi_0$ values are outliers compared with single-star evolutionary models in Fig.~\ref{fig:Pi0_age_mass_bin}. We also detect a deep dip in the period-spacing pattern of KIC\,8264293, suggesting that a recent formation event may have left a detectable signature in its internal structure. Detailed modelling of these pulsating blue stragglers, combined with constraints from their host clusters, may help reveal their progenitors, formation epochs, and formation pathways.

Overall, this work provides a homogeneous sample of g-mode pulsators in coeval stellar populations, highlighting the strong synergy between stellar physics and studies of stellar groups. For stellar physics, our uniformly derived measurements of $f_\mathrm{rot}$ and $\Pi_0$ offer a valuable benchmark for future asteroseismic modelling of internal processes, including AM transport, chemical mixing, and the calibration of 1D and 2D models for early-type main-sequence stars. For studies of stellar groups, our asteroseismic age constraints provide an independent perspective on age dating, while peculiar objects such as pre-main-sequence pulsators and pulsating blue stragglers offer an interior view of star-formation sequences in stellar complexes \citep[e.g.][]{Ratzenbock2023} and of stellar interactions in bound clusters.

Beyond the Kepler and TESS missions, future space-based photometric missions, including PLATO \citep{Rauer2014PLATO} and ET2.0 \citep{Ge_2022_ET}, will continue to advance the synergy between asteroseismology and research on stellar groups. Their longer time span, higher cadence, and wider coverage will provide unparalleled opportunities for ensemble asteroseismic studies of well-characterised stellar populations.

  %% Please use the acknowledgment and contribution environments. This will
  %% be anonomyized when the "anonymous" style option is used.
  \begin{acknowledgments}
    GL acknowledges the support of the Australian Research Council through the DECRA project DE250100773. 
    SJM was supported by the Australian Research Council through Future Fellowship FT210100485. 
    FDM acknowledges support by the National Natural Science Foundation of China (NSFC, Nos. 12288102, 12125303), the Strategic Priority Research Program of the Chinese Academy of Sciences (grant Nos. XDB1160303, XDB1160300, XDB1160000, XDB1160200, XDB1160201), the CAS Project for Young Scientists in Basic Research (YSBR-148), the Yunnan Revitalization Talent Support Program ``YunLing Scholar'' project, International Centre of Supernovae (ICESUN), Yunnan Key Laboratory of Supernova Research (No. 202505AV340004), the New Cornerstone Science Foundation through the XPLORER PRIZE, and the Yunnan Revitalization Talent Support Program-Science \& Technology Champion Project (No. 202305AB350003).
    The research leading to these results has received funding from the Flemish Government under the long-term structural Methusalem funding program by means of the project SOUL: Stellar evolution in full glory, grant METH/24/012 at KU Leuven, as well as from the European Research Council (ERC) under the Horizon Europe programme (Synergy Grant agreement N$^\circ$ 101071505: 4D-STAR). 
    While partially funded by the European Union, views and opinions expressed are however those of the author(s) only and do not necessarily reflect those of the European Union or the European Research Council. Neither the European Union nor the granting authority can be held responsible for them. 
    TRB acknowledges support from the Australian Research Council through Laureate Fellowship FL220100117.
    GL received travel support from the National Natural Science Foundation of China (NSFC) through grant 12273002.
    CYL is supported by the NSFC (grant No. 12233013) and by the CSST project CMS-CSST-2025-A13.

    This work made use of Lightkurve \citep{Lightkurve2018} to download the Kepler light curves and TESSUTILS\footnote{\url{https://github.com/IvS-KULeuven/TESSutils}} \citep{Garcia2022} to extract and process the TESS light curves. We used NumPy \citep{harris2020array} and Astropy \citep{2013A&A...558A..33A,2018AJ....156..123A,2022ApJ...935..167A} for numerical calculations, Matplotlib \citep{Hunter2007Matplotlib} for plotting, and emcee \citep{2013PASP..125..306F} for Markov chain Monte Carlo sampling.

    GL used ChatGPT (OpenAI; \url{https://chatgpt.com/}) to assist with programming, figure preparation, and language editing of an English-language draft originally written by GL. Prism (OpenAI; \url{https://openai.com/prism/}) was used to assist with LaTeX editing and formatting. All AI-assisted content was reviewed and revised by the authors.
  \end{acknowledgments}

  \begin{contribution}
    %%This section gives authors the space to recognize author contributions. The text inside this environment is NOT counted towards the total word quanta. At a minimum, manuscripts are expected to include this text:

    %All authors contributed equally to the Terra Mater collaboration.

    %% But authors are expected to provide more specific details, e.g.
    %%
    %%SC was responsible for writing and submitting the manuscript.
    %%WWM came up with the initial research concept and edited the manuscript.
    %%OTS obtained the funding and edited the manuscript.
    %%EBF provided the formal analysis and validation. He also edited the manuscript.
    %%GEH Supervised the undergraduates, wrote the software and administers the project github and Zenodo repositories.
    %%
    %% Authors can use the Contributor Role Taxonomy (CRediT) at
    %% for ideas on how write a good statement tailored to their needs.

  \end{contribution}

  \bibliography{V2-ligangreference-2}

@ARTICLE{2022ApJ...935..167A,
       author = {{Astropy Collaboration} and {Price-Whelan}, Adrian M. and {Lim}, Pey Lian and {Earl}, Nicholas and {Starkman}, Nathaniel and {Bradley}, Larry and {Shupe}, David L. and {Patil}, Aarya A. and {Corrales}, Lia and {Brasseur}, C.~E. and {N{\"o}the}, Maximilian and {Donath}, Axel and {Tollerud}, Erik and {Morris}, Brett M. and {Ginsburg}, Adam and {Vaher}, Eero and {Weaver}, Benjamin A. and {Tocknell}, James and {Jamieson}, William and {van Kerkwijk}, Marten H. and {Robitaille}, Thomas P. and {Merry}, Bruce and {Bachetti}, Matteo and {G{\"u}nther}, H. Moritz and {Aldcroft}, Thomas L. and {Alvarado-Montes}, Jaime A. and {Archibald}, Anne M. and {B{\'o}di}, Attila and {Bapat}, Shreyas and {Barentsen}, Geert and {Baz{\'a}n}, Juanjo and {Biswas}, Manish and {Boquien}, M{\'e}d{\'e}ric and {Burke}, D.~J. and {Cara}, Daria and {Cara}, Mihai and {Conroy}, Kyle E. and {Conseil}, Simon and {Craig}, Matthew W. and {Cross}, Robert M. and {Cruz}, Kelle L. and {D'Eugenio}, Francesco and {Dencheva}, Nadia and {Devillepoix}, Hadrien A.~R. and {Dietrich}, J{\"o}rg P. and {Eigenbrot}, Arthur Davis and {Erben}, Thomas and {Ferreira}, Leonardo and {Foreman-Mackey}, Daniel and {Fox}, Ryan and {Freij}, Nabil and {Garg}, Suyog and {Geda}, Robel and {Glattly}, Lauren and {Gondhalekar}, Yash and {Gordon}, Karl D. and {Grant}, David and {Greenfield}, Perry and {Groener}, Austen M. and {Guest}, Steve and {Gurovich}, Sebastian and {Handberg}, Rasmus and {Hart}, Akeem and {Hatfield-Dodds}, Zac and {Homeier}, Derek and {Hosseinzadeh}, Griffin and {Jenness}, Tim and {Jones}, Craig K. and {Joseph}, Prajwel and {Kalmbach}, J. Bryce and {Karamehmetoglu}, Emir and {Ka{\l}uszy{\'n}ski}, Miko{\l}aj and {Kelley}, Michael S.~P. and {Kern}, Nicholas and {Kerzendorf}, Wolfgang E. and {Koch}, Eric W. and {Kulumani}, Shankar and {Lee}, Antony and {Ly}, Chun and {Ma}, Zhiyuan and {MacBride}, Conor and {Maljaars}, Jakob M. and {Muna}, Demitri and {Murphy}, N.~A. and {Norman}, Henrik and {O'Steen}, Richard and {Oman}, Kyle A. and {Pacifici}, Camilla and {Pascual}, Sergio and {Pascual-Granado}, J. and {Patil}, Rohit R. and {Perren}, Gabriel I. and {Pickering}, Timothy E. and {Rastogi}, Tanuj and {Roulston}, Benjamin R. and {Ryan}, Daniel F. and {Rykoff}, Eli S. and {Sabater}, Jose and {Sakurikar}, Parikshit and {Salgado}, Jes{\'u}s and {Sanghi}, Aniket and {Saunders}, Nicholas and {Savchenko}, Volodymyr and {Schwardt}, Ludwig and {Seifert-Eckert}, Michael and {Shih}, Albert Y. and {Jain}, Anany Shrey and {Shukla}, Gyanendra and {Sick}, Jonathan and {Simpson}, Chris and {Singanamalla}, Sudheesh and {Singer}, Leo P. and {Singhal}, Jaladh and {Sinha}, Manodeep and {Sip{\H{o}}cz}, Brigitta M. and {Spitler}, Lee R. and {Stansby}, David and {Streicher}, Ole and {{\v{S}}umak}, Jani and {Swinbank}, John D. and {Taranu}, Dan S. and {Tewary}, Nikita and {Tremblay}, Grant R. and {de Val-Borro}, Miguel and {Van Kooten}, Samuel J. and {Vasovi{\'c}}, Zlatan and {Verma}, Shresth and {de Miranda Cardoso}, Jos{\'e} Vin{\'\i}cius and {Williams}, Peter K.~G. and {Wilson}, Tom J. and {Winkel}, Benjamin and {Wood-Vasey}, W.~M. and {Xue}, Rui and {Yoachim}, Peter and {Zhang}, Chen and {Zonca}, Andrea and {Astropy Project Contributors}},
        title = "{The Astropy Project: Sustaining and Growing a Community-oriented Open-source Project and the Latest Major Release (v5.0) of the Core Package}",
      journal = {\apj},
         year = 2022,
        month = aug,
       volume = {935},
       number = {2},
          eid = {167},
        pages = {167},
          doi = {10.3847/1538-4357/ac7c74},
archivePrefix = {arXiv},
       eprint = {2206.14220},
 primaryClass = {astro-ph.IM},
       adsurl = {https://ui.adsabs.harvard.edu/abs/2022ApJ...935..167A}
}

@ARTICLE{Rain2021_BSS,
       author = {{Rain}, M.~J. and {Ahumada}, J.~A. and {Carraro}, G.},
        title = "{A new, Gaia-based, catalogue of blue straggler stars in open clusters}",
      journal = {\aap},
         year = 2021,
        month = jun,
       volume = {650},
          eid = {A67},
        pages = {A67},
          doi = {10.1051/0004-6361/202040072},
archivePrefix = {arXiv},
       eprint = {2103.06004},
 primaryClass = {astro-ph.SR},
       adsurl = {https://ui.adsabs.harvard.edu/abs/2021A&A...650A..67R}
}

@ARTICLE{Ihallaine_2026,
       author = {{Ihallaine}, S. and {Ballot}, J. and {Ligni{\`e}res}, F. and {Ferri{\'e}}, L. and {Charpinet}, S. and {Galoy}, M. and {Li}, G.},
        title = "{Seismic signature of a magnetic field in the {\ensuremath{\gamma}} Doradus star KIC 2309579}",
      journal = {\aap},
         year = 2026,
        month = jun,
       volume = {710},
          eid = {L18},
        pages = {L18},
          doi = {10.1051/0004-6361/202660461},
archivePrefix = {arXiv},
       eprint = {2605.22533},
 primaryClass = {astro-ph.SR},
       adsurl = {https://ui.adsabs.harvard.edu/abs/2026A&A...710L..18I}
}

@ARTICLE{Mombarg2026_calibrating_AM,
       author = {{Mombarg}, J.~S.~G. and {Mathis}, S.},
        title = "{Calibrating angular momentum transport in intermediate-mass stars from gravity-mode asteroseismology: II. Modelling 2937 BAF-type stars}",
      journal = {\aap},
         year = 2026,
        month = jul,
       volume = {711},
          eid = {A190},
        pages = {A190},
          doi = {10.1051/0004-6361/202659483},
archivePrefix = {arXiv},
       eprint = {2606.23819},
 primaryClass = {astro-ph.SR},
       adsurl = {https://ui.adsabs.harvard.edu/abs/2026A&A...711A.190M}
}

@ARTICLE{Armstrong2022_VelaOB2,
       author = {{Armstrong}, Joseph J. and {Wright}, Nicholas J. and {Jeffries}, R.~D. and {Jackson}, R.~J. and {Cantat-Gaudin}, Tristan},
        title = "{The structure and 3D kinematics of vela OB2}",
      journal = {\mnras},
         year = 2022,
        month = dec,
       volume = {517},
       number = {4},
        pages = {5704-5720},
          doi = {10.1093/mnras/stac3101},
archivePrefix = {arXiv},
       eprint = {2208.05277},
 primaryClass = {astro-ph.SR},
       adsurl = {https://ui.adsabs.harvard.edu/abs/2022MNRAS.517.5704A}
}

@ARTICLE{Lamers2006,
       author = {{Lamers}, H.~J.~G.~L.~M. and {Gieles}, M.},
        title = "{Clusters in the solar neighbourhood: how are they destroyed?}",
      journal = {\aap},
         year = 2006,
        month = aug,
       volume = {455},
       number = {2},
        pages = {L17-L20},
          doi = {10.1051/0004-6361:20065567},
archivePrefix = {arXiv},
       eprint = {astro-ph/0606749},
 primaryClass = {astro-ph},
       adsurl = {https://ui.adsabs.harvard.edu/abs/2006A&A...455L..17L}
}

@ARTICLE{Krumholz2019,
       author = {{Krumholz}, Mark R. and {McKee}, Christopher F. and {Bland-Hawthorn}, Joss},
        title = "{Star Clusters Across Cosmic Time}",
      journal = {\araa},
         year = 2019,
        month = aug,
       volume = {57},
        pages = {227-303},
          doi = {10.1146/annurev-astro-091918-104430},
archivePrefix = {arXiv},
       eprint = {1812.01615},
 primaryClass = {astro-ph.GA},
       adsurl = {https://ui.adsabs.harvard.edu/abs/2019ARA&A..57..227K}
}

@ARTICLE{Ratzenbock2023,
       author = {{Ratzenb{\"o}ck}, Sebastian and {Gro{\ss}schedl}, Josefa E. and {Alves}, Jo{\~a}o and {Miret-Roig}, N{\'u}ria and {Bomze}, Immanuel and {Forbes}, John and {Goodman}, Alyssa and {Hacar}, {\'A}lvaro and {Lin}, Doug and {Meingast}, Stefan and {M{\"o}ller}, Torsten and {Piecka}, Martin and {Posch}, Laura and {Rottensteiner}, Alena and {Swiggum}, Cameren and {Zucker}, Catherine},
        title = "{The star formation history of the Sco-Cen association. Coherent star formation patterns in space and time}",
      journal = {\aap},
         year = 2023,
        month = oct,
       volume = {678},
          eid = {A71},
        pages = {A71},
          doi = {10.1051/0004-6361/202346901},
archivePrefix = {arXiv},
       eprint = {2302.07853},
 primaryClass = {astro-ph.SR},
       adsurl = {https://ui.adsabs.harvard.edu/abs/2023A&A...678A..71R}
}

@ARTICLE{Murphy_2026,
       author = {{Murphy}, Simon J. and {Gautam}, Anuj and {Claytor}, Zachary R.},
        title = "{Quantifying isochrone-based age uncertainties for rapidly rotating A-type stars}",
      journal = {\mnras},
         year = 2026,
        month = jul,
          doi = {10.1093/mnras/stag1237},
archivePrefix = {arXiv},
       eprint = {2606.09485},
 primaryClass = {astro-ph.SR},
       adsurl = {https://ui.adsabs.harvard.edu/abs/2026MNRAS.tmp.1171M}
}

@misc{Ligang_2026_NGC3532,
      title={Gravity-mode main-sequence pulsators in the open clusters NGC 3532 and NGC 2516: Instability strip, near-core rotation, and internal structure}, 
      author={Gang Li and Chenyu He and Joey S. G. Mombarg and Dario J. Fritzewski and Conny Aerts},
      year={2026},
      eprint={2606.14359},
      archivePrefix={arXiv},
      primaryClass={astro-ph.SR},
      url={https://arxiv.org/abs/2606.14359}, 
}

@ARTICLE{Tassoul1980,
       author = {{Tassoul}, M.},
        title = "{Asymptotic approximations for stellar nonradial pulsations.}",
      journal = {\apjs},
         year = 1980,
        month = aug,
       volume = {43},
        pages = {469-490},
          doi = {10.1086/190678},
       adsurl = {https://ui.adsabs.harvard.edu/abs/1980ApJS...43..469T}
}

@ARTICLE{Zwintz2013_PMS_gdor_CoRoT,
       author = {{Zwintz}, K. and {Fossati}, L. and {Ryabchikova}, T. and {Kaiser}, A. and {Gruberbauer}, M. and {Barnes}, T.~G. and {Baglin}, A. and {Chaintreuil}, S.},
        title = "{{\ensuremath{\gamma}} Doradus pulsation in two pre-main sequence stars discovered by CoRoT}",
      journal = {\aap},
         year = 2013,
        month = feb,
       volume = {550},
          eid = {A121},
        pages = {A121},
          doi = {10.1051/0004-6361/201220127},
archivePrefix = {arXiv},
       eprint = {1301.0991},
 primaryClass = {astro-ph.SR},
       adsurl = {https://ui.adsabs.harvard.edu/abs/2013A&A...550A.121Z}
}

@ARTICLE{Zwintz201_NGC2264,
       author = {{Zwintz}, K. and {Moravveji}, E. and {P{\'a}pics}, P.~I. and {Tkachenko}, A. and {Przybilla}, N. and {Nieva}, M.-F. and {Kuschnig}, R. and {Antoci}, V. and {Lorenz}, D. and {Theme{\ss}l}, N. and {Fossati}, L. and {Barnes}, T.~G.},
        title = "{A comprehensive study of young B stars in NGC 2264 . I. Space photometry and asteroseismology}",
      journal = {\aap},
         year = 2017,
        month = may,
       volume = {601},
          eid = {A101},
        pages = {A101},
          doi = {10.1051/0004-6361/201630327},
archivePrefix = {arXiv},
       eprint = {1703.06456},
 primaryClass = {astro-ph.SR},
       adsurl = {https://ui.adsabs.harvard.edu/abs/2017A&A...601A.101Z}
}

@ARTICLE{Szewczuk2022_8264293,
       author = {{Szewczuk}, Wojciech and {Walczak}, Przemys{\l}aw and {Daszy{\'n}ska-Daszkiewicz}, Jadwiga and {Mo{\'z}dzierski}, Dawid},
        title = "{Seismic modelling of a very young SPB star - KIC 8264293}",
      journal = {\mnras},
         year = 2022,
        month = mar,
       volume = {511},
       number = {1},
        pages = {1529-1543},
          doi = {10.1093/mnras/stac168},
archivePrefix = {arXiv},
       eprint = {2201.07039},
 primaryClass = {astro-ph.SR},
       adsurl = {https://ui.adsabs.harvard.edu/abs/2022MNRAS.511.1529S}
}

@ARTICLE{IJspeert2024,
       author = {{IJspeert}, L.~W. and {Tkachenko}, A. and {Johnston}, C. and {Pr{\v{s}}a}, A. and {Wells}, M.~A. and {Aerts}, C.},
        title = "{Automated eccentricity measurement from raw eclipsing binary light curves with intrinsic variability}",
      journal = {\aap},
         year = 2024,
        month = may,
       volume = {685},
          eid = {A62},
        pages = {A62},
          doi = {10.1051/0004-6361/202349079},
archivePrefix = {arXiv},
       eprint = {2402.06084},
 primaryClass = {astro-ph.IM},
       adsurl = {https://ui.adsabs.harvard.edu/abs/2024A&A...685A..62I}
}

@INPROCEEDINGS{Ketzer2020_PMS_gdor_K2,
       author = {{Ketzer}, L. and {Zwintz}, K.},
        title = "{Pre-main Sequence Gravity-Mode Pulsators in K2}",
    booktitle = {Stars and their Variability Observed from Space},
         year = 2020,
       editor = {{Neiner}, C. and {Weiss}, W.~W. and {Baade}, D. and {Griffin}, R.~E. and {Lovekin}, C.~C. and {Moffat}, A.~F.~J.},
        month = jan,
        pages = {95-97},
       adsurl = {https://ui.adsabs.harvard.edu/abs/2020svos.conf...95K}
}

@INPROCEEDINGS{Brogaard2015,
       author = {{Brogaard}, Karsten and {Sandquist}, Eric and {Jessen-Hansen}, Jens and {Grundahl}, Frank and {Frandsen}, S{\o}ren},
        title = "{Exploiting the Open Clusters in the Kepler and CoRoT Fields}",
    booktitle = {Asteroseismology of Stellar Populations in the Milky Way},
         year = 2015,
       editor = {{Miglio}, Andrea and {Eggenberger}, Patrick and {Girardi}, L{\'e}o and {Montalb{\'a}n}, Josefina},
       series = {Astrophysics and Space Science Proceedings},
       volume = {39},
        month = jan,
        pages = {51},
          doi = {10.1007/978-3-319-10993-0_6},
archivePrefix = {arXiv},
       eprint = {1409.2271},
 primaryClass = {astro-ph.SR},
       adsurl = {https://ui.adsabs.harvard.edu/abs/2015ASSP...39...51B}
}

@ARTICLE{Piau2005,
       author = {{Piau}, L. and {Ballot}, J. and {Turck-Chi{\`e}ze}, S.},
        title = "{Seismic constraints on open clusters}",
      journal = {\aap},
         year = 2005,
        month = feb,
       volume = {430},
        pages = {571-583},
          doi = {10.1051/0004-6361:20041405},
archivePrefix = {arXiv},
       eprint = {astro-ph/0503157},
 primaryClass = {astro-ph},
       adsurl = {https://ui.adsabs.harvard.edu/abs/2005A&A...430..571P}
}

@ARTICLE{Kraus2009,
       author = {{Kraus}, Adam L. and {Hillenbrand}, Lynne A.},
        title = "{The Coevality of Young Binary Systems}",
      journal = {\apj},
         year = 2009,
        month = oct,
       volume = {704},
       number = {1},
        pages = {531-547},
          doi = {10.1088/0004-637X/704/1/531},
archivePrefix = {arXiv},
       eprint = {0909.0509},
 primaryClass = {astro-ph.SR},
       adsurl = {https://ui.adsabs.harvard.edu/abs/2009ApJ...704..531K}
}

@ARTICLE{Valle2016,
       author = {{Valle}, G. and {Dell'Omodarme}, M. and {Prada Moroni}, P.~G. and {Degl'Innocenti}, S.},
        title = "{A statistical test on the reliability of the non-coevality of stars in binary systems}",
      journal = {\aap},
         year = 2016,
        month = mar,
       volume = {587},
          eid = {A31},
        pages = {A31},
          doi = {10.1051/0004-6361/201526183},
archivePrefix = {arXiv},
       eprint = {1601.02892},
 primaryClass = {astro-ph.SR},
       adsurl = {https://ui.adsabs.harvard.edu/abs/2016A&A...587A..31V}
}

@ARTICLE{Hon2022,
       author = {{Hon}, Marc and {Kuszlewicz}, James S. and {Huber}, Daniel and {Stello}, Dennis and {Reyes}, Claudia},
        title = "{HD-TESS: An Asteroseismic Catalog of Bright Red Giants within TESS Continuous Viewing Zones}",
      journal = {\aj},
         year = 2022,
        month = oct,
       volume = {164},
       number = {4},
          eid = {135},
        pages = {135},
          doi = {10.3847/1538-3881/ac8931},
archivePrefix = {arXiv},
       eprint = {2208.06478},
 primaryClass = {astro-ph.SR},
       adsurl = {https://ui.adsabs.harvard.edu/abs/2022AJ....164..135H}
}

@ARTICLE{Salaris2017,
       author = {{Salaris}, Maurizio and {Cassisi}, Santi},
        title = "{Chemical element transport in stellar evolution models}",
      journal = {Royal Society Open Science},
         year = 2017,
        month = aug,
       volume = {4},
       number = {8},
          eid = {170192},
        pages = {170192},
          doi = {10.1098/rsos.170192},
archivePrefix = {arXiv},
       eprint = {1707.07454},
 primaryClass = {astro-ph.SR},
       adsurl = {https://ui.adsabs.harvard.edu/abs/2017RSOS....470192S}
}

@ARTICLE{Keszthelyi2020,
       author = {{Keszthelyi}, Z. and {Meynet}, G. and {Shultz}, M.~E. and {David-Uraz}, A. and {ud-Doula}, A. and {Townsend}, R.~H.~D. and {Wade}, G.~A. and {Georgy}, C. and {Petit}, V. and {Owocki}, S.~P.},
        title = "{The effects of surface fossil magnetic fields on massive star evolution - II. Implementation of magnetic braking in MESA and implications for the evolution of surface rotation in OB stars}",
      journal = {\mnras},
         year = 2020,
        month = jan,
       volume = {493},
       number = {1},
        pages = {518-535},
          doi = {10.1093/mnras/staa237},
archivePrefix = {arXiv},
       eprint = {2001.06239},
 primaryClass = {astro-ph.SR},
       adsurl = {https://ui.adsabs.harvard.edu/abs/2020MNRAS.493..518K}
}

@INPROCEEDINGS{Bowman2026,
       author = {{Bowman}, Dominic M. and {Bugnet}, Lisa},
        title = "{Asteroseismology}",
    booktitle = {Encyclopedia of Astrophysics},
         year = 2026,
       volume = {2},
        month = jan,
        pages = {133-153},
          doi = {10.1016/B978-0-443-21439-4.00036-5},
       adsurl = {https://ui.adsabs.harvard.edu/abs/2026enap....2..133B}
}

@ARTICLE{Kerr2024_Cep_Her,
       author = {{Kerr}, Ronan and {Kraus}, Adam L. and {Krolikowski}, Daniel and {Bouma}, Luke G. and {Farias}, Juan P.},
        title = "{SPYGLASS. V. Spatially and Temporally Structured Star-forming Environments in the Cepheus-Hercules Complex}",
      journal = {\apj},
         year = 2024,
        month = nov,
       volume = {975},
       number = {1},
          eid = {99},
        pages = {99},
          doi = {10.3847/1538-4357/ad7135},
archivePrefix = {arXiv},
       eprint = {2406.19530},
 primaryClass = {astro-ph.SR},
       adsurl = {https://ui.adsabs.harvard.edu/abs/2024ApJ...975...99K}
}

@ARTICLE{Auvergne2009CoRoT,
       author = {{Auvergne}, M. and {Bodin}, P. and {Boisnard}, L. and {Buey}, J.-T. and {Chaintreuil}, S. and {Epstein}, G. and {Jouret}, M. and {Lam-Trong}, T. and {Levacher}, P. and {Magnan}, A. and {Perez}, R. and {Plasson}, P. and {Plesseria}, J. and {Peter}, G. and {Steller}, M. and {Tiph{\`e}ne}, D. and {Baglin}, A. and {Agogu{\'e}}, P. and {Appourchaux}, T. and {Barbet}, D. and {Beaufort}, T. and {Bellenger}, R. and {Berlin}, R. and {Bernardi}, P. and {Blouin}, D. and {Boumier}, P. and {Bonneau}, F. and {Briet}, R. and {Butler}, B. and {Cautain}, R. and {Chiavassa}, F. and {Costes}, V. and {Cuvilho}, J. and {Cunha-Parro}, V. and {de Oliveira Fialho}, F. and {Decaudin}, M. and {Defise}, J.-M. and {Djalal}, S. and {Docclo}, A. and {Drummond}, R. and {Dupuis}, O. and {Exil}, G. and {Faur{\'e}}, C. and {Gaboriaud}, A. and {Gamet}, P. and {Gavalda}, P. and {Grolleau}, E. and {Gueguen}, L. and {Guivarc'h}, V. and {Guterman}, P. and {Hasiba}, J. and {Huntzinger}, G. and {Hustaix}, H. and {Imbert}, C. and {Jeanville}, G. and {Johlander}, B. and {Jorda}, L. and {Journoud}, P. and {Karioty}, F. and {Kerjean}, L. and {Lafond}, L. and {Lapeyrere}, V. and {Landiech}, P. and {Larqu{\'e}}, T. and {Laudet}, P. and {Le Merrer}, J. and {Leporati}, L. and {Leruyet}, B. and {Levieuge}, B. and {Llebaria}, A. and {Martin}, L. and {Mazy}, E. and {Mesnager}, J.-M. and {Michel}, J.-P. and {Moalic}, J.-P. and {Monjoin}, W. and {Naudet}, D. and {Neukirchner}, S. and {Nguyen-Kim}, K. and {Ollivier}, M. and {Orcesi}, J.-L. and {Ottacher}, H. and {Oulali}, A. and {Parisot}, J. and {Perruchot}, S. and {Piacentino}, A. and {Pinheiro da Silva}, L. and {Platzer}, J. and {Pontet}, B. and {Pradines}, A. and {Quentin}, C. and {Rohbeck}, U. and {Rolland}, G. and {Rollenhagen}, F. and {Romagnan}, R. and {Russ}, N. and {Samadi}, R. and {Schmidt}, R. and {Schwartz}, N. and {Sebbag}, I. and {Smit}, H. and {Sunter}, W. and {Tello}, M. and {Toulouse}, P. and {Ulmer}, B. and {Vandermarcq}, O. and {Vergnault}, E. and {Wallner}, R. and {Waultier}, G. and {Zanatta}, P.},
        title = "{The CoRoT satellite in flight: description and performance}",
      journal = {\aap},
         year = 2009,
        month = oct,
       volume = {506},
       number = {1},
        pages = {411-424},
          doi = {10.1051/0004-6361/200810860},
archivePrefix = {arXiv},
       eprint = {0901.2206},
 primaryClass = {astro-ph.SR},
       adsurl = {https://ui.adsabs.harvard.edu/abs/2009A&A...506..411A}
}

@ARTICLE{Fuller2021,
       author = {{Fuller}, Jim},
        title = "{Inverse tides in pulsating binary stars}",
      journal = {\mnras},
         year = 2021,
        month = feb,
       volume = {501},
       number = {1},
        pages = {483-490},
          doi = {10.1093/mnras/staa3636},
archivePrefix = {arXiv},
       eprint = {2011.06613},
 primaryClass = {astro-ph.SR},
       adsurl = {https://ui.adsabs.harvard.edu/abs/2021MNRAS.501..483F}
}

@ARTICLE{Fellay2026,
       author = {{Fellay}, L. and {Dupret}, M.-A. and {Ko{\l}aczek-Szyma{\'n}ski}, P.~A.},
        title = "{Resonance locking and tidal evolution in rotating {\ensuremath{\gamma}}-Doradus binaries}",
      journal = {\aap},
         year = 2026,
        month = mar,
       volume = {708},
          eid = {A109},
        pages = {A109},
          doi = {10.1051/0004-6361/202555832},
archivePrefix = {arXiv},
       eprint = {2511.21347},
 primaryClass = {astro-ph.SR},
       adsurl = {https://ui.adsabs.harvard.edu/abs/2026A&A...708A.109F}
}

@ARTICLE{Hills1975,
       author = {{Hills}, J.~G.},
        title = "{Encounters between binary and single stars and their effect on the dynamical evolution of stellar systems.}",
      journal = {\aj},
         year = 1975,
        month = oct,
       volume = {80},
        pages = {809-825},
          doi = {10.1086/111815},
       adsurl = {https://ui.adsabs.harvard.edu/abs/1975AJ.....80..809H}
}

@ARTICLE{Heggie1975,
       author = {{Heggie}, D.~C.},
        title = "{Binary evolution in stellar dynamics.}",
      journal = {\mnras},
         year = 1975,
        month = dec,
       volume = {173},
        pages = {729-787},
          doi = {10.1093/mnras/173.3.729},
       adsurl = {https://ui.adsabs.harvard.edu/abs/1975MNRAS.173..729H}
}

@ARTICLE{Fuller2024,
       author = {{Fuller}, Jim and {Felce}, Catherine},
        title = "{Super slowly spinning stars in close binaries}",
      journal = {\mnras},
         year = 2024,
        month = jan,
       volume = {527},
       number = {1},
        pages = {L103-L109},
          doi = {10.1093/mnrasl/slad150},
archivePrefix = {arXiv},
       eprint = {2309.08505},
 primaryClass = {astro-ph.SR},
       adsurl = {https://ui.adsabs.harvard.edu/abs/2024MNRAS.527L.103F}
}

@ARTICLE{2018AJ....156..123A,
       author = {{Astropy Collaboration} and {Price-Whelan}, A.~M. and {Sip{\H{o}}cz}, B.~M. and {G{\"u}nther}, H.~M. and {Lim}, P.~L. and {Crawford}, S.~M. and {Conseil}, S. and {Shupe}, D.~L. and {Craig}, M.~W. and {Dencheva}, N. and {Ginsburg}, A. and {VanderPlas}, J.~T. and {Bradley}, L.~D. and {P{\'e}rez-Su{\'a}rez}, D. and {de Val-Borro}, M. and {Aldcroft}, T.~L. and {Cruz}, K.~L. and {Robitaille}, T.~P. and {Tollerud}, E.~J. and {Ardelean}, C. and {Babej}, T. and {Bach}, Y.~P. and {Bachetti}, M. and {Bakanov}, A.~V. and {Bamford}, S.~P. and {Barentsen}, G. and {Barmby}, P. and {Baumbach}, A. and {Berry}, K.~L. and {Biscani}, F. and {Boquien}, M. and {Bostroem}, K.~A. and {Bouma}, L.~G. and {Brammer}, G.~B. and {Bray}, E.~M. and {Breytenbach}, H. and {Buddelmeijer}, H. and {Burke}, D.~J. and {Calderone}, G. and {Cano Rodr{\'\i}guez}, J.~L. and {Cara}, M. and {Cardoso}, J.~V.~M. and {Cheedella}, S. and {Copin}, Y. and {Corrales}, L. and {Crichton}, D. and {D'Avella}, D. and {Deil}, C. and {Depagne}, {\'E}. and {Dietrich}, J.~P. and {Donath}, A. and {Droettboom}, M. and {Earl}, N. and {Erben}, T. and {Fabbro}, S. and {Ferreira}, L.~A. and {Finethy}, T. and {Fox}, R.~T. and {Garrison}, L.~H. and {Gibbons}, S.~L.~J. and {Goldstein}, D.~A. and {Gommers}, R. and {Greco}, J.~P. and {Greenfield}, P. and {Groener}, A.~M. and {Grollier}, F. and {Hagen}, A. and {Hirst}, P. and {Homeier}, D. and {Horton}, A.~J. and {Hosseinzadeh}, G. and {Hu}, L. and {Hunkeler}, J.~S. and {Ivezi{\'c}}, {\v{Z}}. and {Jain}, A. and {Jenness}, T. and {Kanarek}, G. and {Kendrew}, S. and {Kern}, N.~S. and {Kerzendorf}, W.~E. and {Khvalko}, A. and {King}, J. and {Kirkby}, D. and {Kulkarni}, A.~M. and {Kumar}, A. and {Lee}, A. and {Lenz}, D. and {Littlefair}, S.~P. and {Ma}, Z. and {Macleod}, D.~M. and {Mastropietro}, M. and {McCully}, C. and {Montagnac}, S. and {Morris}, B.~M. and {Mueller}, M. and {Mumford}, S.~J. and {Muna}, D. and {Murphy}, N.~A. and {Nelson}, S. and {Nguyen}, G.~H. and {Ninan}, J.~P. and {N{\"o}the}, M. and {Ogaz}, S. and {Oh}, S. and {Parejko}, J.~K. and {Parley}, N. and {Pascual}, S. and {Patil}, R. and {Patil}, A.~A. and {Plunkett}, A.~L. and {Prochaska}, J.~X. and {Rastogi}, T. and {Reddy Janga}, V. and {Sabater}, J. and {Sakurikar}, P. and {Seifert}, M. and {Sherbert}, L.~E. and {Sherwood-Taylor}, H. and {Shih}, A.~Y. and {Sick}, J. and {Silbiger}, M.~T. and {Singanamalla}, S. and {Singer}, L.~P. and {Sladen}, P.~H. and {Sooley}, K.~A. and {Sornarajah}, S. and {Streicher}, O. and {Teuben}, P. and {Thomas}, S.~W. and {Tremblay}, G.~R. and {Turner}, J.~E.~H. and {Terr{\'o}n}, V. and {van Kerkwijk}, M.~H. and {de la Vega}, A. and {Watkins}, L.~L. and {Weaver}, B.~A. and {Whitmore}, J.~B. and {Woillez}, J. and {Zabalza}, V. and {Astropy Contributors}},
        title = "{The Astropy Project: Building an Open-science Project and Status of the v2.0 Core Package}",
      journal = {\aj},
         year = 2018,
        month = sep,
       volume = {156},
       number = {3},
          eid = {123},
        pages = {123},
          doi = {10.3847/1538-3881/aabc4f},
archivePrefix = {arXiv},
       eprint = {1801.02634},
 primaryClass = {astro-ph.IM},
       adsurl = {https://ui.adsabs.harvard.edu/abs/2018AJ....156..123A}
}

@ARTICLE{Bouabid2011_PMS,
       author = {{Bouabid}, M.-P. and {Montalb{\'a}n}, J. and {Miglio}, A. and {Dupret}, M.-A. and {Grigahc{\`e}ne}, A. and {Noels}, A.},
        title = "{Theoretical seismic properties of pre-main sequence {\ensuremath{\gamma}} Doradus pulsators}",
      journal = {\aap},
         year = 2011,
        month = jul,
       volume = {531},
          eid = {A145},
        pages = {A145},
          doi = {10.1051/0004-6361/201116440},
archivePrefix = {arXiv},
       eprint = {1103.4389},
 primaryClass = {astro-ph.SR},
       adsurl = {https://ui.adsabs.harvard.edu/abs/2011A&A...531A.145B}
}

@ARTICLE{2013A&A...558A..33A,
       author = {{Astropy Collaboration} and {Robitaille}, Thomas P. and
         {Tollerud}, Erik J. and {Greenfield}, Perry and {Droettboom}, Michael and
         {Bray}, Erik and {Aldcroft}, Tom and {Davis}, Matt and
         {Ginsburg}, Adam and {Price-Whelan}, Adrian M. and
         {Kerzendorf}, Wolfgang E. and {Conley}, Alexander and {Crighton}, Neil and
         {Barbary}, Kyle and {Muna}, Demitri and {Ferguson}, Henry and
         {Grollier}, Fr{\'e}d{\'e}ric and {Parikh}, Madhura M. and
         {Nair}, Prasanth H. and {Unther}, Hans M. and {Deil}, Christoph and
         {Woillez}, Julien and {Conseil}, Simon and {Kramer}, Roban and
         {Turner}, James E.~H. and {Singer}, Leo and {Fox}, Ryan and
         {Weaver}, Benjamin A. and {Zabalza}, Victor and {Edwards}, Zachary I. and
         {Azalee Bostroem}, K. and {Burke}, D.~J. and {Casey}, Andrew R. and
         {Crawford}, Steven M. and {Dencheva}, Nadia and {Ely}, Justin and
         {Jenness}, Tim and {Labrie}, Kathleen and {Lim}, Pey Lian and
         {Pierfederici}, Francesco and {Pontzen}, Andrew and {Ptak}, Andy and
         {Refsdal}, Brian and {Servillat}, Mathieu and {Streicher}, Ole},
        title = "{Astropy: A community Python package for astronomy}",
      journal = {\aap},
         year = "2013",
        month = "Oct",
       volume = {558},
          eid = {A33},
        pages = {A33},
          doi = {10.1051/0004-6361/201322068},
archivePrefix = {arXiv},
       eprint = {1307.6212},
 primaryClass = {astro-ph.IM},
       adsurl = {https://ui.adsabs.harvard.edu/abs/2013A&A...558A..33A}
}

@ARTICLE{Breger1993,
       author = {{Breger}, M. and {Stich}, J. and {Garrido}, R. and {Martin}, B. and {Jiang}, S.-Y. and {Li}, Z.-P. and {Hube}, D.~P. and {Ostermann}, W. and {Paparo}, M. and {Scheck}, M.},
        title = "{Nonradial pulsation of the delta Scuti star BU CANCRI in the Praesepe cluster.}",
      journal = {\aap},
         year = 1993,
        month = apr,
       volume = {271},
        pages = {482-486},
       adsurl = {https://ui.adsabs.harvard.edu/abs/1993A&A...271..482B}
}

@ARTICLE{VanReeth2015-method,
       author = {{Van Reeth}, T. and {Tkachenko}, A. and {Aerts}, C. and {P{\'a}pics}, P.~I. and {Degroote}, P. and {Debosscher}, J. and {Zwintz}, K. and {Bloemen}, S. and {De Smedt}, K. and {Hrudkova}, M. and {Raskin}, G. and {Van Winckel}, H.},
        title = "{Detecting non-uniform period spacings in the Kepler photometry of {\ensuremath{\gamma}} Doradus stars: methodology and case studies}",
      journal = {\aap},
         year = 2015,
        month = feb,
       volume = {574},
          eid = {A17},
        pages = {A17},
          doi = {10.1051/0004-6361/201424585},
archivePrefix = {arXiv},
       eprint = {1410.8178},
 primaryClass = {astro-ph.SR},
       adsurl = {https://ui.adsabs.harvard.edu/abs/2015A&A...574A..17V}
}

@ARTICLE{Fritzewski2020-NGC2516,
       author = {{Fritzewski}, D.~J. and {Barnes}, S.~A. and {James}, D.~J. and {Strassmeier}, K.~G.},
        title = "{The rotation period distribution of the rich Pleiades-age southern open cluster NGC 2516. Existence of a representative zero-age main sequence distribution}",
      journal = {\aap},
         year = 2020,
        month = sep,
       volume = {641},
          eid = {A51},
        pages = {A51},
          doi = {10.1051/0004-6361/201936860},
archivePrefix = {arXiv},
       eprint = {2112.03299},
 primaryClass = {astro-ph.SR},
       adsurl = {https://ui.adsabs.harvard.edu/abs/2020A&A...641A..51F}
}

@ARTICLE{Fritzewski-Pleiades,
       author = {{Fritzewski}, D.~J. and {Kemp}, A. and {Li}, G. and {Aerts}, C.},
        title = "{Probing stellar rotation in the Pleiades with gravity-mode pulsators}",
      journal = {\aap, in press},
         year = 2026,
        month = dec,
          eid = {arXiv:2512.09395},
        pages = {arXiv:2512.09395},
          doi = {10.48550/arXiv.2512.09395},
archivePrefix = {arXiv},
       eprint = {2512.09395},
 primaryClass = {astro-ph.SR},
       adsurl = {https://ui.adsabs.harvard.edu/abs/2025arXiv251209395F}
}

@ARTICLE{Aerts2026,
       author = {{Aerts}, Conny},
        title = "{Distributions and evolution of the equatorial rotation velocities of 2937 BAF-type main-sequence stars from asteroseismology: A break in the specific angular momentum at M ≃ 2.5 M$_{{\ensuremath{\odot}}}$}",
      journal = {\aap},
         year = 2025,
        month = dec,
       volume = {704},
          eid = {A332},
        pages = {A332},
          doi = {10.1051/0004-6361/202556794},
archivePrefix = {arXiv},
       eprint = {2511.02909},
 primaryClass = {astro-ph.SR},
       adsurl = {https://ui.adsabs.harvard.edu/abs/2025A&A...704A.332A}
}

@ARTICLE{Berry2025_dsct_in_NGC3532,
       author = {{Berry}, Ian and {Huber}, Daniel and {Li}, Yaguang and {Hey}, Daniel and {Bedding}, Timothy R. and {Murphy}, Simon J.},
        title = "{Discovery of 79 $δ$ Scuti Stars in NGC 3532 Suggests a Decrease of Pulsator Occurrence with Age}",
      journal = {arXiv e-prints},
         year = 2025,
        month = oct,
          eid = {arXiv:2510.20048},
        pages = {arXiv:2510.20048},
          doi = {10.48550/arXiv.2510.20048},
archivePrefix = {arXiv},
       eprint = {2510.20048},
 primaryClass = {astro-ph.SR},
       adsurl = {https://ui.adsabs.harvard.edu/abs/2025arXiv251020048B}
}

@ARTICLE{Wang_Wang_Ong_2026ApJ,
       author = {{Wang}, Xian-Yu and {Wang}, Songhu and {Ong}, J.~M. Joel},
        title = "{Unified Kraft Break at {\ensuremath{\sim}}6500 K: A Newly Identified Single-star Obliquity Transition Matches the Classical Rotation Break}",
      journal = {\apjl},
         year = 2026,
        month = jan,
       volume = {996},
       number = {1},
          eid = {L7},
        pages = {L7},
          doi = {10.3847/2041-8213/ae21c5},
archivePrefix = {arXiv},
       eprint = {2511.15610},
 primaryClass = {astro-ph.EP},
       adsurl = {https://ui.adsabs.harvard.edu/abs/2026ApJ...996L...7W}
}

@ARTICLE{Pang2022,
       author = {{Pang}, Xiaoying and {Tang}, Shih-Yun and {Li}, Yuqian and {Yu}, Zeqiu and {Wang}, Long and {Li}, Jiayu and {Li}, Yezhang and {Wang}, Yifan and {Wang}, Yanshu and {Zhang}, Teng and {Pasquato}, Mario and {Kouwenhoven}, M.~B.~N.},
        title = "{3D Morphology of Open Clusters in the Solar Neighborhood with Gaia EDR 3. II. Hierarchical Star Formation Revealed by Spatial and Kinematic Substructures}",
      journal = {\apj},
         year = 2022,
        month = jun,
       volume = {931},
       number = {2},
          eid = {156},
        pages = {156},
          doi = {10.3847/1538-4357/ac674e},
archivePrefix = {arXiv},
       eprint = {2204.06000},
 primaryClass = {astro-ph.GA},
       adsurl = {https://ui.adsabs.harvard.edu/abs/2022ApJ...931..156P}
}

@ARTICLE{Varghese2023,
       author = {{Varghese}, A. and {Ratnasingam}, R.~P. and {Vanon}, R. and {Edelmann}, P.~V.~F. and {Rogers}, T.~M.},
        title = "{Chemical Mixing Induced by Internal Gravity Waves in Intermediate-mass Stars}",
      journal = {\apj},
         year = 2023,
        month = jan,
       volume = {942},
       number = {1},
          eid = {53},
        pages = {53},
          doi = {10.3847/1538-4357/aca092},
archivePrefix = {arXiv},
       eprint = {2211.06432},
 primaryClass = {astro-ph.SR},
       adsurl = {https://ui.adsabs.harvard.edu/abs/2023ApJ...942...53V}
}

@ARTICLE{Varghese2025,
       author = {{Varghese}, A. and {Ratnasingam}, R.~P. and {Ram{\'\i}rez-Galeano}, L. and {Mathis}, S. and {Rogers}, T.~M.},
        title = "{Numerical Simulations Confirm Wave-induced Shear Mixing in Stellar Interiors}",
      journal = {\apjl},
         year = 2025,
        month = oct,
       volume = {992},
       number = {1},
          eid = {L5},
        pages = {L5},
          doi = {10.3847/2041-8213/ae09ae},
archivePrefix = {arXiv},
       eprint = {2509.18244},
 primaryClass = {astro-ph.SR},
       adsurl = {https://ui.adsabs.harvard.edu/abs/2025ApJ...992L...5V}
}

@ARTICLE{Dupret2005,
       author = {{Dupret}, M. -A. and {Grigahc{\`e}ne}, A. and {Garrido}, R. and {Gabriel}, M. and {Scuflaire}, R.},
        title = "{Convection-pulsation coupling. II. Excitation and stabilization mechanisms in {\ensuremath{\delta}} Sct and {\ensuremath{\gamma}} Dor stars}",
      journal = {\aap},
         year = 2005,
        month = jun,
       volume = {435},
       number = {3},
        pages = {927-939},
          doi = {10.1051/0004-6361:20041817},
       adsurl = {https://ui.adsabs.harvard.edu/abs/2005A&A...435..927D}
}

@ARTICLE{Aerts2018ApJS,
       author = {{Aerts}, C. and {Molenberghs}, G. and {Michielsen}, M. and {Pedersen}, M.~G. and {Bj{\"o}rklund}, R. and {Johnston}, C. and {Mombarg}, J.~S.~G. and {Bowman}, D.~M. and {Buysschaert}, B. and {P{\'a}pics}, P.~I. and {Sekaran}, S. and {Sundqvist}, J.~O. and {Tkachenko}, A. and {Truyaert}, K. and {Van Reeth}, T. and {Vermeyen}, E.},
        title = "{Forward Asteroseismic Modeling of Stars with a Convective Core from Gravity-mode Oscillations: Parameter Estimation and Stellar Model Selection}",
      journal = {\apjs},
         year = 2018,
        month = jul,
       volume = {237},
       number = {1},
          eid = {15},
        pages = {15},
          doi = {10.3847/1538-4365/aaccfb},
archivePrefix = {arXiv},
       eprint = {1806.06869},
 primaryClass = {astro-ph.SR},
       adsurl = {https://ui.adsabs.harvard.edu/abs/2018ApJS..237...15A}
}

@ARTICLE{Rauer2014PLATO,
       author = {{Rauer}, H. and {Catala}, C. and {Aerts}, C. and {Appourchaux}, T. and {Benz}, W. and {Brandeker}, A. and {Christensen-Dalsgaard}, J. and {Deleuil}, M. and {Gizon}, L. and {Goupil}, M. -J. and {G{\"u}del}, M. and {Janot-Pacheco}, E. and {Mas-Hesse}, M. and {Pagano}, I. and {Piotto}, G. and {Pollacco}, D. and {Santos}, {\.{C}}. and {Smith}, A. and {Su{\'a}rez}, J. -C. and {Szab{\'o}}, R. and {Udry}, S. and {Adibekyan}, V. and {Alibert}, Y. and {Almenara}, J. -M. and {Amaro-Seoane}, P. and {Eiff}, M. Ammler-von and {Asplund}, M. and {Antonello}, E. and {Barnes}, S. and {Baudin}, F. and {Belkacem}, K. and {Bergemann}, M. and {Bihain}, G. and {Birch}, A.~C. and {Bonfils}, X. and {Boisse}, I. and {Bonomo}, A.~S. and {Borsa}, F. and {Brand{\~a}o}, I.~M. and {Brocato}, E. and {Brun}, S. and {Burleigh}, M. and {Burston}, R. and {Cabrera}, J. and {Cassisi}, S. and {Chaplin}, W. and {Charpinet}, S. and {Chiappini}, C. and {Church}, R.~P. and {Csizmadia}, Sz. and {Cunha}, M. and {Damasso}, M. and {Davies}, M.~B. and {Deeg}, H.~J. and {D{\'\i}az}, R.~F. and {Dreizler}, S. and {Dreyer}, C. and {Eggenberger}, P. and {Ehrenreich}, D. and {Eigm{\"u}ller}, P. and {Erikson}, A. and {Farmer}, R. and {Feltzing}, S. and {de Oliveira Fialho}, F. and {Figueira}, P. and {Forveille}, T. and {Fridlund}, M. and {Garc{\'\i}a}, R.~A. and {Giommi}, P. and {Giuffrida}, G. and {Godolt}, M. and {Gomes da Silva}, J. and {Granzer}, T. and {Grenfell}, J.~L. and {Grotsch-Noels}, A. and {G{\"u}nther}, E. and {Haswell}, C.~A. and {Hatzes}, A.~P. and {H{\'e}brard}, G. and {Hekker}, S. and {Helled}, R. and {Heng}, K. and {Jenkins}, J.~M. and {Johansen}, A. and {Khodachenko}, M.~L. and {Kislyakova}, K.~G. and {Kley}, W. and {Kolb}, U. and {Krivova}, N. and {Kupka}, F. and {Lammer}, H. and {Lanza}, A.~F. and {Lebreton}, Y. and {Magrin}, D. and {Marcos-Arenal}, P. and {Marrese}, P.~M. and {Marques}, J.~P. and {Martins}, J. and {Mathis}, S. and {Mathur}, S. and {Messina}, S. and {Miglio}, A. and {Montalban}, J. and {Montalto}, M. and {Monteiro}, M.~J.~P.~F.~G. and {Moradi}, H. and {Moravveji}, E. and {Mordasini}, C. and {Morel}, T. and {Mortier}, A. and {Nascimbeni}, V. and {Nelson}, R.~P. and {Nielsen}, M.~B. and {Noack}, L. and {Norton}, A.~J. and {Ofir}, A. and {Oshagh}, M. and {Ouazzani}, R. -M. and {P{\'a}pics}, P. and {Parro}, V.~C. and {Petit}, P. and {Plez}, B. and {Poretti}, E. and {Quirrenbach}, A. and {Ragazzoni}, R. and {Raimondo}, G. and {Rainer}, M. and {Reese}, D.~R. and {Redmer}, R. and {Reffert}, S. and {Rojas-Ayala}, B. and {Roxburgh}, I.~W. and {Salmon}, S. and {Santerne}, A. and {Schneider}, J. and {Schou}, J. and {Schuh}, S. and {Schunker}, H. and {Silva-Valio}, A. and {Silvotti}, R. and {Skillen}, I. and {Snellen}, I. and {Sohl}, F. and {Sousa}, S.~G. and {Sozzetti}, A. and {Stello}, D. and {Strassmeier}, K.~G. and {{\v{S}}vanda}, M. and {Szab{\'o}}, Gy. M. and {Tkachenko}, A. and {Valencia}, D. and {Van Grootel}, V. and {Vauclair}, S.~D. and {Ventura}, P. and {Wagner}, F.~W. and {Walton}, N.~A. and {Weingrill}, J. and {Werner}, S.~C. and {Wheatley}, P.~J. and {Zwintz}, K.},
        title = "{The PLATO 2.0 mission}",
      journal = {Experimental Astronomy},
         year = 2014,
        month = nov,
       volume = {38},
       number = {1-2},
        pages = {249-330},
          doi = {10.1007/s10686-014-9383-4},
archivePrefix = {arXiv},
       eprint = {1310.0696},
 primaryClass = {astro-ph.EP},
       adsurl = {https://ui.adsabs.harvard.edu/abs/2014ExA....38..249R}
}

@ARTICLE{Ge_2022_ET,
       author = {{Ge}, Jian and {Zhang}, Hui and {Zang}, Weicheng and {Deng}, Hongping and {Mao}, Shude and {Xie}, Ji-Wei and {Liu}, Hui-Gen and {Zhou}, Ji-Lin and {Willis}, Kevin and {Huang}, Chelsea and {Howell}, Steve B. and {Feng}, Fabo and {Zhu}, Jiapeng and {Yao}, Xinyu and {Liu}, Beibei and {Aizawa}, Masataka and {Zhu}, Wei and {Li}, Ya-Ping and {Ma}, Bo and {Ye}, Quanzhi and {Yu}, Jie and {Xiang}, Maosheng and {Yu}, Cong and {Liu}, Shangfei and {Yang}, Ming and {Wang}, Mu-Tian and {Shi}, Xian and {Fang}, Tong and {Zong}, Weikai and {Liu}, Jinzhong and {Zhang}, Yu and {Zhang}, Liyun and {El-Badry}, Kareem and {Shen}, Rongfeng and {Tam}, Pak-Hin Thomas and {Hu}, Zhecheng and {Yang}, Yanlv and {Zou}, Yuan-Chuan and {Wu}, Jia-Li and {Lei}, Wei-Hua and {Wei}, Jun-Jie and {Wu}, Xue-Feng and {Sun}, Tian-Rui and {Wang}, Fa-Yin and {Zhang}, Bin-Bin and {Xu}, Dong and {Yang}, Yuan-Pei and {Li}, Wen-Xiong and {Xiang}, Dan-Feng and {Wang}, Xiaofeng and {Wang}, Tinggui and {Zhang}, Bing and {Jia}, Peng and {Yuan}, Haibo and {Zhang}, Jinghua and {Xuesong Wang}, Sharon and {Gan}, Tianjun and {Wang}, Wei and {Zhao}, Yinan and {Liu}, Yujuan and {Wei}, Chuanxin and {Kang}, Yanwu and {Yang}, Baoyu and {Qi}, Chao and {Liu}, Xiaohua and {Zhang}, Quan and {Zhu}, Yuji and {Zhou}, Dan and {Zhang}, Congcong and {Yu}, Yong and {Zhang}, Yongshuai and {Li}, Yan and {Tang}, Zhenghong and {Wang}, Chaoyan and {Wang}, Fengtao and {Li}, Wei and {Cheng}, Pengfei and {Shen}, Chao and {Li}, Baopeng and {Pan}, Yue and {Yang}, Sen and {Gao}, Wei and {Song}, Zongxi and {Wang}, Jian and {Zhang}, Hongfei and {Chen}, Cheng and {Wang}, Hui and {Zhang}, Jun and {Wang}, Zhiyue and {Zeng}, Feng and {Zheng}, Zhenhao and {Zhu}, Jie and {Guo}, Yingfan and {Zhang}, Yihao and {Li}, Yudong and {Wen}, Lin and {Feng}, Jie and {Chen}, Wen and {Chen}, Kun and {Han}, Xingbo and {Yang}, Yingquan and {Wang}, Haoyu and {Duan}, Xuliang and {Huang}, Jiangjiang and {Liang}, Hong and {Bi}, Shaolan and {Gai}, Ning and {Ge}, Zhishuai and {Guo}, Zhao and {Huang}, Yang and {Li}, Gang and {Li}, Haining and {Li}, Tanda and {Yuxi} and {Lu} and {Rix}, Hans-Walter and {Shi}, Jianrong and {Song}, Fen and {Tang}, Yanke and {Ting}, Yuan-Sen and {Wu}, Tao and {Wu}, Yaqian and {Yang}, Taozhi and {Yin}, Qing-Zhu and {Gould}, Andrew and {Lee}, Chung-Uk and {Dong}, Subo and {Yee}, Jennifer C. and {Shvartzvald}, Yossi and {Yang}, Hongjing and {Kuang}, Renkun and {Zhang}, Jiyuan and {Liao}, Shilong and {Qi}, Zhaoxiang and {Yang}, Jun and {Zhang}, Ruisheng and {Jiang}, Chen and {Ou}, Jian-Wen and {Li}, Yaguang and {Beck}, Paul and {Bedding}, Timothy R. and {Campante}, Tiago L. and {Chaplin}, William J. and {Christensen-Dalsgaard}, J{\o}rgen and {Garc{\'\i}a}, Rafael A. and {Gaulme}, Patrick and {Gizon}, Laurent and {Hekker}, Saskia and {Huber}, Daniel and {Khanna}, Shourya and {Li}, Yan and {Mathur}, Savita and {Miglio}, Andrea and {Mosser}, Beno{\^\i}t and {Ong}, J.~M. Joel and {Santos}, {\^A}ngela R.~G. and {Stello}, Dennis and {Bowman}, Dominic M. and {Lares-Martiz}, Mariel and {Murphy}, Simon and {Niu}, Jia-Shu and {Ma}, Xiao-Yu and {Moln{\'a}r}, L{\'a}szl{\'o} and {Fu}, Jian-Ning and {De Cat}, Peter and {Su}, Jie and {consortium}, the ET},
        title = "{ET White Paper: To Find the First Earth 2.0}",
      journal = {arXiv e-prints},
         year = 2022,
        month = jun,
          eid = {arXiv:2206.06693},
        pages = {arXiv:2206.06693},
          doi = {10.48550/arXiv.2206.06693},
archivePrefix = {arXiv},
       eprint = {2206.06693},
 primaryClass = {astro-ph.IM},
       adsurl = {https://ui.adsabs.harvard.edu/abs/2022arXiv220606693G}
}

@ARTICLE{Kerr2023SPYGLASSIV,
       author = {{Kerr}, Ronan and {Kraus}, Adam L. and {Rizzuto}, Aaron C.},
        title = "{SPYGLASS. IV. New Stellar Survey of Recent Star Formation within 1 kpc}",
      journal = {\apj},
         year = 2023,
        month = sep,
       volume = {954},
       number = {2},
          eid = {134},
        pages = {134},
          doi = {10.3847/1538-4357/ace5b3},
archivePrefix = {arXiv},
       eprint = {2306.08150},
 primaryClass = {astro-ph.GA},
       adsurl = {https://ui.adsabs.harvard.edu/abs/2023ApJ...954..134K}
}

@ARTICLE{Dileep2025_NGC2126,
       author = {{Dileep}, Athul and {Joshi}, Santosh and {Lata}, Sneh and {Lampens}, Patricia and {De Cat}, Peter and {Z{\'u}{\~n}iga-Fern{\'a}ndez}, Sebastian and {Mkrtichian}, David and {Kumar S.}, Pramod and {Sarkar}, Mrinmoy and {Panchal}, Alaxender and {Joshi}, Yogesh C. and {Rishi}, C. and {Panwar}, Neelam and {Jain}, Arjav and {Rathore}, Neeraj},
        title = "{Investigating pulsating variables and eclipsing binaries in NGC 2126 using ground- and space-based photometry, astrometry, spectroscopy, and modelling}",
      journal = {\mnras},
         year = 2025,
        month = apr,
       volume = {538},
       number = {3},
        pages = {1747-1772},
          doi = {10.1093/mnras/staf361},
archivePrefix = {arXiv},
       eprint = {2502.18112},
 primaryClass = {astro-ph.SR},
       adsurl = {https://ui.adsabs.harvard.edu/abs/2025MNRAS.538.1747D}
}

@ARTICLE{Bouma2023_gyrointerp,
       author = {{Bouma}, Luke G. and {Palumbo}, Elsa K. and {Hillenbrand}, Lynne A.},
        title = "{The Empirical Limits of Gyrochronology}",
      journal = {\apjl},
         year = 2023,
        month = apr,
       volume = {947},
       number = {1},
          eid = {L3},
        pages = {L3},
          doi = {10.3847/2041-8213/acc589},
archivePrefix = {arXiv},
       eprint = {2303.08830},
 primaryClass = {astro-ph.SR},
       adsurl = {https://ui.adsabs.harvard.edu/abs/2023ApJ...947L...3B}
}

@ARTICLE{Kos2018_NGC1901_age,
       author = {{Kos}, Janez and {de Silva}, Gayandhi and {Buder}, Sven and {Bland-Hawthorn}, Joss and {Sharma}, Sanjib and {Asplund}, Martin and {D'Orazi}, Valentina and {Duong}, Ly and {Freeman}, Ken and {Lewis}, Geraint F. and {Lin}, Jane and {Lind}, Karin and {Martell}, Sarah L. and {Schlesinger}, Katharine J. and {Simpson}, Jeffrey D. and {Zucker}, Daniel B. and {Zwitter}, Toma{\v{z}} and {Bedding}, Timothy R. and {{\v{C}}otar}, Klemen and {Horner}, Jonathan and {Nordlander}, Thomas and {Stello}, Denis and {Ting}, Yuan-Sen and {Traven}, Gregor},
        title = "{The GALAH survey and Gaia DR2: (non-)existence of five sparse high-latitude open clusters}",
      journal = {\mnras},
         year = 2018,
        month = nov,
       volume = {480},
       number = {4},
        pages = {5242-5259},
          doi = {10.1093/mnras/sty2171},
archivePrefix = {arXiv},
       eprint = {1807.00822},
 primaryClass = {astro-ph.GA},
       adsurl = {https://ui.adsabs.harvard.edu/abs/2018MNRAS.480.5242K}
}

@ARTICLE{Malhotra2026_NGC1901,
       author = {{Malhotra}, Sagar and {Castro-Ginard}, Alfred and {Anders}, Friedrich and {Jordi}, Carme and {Donada}, Judit and {Luri}, Xavier and {Balaguer-N{\'u}{\~n}ez}, Lola and {Qin}, Songmei and {Jiang}, Yueyue and {{\v{Z}}upi{\'c}}, Andrija},
        title = "{Stellar masses and mass ratios for Gaia open cluster members}",
      journal = {\aap},
         year = 2026,
        month = feb,
       volume = {706},
          eid = {A62},
        pages = {A62},
          doi = {10.1051/0004-6361/202557529},
archivePrefix = {arXiv},
       eprint = {2512.10616},
 primaryClass = {astro-ph.GA},
       adsurl = {https://ui.adsabs.harvard.edu/abs/2026A&A...706A..62M}
}

@ARTICLE{Bernizzoni2025_NGC1901,
       author = {{Bernizzoni}, Mara and {Bedding}, Timothy R. and {Mani}, Prasad and {Bouma}, Luke G. and {Crawford}, Courtney L. and {Curtis}, Jason Lee and {Montet}, Benjamin T.},
        title = "{NGC 1901: A new gyrochronal benchmark cluster in the TESS Southern continuous viewing zone}",
         year = 2026,
      journal = {submitted}
}

@ARTICLE{Murphy2024Cep_Her_complex,
       author = {{Murphy}, Simon J. and {Bedding}, Timothy R. and {Gautam}, Anuj and {Kerr}, Ronan P. and {Mani}, Prasad},
        title = "{The {\ensuremath{\delta}} Scuti stars of the Cep-Her Complex - I. Pulsator fraction, rotation, asteroseismic large spacings, and the {\ensuremath{\nu}}$_{max}$ relation}",
      journal = {\mnras},
         year = 2024,
        month = nov,
       volume = {534},
       number = {4},
        pages = {3022-3039},
          doi = {10.1093/mnras/stae2226},
archivePrefix = {arXiv},
       eprint = {2409.13135},
 primaryClass = {astro-ph.SR},
       adsurl = {https://ui.adsabs.harvard.edu/abs/2024MNRAS.534.3022M}
}

@ARTICLE{Joyce2023,
       author = {{Joyce}, Meridith and {Tayar}, Jamie},
        title = "{A Review of the Mixing Length Theory of Convection in 1D Stellar Modeling}",
      journal = {Galaxies},
         year = 2023,
        month = jun,
       volume = {11},
       number = {3},
          eid = {75},
        pages = {75},
          doi = {10.3390/galaxies11030075},
archivePrefix = {arXiv},
       eprint = {2303.09596},
 primaryClass = {astro-ph.SR},
       adsurl = {https://ui.adsabs.harvard.edu/abs/2023Galax..11...75J}
}

@ARTICLE{HeChenyu2025_NGC332,
       author = {{He}, Chenyu and {Li}, Chengyuan and {Li}, Gang},
        title = "{Fast-rotating A- and F-type Stars with H{\ensuremath{\alpha}} Emissions in NGC 3532: Candidate UV-dim Stars?}",
      journal = {\apj},
         year = 2025,
        month = feb,
       volume = {979},
       number = {2},
          eid = {246},
        pages = {246},
          doi = {10.3847/1538-4357/ad9de3},
archivePrefix = {arXiv},
       eprint = {2412.09217},
 primaryClass = {astro-ph.SR},
       adsurl = {https://ui.adsabs.harvard.edu/abs/2025ApJ...979..246H}
}

@ARTICLE{DeCat2002A&A,
       author = {{De Cat}, P. and {Aerts}, C.},
        title = "{A study of bright southern slowly pulsating B stars. II. The intrinsic frequencies}",
      journal = {\aap},
         year = 2002,
        month = oct,
       volume = {393},
        pages = {965-981},
          doi = {10.1051/0004-6361:20021068},
       adsurl = {https://ui.adsabs.harvard.edu/abs/2002A&A...393..965D}
}

@ARTICLE{Paxton2013ApJS,
       author = {{Paxton}, Bill and {Cantiello}, Matteo and {Arras}, Phil and {Bildsten}, Lars and {Brown}, Edward F. and {Dotter}, Aaron and {Mankovich}, Christopher and {Montgomery}, M.~H. and {Stello}, Dennis and {Timmes}, F.~X. and {Townsend}, Richard},
        title = "{Modules for Experiments in Stellar Astrophysics (MESA): Planets, Oscillations, Rotation, and Massive Stars}",
      journal = {\apjs},
         year = 2013,
        month = sep,
       volume = {208},
       number = {1},
          eid = {4},
        pages = {4},
          doi = {10.1088/0067-0049/208/1/4},
archivePrefix = {arXiv},
       eprint = {1301.0319},
 primaryClass = {astro-ph.SR},
       adsurl = {https://ui.adsabs.harvard.edu/abs/2013ApJS..208....4P}
}

@ARTICLE{Mombarg2025,
       author = {{Mombarg}, J.~S.~G. and {Varghese}, A. and {Ratnasingam}, R.~P.},
        title = "{Calibrating chemical mixing induced by internal gravity waves based on hydrodynamical simulations: The chemical evolution of OB-type stars}",
      journal = {\aap},
         year = 2025,
        month = mar,
       volume = {695},
          eid = {A255},
        pages = {A255},
          doi = {10.1051/0004-6361/202452956},
archivePrefix = {arXiv},
       eprint = {2502.13204},
 primaryClass = {astro-ph.SR},
       adsurl = {https://ui.adsabs.harvard.edu/abs/2025A&A...695A.255M}
}

@ARTICLE{Bedding2023ApJ,
       author = {{Bedding}, Timothy R. and {Murphy}, Simon J. and {Crawford}, Courtney and {Hey}, Daniel R. and {Huber}, Daniel and {Kjeldsen}, Hans and {Li}, Yaguang and {Mann}, Andrew W. and {Torres}, Guillermo and {White}, Timothy R. and {Zhou}, George},
        title = "{TESS Observations of the Pleiades Cluster: A Nursery for {\ensuremath{\delta}} Scuti Stars}",
      journal = {\apjl},
         year = 2023,
        month = mar,
       volume = {946},
       number = {1},
          eid = {L10},
        pages = {L10},
          doi = {10.3847/2041-8213/acc17a},
archivePrefix = {arXiv},
       eprint = {2212.12087},
 primaryClass = {astro-ph.SR},
       adsurl = {https://ui.adsabs.harvard.edu/abs/2023ApJ...946L..10B}
}

@ARTICLE{Li2020MNRAS_611,
       author = {{Li}, Gang and {Van Reeth}, Timothy and {Bedding}, Timothy R. and {Murphy}, Simon J. and {Antoci}, Victoria and {Ouazzani}, Rhita-Maria and {Barbara}, Nicholas H.},
        title = "{Gravity-mode period spacings and near-core rotation rates of 611 {\ensuremath{\gamma}} Doradus stars with Kepler}",
      journal = {\mnras},
         year = 2020,
        month = jan,
       volume = {491},
       number = {3},
        pages = {3586-3605},
          doi = {10.1093/mnras/stz2906},
archivePrefix = {arXiv},
       eprint = {1910.06634},
 primaryClass = {astro-ph.SR},
       adsurl = {https://ui.adsabs.harvard.edu/abs/2020MNRAS.491.3586L}
}

@ARTICLE{Miglio2008MNRAS,
       author = {{Miglio}, Andrea and {Montalb{\'a}n}, Josefina and {Noels}, Arlette and {Eggenberger}, Patrick},
        title = "{Probing the properties of convective cores through g modes: high-order g modes in SPB and {\ensuremath{\gamma}} Doradus stars}",
      journal = {\mnras},
         year = 2008,
        month = may,
       volume = {386},
       number = {3},
        pages = {1487-1502},
          doi = {10.1111/j.1365-2966.2008.13112.x},
archivePrefix = {arXiv},
       eprint = {0802.2057},
 primaryClass = {astro-ph},
       adsurl = {https://ui.adsabs.harvard.edu/abs/2008MNRAS.386.1487M}
}

@ARTICLE{Gagne2018,
       author = {{Gagn{\'e}}, Jonathan and {Faherty}, Jacqueline K.},
        title = "{BANYAN. XIII. A First Look at Nearby Young Associations with Gaia Data Release 2}",
      journal = {\apj},
         year = 2018,
        month = aug,
       volume = {862},
       number = {2},
          eid = {138},
        pages = {138},
          doi = {10.3847/1538-4357/aaca2e},
archivePrefix = {arXiv},
       eprint = {1805.11715},
 primaryClass = {astro-ph.SR},
       adsurl = {https://ui.adsabs.harvard.edu/abs/2018ApJ...862..138G}
}

@ARTICLE{Pang2021_huluwa,
       author = {{Pang}, Xiaoying and {Yu}, Zeqiu and {Tang}, Shih-Yun and {Hong}, Jongsuk and {Yuan}, Zhen and {Pasquato}, Mario and {Kouwenhoven}, M.~B.~N.},
        title = "{Disruption of Hierarchical Clustering in the Vela OB2 Complex and the Cluster Pair Collinder 135 and UBC 7 with Gaia EDR3: Evidence of Supernova Quenching}",
      journal = {\apj},
         year = 2021,
        month = dec,
       volume = {923},
       number = {1},
          eid = {20},
        pages = {20},
          doi = {10.3847/1538-4357/ac2838},
archivePrefix = {arXiv},
       eprint = {2106.07658},
 primaryClass = {astro-ph.GA},
       adsurl = {https://ui.adsabs.harvard.edu/abs/2021ApJ...923...20P}
}

@ARTICLE{Meingast2021,
       author = {{Meingast}, Stefan and {Alves}, Jo{\~a}o and {Rottensteiner}, Alena},
        title = "{Extended stellar systems in the solar neighborhood. V. Discovery of coronae of nearby star clusters}",
      journal = {\aap},
         year = 2021,
        month = jan,
       volume = {645},
          eid = {A84},
        pages = {A84},
          doi = {10.1051/0004-6361/202038610},
archivePrefix = {arXiv},
       eprint = {2010.06591},
 primaryClass = {astro-ph.GA},
       adsurl = {https://ui.adsabs.harvard.edu/abs/2021A&A...645A..84M}
}

@BOOK{Maeder2009,
       author = {{Maeder}, Andr{\'e}},
        title = "{Physics, Formation and Evolution of Rotating Stars, Springer-Verlag, Heidelberg}",
         year = 2009,
          doi = {10.1007/978-3-540-76949-1},
       adsurl = {https://ui.adsabs.harvard.edu/abs/2009pfer.book.....M}
}

@ARTICLE{Rogers2017,
       author = {{Rogers}, T.~M. and {McElwaine}, J.~N.},
        title = "{On the Chemical Mixing Induced by Internal Gravity Waves}",
      journal = {\apjl},
         year = 2017,
        month = oct,
       volume = {848},
       number = {1},
          eid = {L1},
        pages = {L1},
          doi = {10.3847/2041-8213/aa8d13},
archivePrefix = {arXiv},
       eprint = {1709.04920},
 primaryClass = {astro-ph.SR},
       adsurl = {https://ui.adsabs.harvard.edu/abs/2017ApJ...848L...1R}
}

@ARTICLE{Zahn1992,
       author = {{Zahn}, J. -P.},
        title = "{Circulation and turbulence in rotating stars.}",
      journal = {\aap},
         year = 1992,
        month = nov,
       volume = {265},
        pages = {115-132},
       adsurl = {https://ui.adsabs.harvard.edu/abs/1992A&A...265..115Z}
}

@ARTICLE{Fuller2019,
       author = {{Fuller}, Jim and {Piro}, Anthony L. and {Jermyn}, Adam S.},
        title = "{Slowing the spins of stellar cores}",
      journal = {\mnras},
         year = 2019,
        month = may,
       volume = {485},
       number = {3},
        pages = {3661-3680},
          doi = {10.1093/mnras/stz514},
archivePrefix = {arXiv},
       eprint = {1902.08227},
 primaryClass = {astro-ph.SR},
       adsurl = {https://ui.adsabs.harvard.edu/abs/2019MNRAS.485.3661F}
}

@ARTICLE{Ratnasingam2019,
       author = {{Ratnasingam}, R.~P. and {Edelmann}, P.~V.~F. and {Rogers}, T.~M.},
        title = "{Onset of non-linear internal gravity waves in intermediate-mass stars}",
      journal = {\mnras},
         year = 2019,
        month = feb,
       volume = {482},
       number = {4},
        pages = {5500-5512},
          doi = {10.1093/mnras/sty3086},
archivePrefix = {arXiv},
       eprint = {1812.01046},
 primaryClass = {astro-ph.SR},
       adsurl = {https://ui.adsabs.harvard.edu/abs/2019MNRAS.482.5500R}
}

@ARTICLE{Bouchaud2020,
       author = {{Bouchaud}, K. and {Domiciano de Souza}, A. and {Rieutord}, M. and {Reese}, D.~R. and {Kervella}, P.},
        title = "{A realistic two-dimensional model of Altair}",
      journal = {\aap},
         year = 2020,
        month = jan,
       volume = {633},
          eid = {A78},
        pages = {A78},
          doi = {10.1051/0004-6361/201936830},
archivePrefix = {arXiv},
       eprint = {1912.03138},
 primaryClass = {astro-ph.SR},
       adsurl = {https://ui.adsabs.harvard.edu/abs/2020A&A...633A..78B}
}

@ARTICLE{Royer2007,
       author = {{Royer}, F. and {Zorec}, J. and {G{\'o}mez}, A.~E.},
        title = "{Rotational velocities of A-type stars. III. Velocity distributions}",
      journal = {\aap},
         year = 2007,
        month = feb,
       volume = {463},
       number = {2},
        pages = {671-682},
          doi = {10.1051/0004-6361:20065224},
archivePrefix = {arXiv},
       eprint = {astro-ph/0610785},
 primaryClass = {astro-ph},
       adsurl = {https://ui.adsabs.harvard.edu/abs/2007A&A...463..671R}
}

@ARTICLE{Aerts2019ARAA,
       author = {{Aerts}, Conny and {Mathis}, St{\'e}phane and {Rogers}, Tamara M.},
        title = "{Angular Momentum Transport in Stellar Interiors}",
      journal = {\araa},
         year = 2019,
        month = aug,
       volume = {57},
        pages = {35-78},
          doi = {10.1146/annurev-astro-091918-104359},
archivePrefix = {arXiv},
       eprint = {1809.07779},
 primaryClass = {astro-ph.SR},
       adsurl = {https://ui.adsabs.harvard.edu/abs/2019ARA&A..57...35A}
}

@ARTICLE{Aerts2021RvMP,
       author = {{Aerts}, C.},
        title = "{Probing the interior physics of stars through asteroseismology}",
      journal = {Reviews of Modern Physics},
         year = 2021,
        month = jan,
       volume = {93},
       number = {1},
          eid = {015001},
        pages = {015001},
          doi = {10.1103/RevModPhys.93.015001},
archivePrefix = {arXiv},
       eprint = {1912.12300},
 primaryClass = {astro-ph.SR},
       adsurl = {https://ui.adsabs.harvard.edu/abs/2021RvMP...93a5001A}
}

@BOOK{Aerts2010book,
       author = {{Aerts}, Conny and {Christensen-Dalsgaard}, J{\o}rgen and {Kurtz}, Donald W.},
        title = "{Asteroseismology, Springer-Verlag, Heidelberg}",
         year = 2010,
          doi = {10.1007/978-1-4020-5803-5},
       adsurl = {https://ui.adsabs.harvard.edu/abs/2010aste.book.....A}
}

@ARTICLE{Gallet2013,
       author = {{Gallet}, F. and {Bouvier}, J.},
        title = "{Improved angular momentum evolution model for solar-like stars}",
      journal = {\aap},
         year = 2013,
        month = aug,
       volume = {556},
          eid = {A36},
        pages = {A36},
          doi = {10.1051/0004-6361/201321302},
archivePrefix = {arXiv},
       eprint = {1306.2130},
 primaryClass = {astro-ph.SR},
       adsurl = {https://ui.adsabs.harvard.edu/abs/2013A&A...556A..36G}
}

@ARTICLE{Bouabid2013,
       author = {{Bouabid}, M. -P. and {Dupret}, M. -A. and {Salmon}, S. and {Montalb{\'a}n}, J. and {Miglio}, A. and {Noels}, A.},
        title = "{Effects of the Coriolis force on high-order g modes in {\ensuremath{\gamma}} Doradus stars}",
      journal = {\mnras},
         year = 2013,
        month = mar,
       volume = {429},
       number = {3},
        pages = {2500-2514},
          doi = {10.1093/mnras/sts517},
       adsurl = {https://ui.adsabs.harvard.edu/abs/2013MNRAS.429.2500B}
}

@ARTICLE{Kaye1999,
       author = {{Kaye}, Anthony B. and {Handler}, Gerald and {Krisciunas}, Kevin and {Poretti}, Ennio and {Zerbi}, Filippo M.},
        title = "{Gamma Doradus Stars: Defining a New Class of Pulsating Variables}",
      journal = {\pasp},
         year = 1999,
        month = jul,
       volume = {111},
       number = {761},
        pages = {840-844},
          doi = {10.1086/316399},
archivePrefix = {arXiv},
       eprint = {astro-ph/9905042},
 primaryClass = {astro-ph},
       adsurl = {https://ui.adsabs.harvard.edu/abs/1999PASP..111..840K}
}

@ARTICLE{Balona1994,
       author = {{Balona}, L.~A. and {Krisciunas}, K. and {Cousins}, A.~W.~J.},
        title = "{Gamma Doradus : evidence for a new class of pulsating star.}",
      journal = {\mnras},
         year = 1994,
        month = oct,
       volume = {270},
        pages = {905-913},
          doi = {10.1093/mnras/270.4.905},
       adsurl = {https://ui.adsabs.harvard.edu/abs/1994MNRAS.270..905B}
}

@ARTICLE{Schmid2016_10080943_modelling,
       author = {{Schmid}, V.~S. and {Aerts}, C.},
        title = "{Asteroseismic modelling of the two F-type hybrid pulsators KIC 10080943A and KIC 10080943B}",
      journal = {\aap},
         year = 2016,
        month = aug,
       volume = {592},
          eid = {A116},
        pages = {A116},
          doi = {10.1051/0004-6361/201628617},
archivePrefix = {arXiv},
       eprint = {1605.07958},
 primaryClass = {astro-ph.SR},
       adsurl = {https://ui.adsabs.harvard.edu/abs/2016A&A...592A.116S}
}

@ARTICLE{LiYaGuang2018_solarlike_binary,
       author = {{Li}, Yaguang and {Bedding}, Timothy R. and {Li}, Tanda and {Bi}, Shaolan and {Murphy}, Simon J. and {Corsaro}, Enrico and {Chen}, Li and {Tian}, Zhijia},
        title = "{Asteroseismology of KIC 7107778: a binary comprising almost identical subgiants}",
      journal = {\mnras},
         year = 2018,
        month = may,
       volume = {476},
       number = {1},
        pages = {470-481},
          doi = {10.1093/mnras/sty222},
archivePrefix = {arXiv},
       eprint = {1801.09023},
 primaryClass = {astro-ph.SR},
       adsurl = {https://ui.adsabs.harvard.edu/abs/2018MNRAS.476..470L}
}

@ARTICLE{Steindl2021_PMS_IS,
       author = {{Steindl}, T. and {Zwintz}, K. and {Barnes}, T.~G. and {M{\"u}llner}, M. and {Vorobyov}, E.~I.},
        title = "{Pulsational instability of pre-main-sequence models from accreting protostars. I. Constraining the input physics for accretion with spectroscopic parameters and stellar pulsations}",
      journal = {\aap},
         year = 2021,
        month = oct,
       volume = {654},
          eid = {A36},
        pages = {A36},
          doi = {10.1051/0004-6361/202140818},
archivePrefix = {arXiv},
       eprint = {2107.07568},
 primaryClass = {astro-ph.SR},
       adsurl = {https://ui.adsabs.harvard.edu/abs/2021A&A...654A..36S}
}

@ARTICLE{Sekaran2021_9850387,
       author = {{Sekaran}, S. and {Tkachenko}, A. and {Johnston}, C. and {Aerts}, C.},
        title = "{A comparison of the dynamical and model-derived parameters of the pulsating eclipsing binary KIC 9850387}",
      journal = {\aap},
         year = 2021,
        month = apr,
       volume = {648},
          eid = {A91},
        pages = {A91},
          doi = {10.1051/0004-6361/202040154},
archivePrefix = {arXiv},
       eprint = {2102.11272},
 primaryClass = {astro-ph.SR},
       adsurl = {https://ui.adsabs.harvard.edu/abs/2021A&A...648A..91S}
}

@ARTICLE{Zhang2020_KIC9850387,
       author = {{Zhang}, Xiaobin and {Chen}, Xinghao and {Zhang}, Haotong and {Fu}, Jianning and {Li}, Yan},
        title = "{A Pre-main-sequence {\ensuremath{\gamma}} Dor-{\ensuremath{\delta}} Sct Hybrid with Extremely Slow Internal Rotation in a Short-period Eclipsing Binary KIC 9850387 Revealed by Asteroseismology}",
      journal = {\apj},
         year = 2020,
        month = jun,
       volume = {895},
       number = {2},
          eid = {124},
        pages = {124},
          doi = {10.3847/1538-4357/ab8ad5},
       adsurl = {https://ui.adsabs.harvard.edu/abs/2020ApJ...895..124Z}
}

@ARTICLE{Corsaro_2012,
       author = {{Corsaro}, Enrico and {Stello}, Dennis and {Huber}, Daniel and {Bedding}, Timothy R. and {Bonanno}, Alfio and {Brogaard}, Karsten and {Kallinger}, Thomas and {Benomar}, Othman and {White}, Timothy R. and {Mosser}, Benoit and {Basu}, Sarbani and {Chaplin}, William J. and {Christensen-Dalsgaard}, J{\o}rgen and {Elsworth}, Yvonne P. and {Garc{\'\i}a}, Rafael A. and {Hekker}, Saskia and {Kjeldsen}, Hans and {Mathur}, Savita and {Meibom}, S{\o}ren and {Hall}, Jennifer R. and {Ibrahim}, Khadeejah A. and {Klaus}, Todd C.},
        title = "{Asteroseismology of the Open Clusters NGC 6791, NGC 6811, and NGC 6819 from 19 Months of Kepler Photometry}",
      journal = {\apj},
         year = 2012,
        month = oct,
       volume = {757},
       number = {2},
          eid = {190},
        pages = {190},
          doi = {10.1088/0004-637X/757/2/190},
archivePrefix = {arXiv},
       eprint = {1205.4023},
 primaryClass = {astro-ph.SR},
       adsurl = {https://ui.adsabs.harvard.edu/abs/2012ApJ...757..190C}
}

@ARTICLE{Balona2013,
       author = {{Balona}, L.~A. and {Joshi}, S. and {Joshi}, Y.~C. and {Sagar}, R.},
        title = "{Pulsation and rotation of Kepler stars in the NGC 6866 field}",
      journal = {\mnras},
         year = 2013,
        month = feb,
       volume = {429},
       number = {2},
        pages = {1466-1478},
          doi = {10.1093/mnras/sts429},
       adsurl = {https://ui.adsabs.harvard.edu/abs/2013MNRAS.429.1466B}
}

@ARTICLE{Balona2013_NGC6819,
       author = {{Balona}, L.~A. and {Medupe}, T. and {Abedigamba}, O.~P. and {Ayane}, G. and {Keeley}, L. and {Matsididi}, M. and {Mekonnen}, G. and {Nhlapo}, M.~D. and {Sithole}, N.},
        title = "{Kepler observations of the open cluster NGC 6819}",
      journal = {\mnras},
         year = 2013,
        month = apr,
       volume = {430},
       number = {4},
        pages = {3472-3482},
          doi = {10.1093/mnras/stt148},
       adsurl = {https://ui.adsabs.harvard.edu/abs/2013MNRAS.430.3472B}
}

@ARTICLE{Stello2011,
       author = {{Stello}, Dennis and {Meibom}, S{\o}ren and {Gilliland}, Ronald L. and {Grundahl}, Frank and {Hekker}, Saskia and {Mosser}, Beno{\^\i}t and {Kallinger}, Thomas and {Mathur}, Savita and {Garc{\'\i}a}, Rafael A. and {Huber}, Daniel and {Basu}, Sarbani and {Bedding}, Timothy R. and {Brogaard}, Karsten and {Chaplin}, William J. and {Elsworth}, Yvonne P. and {Molenda-{\.Z}akowicz}, Joanna and {Szab{\'o}}, Robert and {Still}, Martin and {Jenkins}, Jon M. and {Christensen-Dalsgaard}, J{\o}rgen and {Kjeldsen}, Hans and {Serenelli}, Aldo M. and {Wohler}, Bill},
        title = "{An Asteroseismic Membership Study of the Red Giants in Three Open Clusters Observed by Kepler: NGC 6791, NGC 6819, and NGC 6811}",
      journal = {\apj},
         year = 2011,
        month = sep,
       volume = {739},
       number = {1},
          eid = {13},
        pages = {13},
          doi = {10.1088/0004-637X/739/1/13},
archivePrefix = {arXiv},
       eprint = {1107.1234},
 primaryClass = {astro-ph.SR},
       adsurl = {https://ui.adsabs.harvard.edu/abs/2011ApJ...739...13S}
}

@ARTICLE{Stello2016_M67,
       author = {{Stello}, Dennis and {Vanderburg}, Andrew and {Casagrande}, Luca and {Gilliland}, Ron and {Silva Aguirre}, Victor and {Sandquist}, Eric and {Leiner}, Emily and {Mathieu}, Robert and {Soderblom}, David R.},
        title = "{The K2 M67 Study: Revisiting Old Friends with K2 Reveals Oscillating Red Giants in the Open Cluster M67}",
      journal = {\apj},
         year = 2016,
        month = dec,
       volume = {832},
       number = {2},
          eid = {133},
        pages = {133},
          doi = {10.3847/0004-637X/832/2/133},
archivePrefix = {arXiv},
       eprint = {1610.03060},
 primaryClass = {astro-ph.SR},
       adsurl = {https://ui.adsabs.harvard.edu/abs/2016ApJ...832..133S}
}

@ARTICLE{Schimak2026,
       author = {{Schimak}, Lea S. and {Bedding}, Timothy R. and {Crawford}, Courtney L. and {Beck}, Paul G. and {Li}, Yaguang and {Huber}, Daniel and {Ong}, Joel and {Montet}, Benjamin T. and {Pedersen}, May Gade and {Grossmann}, Desmond H. and {Mathur}, Savita and {Garc{\'\i}a}, Rafael A.},
        title = "{Testing Red Clump Models with the Asteroseismic Binary KIC 10841730}",
      journal = {\mnras},
         year = 2026,
        month = jan,
          doi = {10.1093/mnras/stag151},
       adsurl = {https://ui.adsabs.harvard.edu/abs/2026MNRAS.tmp..189S}
}

@ARTICLE{Beck2024,
       author = {{Beck}, P.~G. and {Grossmann}, D.~H. and {Steinwender}, L. and {Schimak}, L.~S. and {Muntean}, N. and {Vrard}, M. and {Patton}, R.~A. and {Merc}, J. and {Mathur}, S. and {Garcia}, R.~A. and {Pinsonneault}, M.~H. and {Rowan}, D.~M. and {Gaulme}, P. and {Allende Prieto}, C. and {Arellano-C{\'o}rdova}, K.~Z. and {Cao}, L. and {Corsaro}, E. and {Creevey}, O. and {Hambleton}, K.~M. and {Hanslmeier}, A. and {Holl}, B. and {Johnson}, J. and {Mathis}, S. and {Godoy-Rivera}, D. and {S{\'\i}mon-D{\'\i}az}, S. and {Zinn}, J.~C.},
        title = "{Constraining stellar and orbital co-evolution through ensemble seismology of solar-like oscillators in binary systems. A census of oscillating red giants and dwarf stars in Gaia DR3 binaries}",
      journal = {\aap},
         year = 2024,
        month = feb,
       volume = {682},
          eid = {A7},
        pages = {A7},
          doi = {10.1051/0004-6361/202346810},
archivePrefix = {arXiv},
       eprint = {2307.10812},
 primaryClass = {astro-ph.SR},
       adsurl = {https://ui.adsabs.harvard.edu/abs/2024A&A...682A...7B}
}

@ARTICLE{Guo2017,
       author = {{Guo}, Z. and {Gies}, D.~R. and {Matson}, R.~A.},
        title = "{Gravity Modes Reveal the Internal Rotation of a Post-mass-transfer Gamma Doradus/Delta Scuti Hybrid Pulsator in Kepler  Eclipsing Binary KIC 9592855}",
      journal = {\apj},
         year = 2017,
        month = dec,
       volume = {851},
       number = {1},
          eid = {39},
        pages = {39},
          doi = {10.3847/1538-4357/aa978c},
archivePrefix = {arXiv},
       eprint = {1704.03789},
 primaryClass = {astro-ph.SR},
       adsurl = {https://ui.adsabs.harvard.edu/abs/2017ApJ...851...39G}
}

@ARTICLE{Johnston2019,
       author = {{Johnston}, C. and {Tkachenko}, A. and {Aerts}, C. and {Molenberghs}, G. and {Bowman}, D.~M. and {Pedersen}, M.~G. and {Buysschaert}, B. and {P{\'a}pics}, P.~I.},
        title = "{Binary asteroseismic modelling: isochrone-cloud methodology and application to Kepler gravity mode pulsators}",
      journal = {\mnras},
         year = 2019,
        month = jan,
       volume = {482},
       number = {1},
        pages = {1231-1246},
          doi = {10.1093/mnras/sty2671},
archivePrefix = {arXiv},
       eprint = {1810.00780},
 primaryClass = {astro-ph.SR},
       adsurl = {https://ui.adsabs.harvard.edu/abs/2019MNRAS.482.1231J}
}

@ARTICLE{Sekaran2020,
       author = {{Sekaran}, S. and {Tkachenko}, A. and {Abdul-Masih}, M. and {Pr{\v{s}}a}, A. and {Johnston}, C. and {Huber}, D. and {Murphy}, S.~J. and {Banyard}, G. and {Howard}, A.~W. and {Isaacson}, H. and {Bowman}, D.~M. and {Aerts}, C.},
        title = "{Tango of celestial dancers: A sample of detached eclipsing binary systems containing g-mode pulsating components. A case study of KIC9850387}",
      journal = {\aap},
         year = 2020,
        month = nov,
       volume = {643},
          eid = {A162},
        pages = {A162},
          doi = {10.1051/0004-6361/202038989},
archivePrefix = {arXiv},
       eprint = {2010.02187},
 primaryClass = {astro-ph.SR},
       adsurl = {https://ui.adsabs.harvard.edu/abs/2020A&A...643A.162S}
}

@ARTICLE{Colman2022,
       author = {{Colman}, Isabel L. and {Bedding}, Timothy R. and {Huber}, Daniel and {Kjeldsen}, Hans},
        title = "{The Kepler IRIS Catalog: Image Subtraction Light Curves for 9150 Stars in and around the Open Clusters NGC 6791 and NGC 6819}",
      journal = {\apjs},
         year = 2022,
        month = feb,
       volume = {258},
       number = {2},
          eid = {39},
        pages = {39},
          doi = {10.3847/1538-4365/ac3a11},
archivePrefix = {arXiv},
       eprint = {2112.05174},
 primaryClass = {astro-ph.SR},
       adsurl = {https://ui.adsabs.harvard.edu/abs/2022ApJS..258...39C}
}

@ARTICLE{Waelkens1991,
       author = {{Waelkens}, C.},
        title = "{Slowly pulsating B stars.}",
      journal = {\aap},
         year = 1991,
        month = jun,
       volume = {246},
        pages = {453},
       adsurl = {https://ui.adsabs.harvard.edu/abs/1991A&A...246..453W}
}

@ARTICLE{Pedersen2021,
       author = {{Pedersen}, May G. and {Aerts}, Conny and {P{\'a}pics}, P{\'e}ter I. and {Michielsen}, Mathias and {Gebruers}, Sarah and {Rogers}, Tamara M. and {Molenberghs}, Geert and {Burssens}, Siemen and {Garcia}, Stefano and {Bowman}, Dominic M.},
        title = "{Internal mixing of rotating stars inferred from dipole gravity modes}",
      journal = {Nature Astronomy},
         year = 2021,
        month = jan,
       volume = {5},
        pages = {715-722},
          doi = {10.1038/s41550-021-01351-x},
archivePrefix = {arXiv},
       eprint = {2105.04533},
 primaryClass = {astro-ph.SR},
       adsurl = {https://ui.adsabs.harvard.edu/abs/2021NatAs...5..715P}
}

@ARTICLE{Saio2018_r_modes,
       author = {{Saio}, Hideyuki and {Kurtz}, Donald W. and {Murphy}, Simon J. and {Antoci}, Victoria L. and {Lee}, Umin},
        title = "{Theory and evidence of global Rossby waves in upper main-sequence stars: r-mode oscillations in many Kepler stars}",
      journal = {\mnras},
         year = 2018,
        month = feb,
       volume = {474},
       number = {2},
        pages = {2774-2786},
          doi = {10.1093/mnras/stx2962},
archivePrefix = {arXiv},
       eprint = {1711.04908},
 primaryClass = {astro-ph.SR},
       adsurl = {https://ui.adsabs.harvard.edu/abs/2018MNRAS.474.2774S}
}

@ARTICLE{Mombarg2024AA_14000Gaia,
       author = {{Mombarg}, Joey S.~G. and {Aerts}, Conny and {Van Reeth}, Timothy and {Hey}, Daniel},
        title = "{Estimates of (convective core) masses, radii, and relative ages for {\ensuremath{\sim}}14 000 Gaia-discovered gravity-mode pulsators monitored by TESS}",
      journal = {\aap},
         year = 2024,
        month = nov,
       volume = {691},
          eid = {A131},
        pages = {A131},
          doi = {10.1051/0004-6361/202451651},
archivePrefix = {arXiv},
       eprint = {2410.05367},
 primaryClass = {astro-ph.SR},
       adsurl = {https://ui.adsabs.harvard.edu/abs/2024A&A...691A.131M}
}

@ARTICLE{BeyerWhite2024,
       author = {{Beyer}, Alexa C. and {White}, Russel J.},
        title = "{The Kraft Break Sharply Divides Low-mass and Intermediate-mass Stars}",
      journal = {\apj},
         year = 2024,
        month = sep,
       volume = {973},
       number = {1},
          eid = {28},
        pages = {28},
          doi = {10.3847/1538-4357/ad6b0d},
archivePrefix = {arXiv},
       eprint = {2408.02638},
 primaryClass = {astro-ph.SR},
       adsurl = {https://ui.adsabs.harvard.edu/abs/2024ApJ...973...28B}
}

@ARTICLE{Perren2023,
       author = {{Perren}, Gabriel I. and {Pera}, Mar{\'\i}a S. and {Navone}, Hugo D. and {V{\'a}zquez}, Rub{\'e}n A.},
        title = "{The Unified Cluster Catalogue: towards a comprehensive and homogeneous data base of stellar clusters}",
      journal = {\mnras},
         year = 2023,
        month = dec,
       volume = {526},
       number = {3},
        pages = {4107-4119},
          doi = {10.1093/mnras/stad2826},
archivePrefix = {arXiv},
       eprint = {2308.04546},
 primaryClass = {astro-ph.GA},
       adsurl = {https://ui.adsabs.harvard.edu/abs/2023MNRAS.526.4107P}
}

@ARTICLE{VanReeth2016_TAR,
       author = {{Van Reeth}, T. and {Tkachenko}, A. and {Aerts}, C.},
        title = "{Interior rotation of a sample of {\ensuremath{\gamma}} Doradus stars from ensemble modelling of their gravity-mode period spacings}",
      journal = {\aap},
         year = 2016,
        month = oct,
       volume = {593},
          eid = {A120},
        pages = {A120},
          doi = {10.1051/0004-6361/201628616},
archivePrefix = {arXiv},
       eprint = {1607.00820},
 primaryClass = {astro-ph.SR},
       adsurl = {https://ui.adsabs.harvard.edu/abs/2016A&A...593A.120V}
}

@ARTICLE{Saio2018,
       author = {{Saio}, Hideyuki and {Bedding}, Timothy R. and {Kurtz}, Donald W. and {Murphy}, Simon J. and {Antoci}, Victoria and {Shibahashi}, Hiromoto and {Li}, Gang and {Takata}, Masao},
        title = "{An astrophysical interpretation of the remarkable g-mode frequency groups of the rapidly rotating {\ensuremath{\gamma}} Dor star, KIC 5608334}",
      journal = {\mnras},
         year = 2018,
        month = jun,
       volume = {477},
       number = {2},
        pages = {2183-2195},
          doi = {10.1093/mnras/sty784},
archivePrefix = {arXiv},
       eprint = {1803.08677},
 primaryClass = {astro-ph.SR},
       adsurl = {https://ui.adsabs.harvard.edu/abs/2018MNRAS.477.2183S}
}

@ARTICLE{Shibahashi1979,
       author = {{Shibahashi}, H.},
        title = "{Modal Analysis of Stellar Nonradial Oscillations by an Asymptotic Method}",
      journal = {\pasj},
         year = 1979,
        month = jan,
       volume = {31},
        pages = {87-104},
       adsurl = {https://ui.adsabs.harvard.edu/abs/1979PASJ...31...87S}
}

@ARTICLE{Ouazzani2017,
       author = {{Ouazzani}, Rhita-Maria and {Salmon}, S.~J.~A.~J. and {Antoci}, V. and {Bedding}, T.~R. and {Murphy}, S.~J. and {Roxburgh}, I.~W.},
        title = "{A new asteroseismic diagnostic for internal rotation in {\ensuremath{\gamma}} Doradus stars}",
      journal = {\mnras},
         year = 2017,
        month = feb,
       volume = {465},
       number = {2},
        pages = {2294-2309},
          doi = {10.1093/mnras/stw2717},
archivePrefix = {arXiv},
       eprint = {1610.06184},
 primaryClass = {astro-ph.SR},
       adsurl = {https://ui.adsabs.harvard.edu/abs/2017MNRAS.465.2294O}
}

@ARTICLE{Rogers2025,
       author = {{Rogers}, T.~M. and {Ratnasingam}, R.~P.},
        title = "{Angular Momentum Transport by Internal Gravity Waves across Age}",
      journal = {\apjl},
         year = 2025,
        month = apr,
       volume = {983},
       number = {2},
          eid = {L38},
        pages = {L38},
          doi = {10.3847/2041-8213/adc45a},
       adsurl = {https://ui.adsabs.harvard.edu/abs/2025ApJ...983L..38R}
}

@ARTICLE{Mombarg2021,
       author = {{Mombarg}, J.~S.~G. and {Van Reeth}, T. and {Aerts}, C.},
        title = "{Constraining stellar evolution theory with asteroseismology of {\ensuremath{\gamma}} Doradus stars using deep learning. Stellar masses, ages, and core-boundary mixing}",
      journal = {\aap},
         year = 2021,
        month = jun,
       volume = {650},
          eid = {A58},
        pages = {A58},
          doi = {10.1051/0004-6361/202039543},
archivePrefix = {arXiv},
       eprint = {2103.13394},
 primaryClass = {astro-ph.SR},
       adsurl = {https://ui.adsabs.harvard.edu/abs/2021A&A...650A..58M}
}

@ARTICLE{Mombarg2024,
       author = {{Mombarg}, J.~S.~G. and {Aerts}, C. and {Molenberghs}, G.},
        title = "{Probability distributions of initial rotation velocities and core-boundary mixing efficiencies of {\ensuremath{\gamma}} Doradus stars}",
      journal = {\aap},
         year = 2024,
        month = may,
       volume = {685},
          eid = {A21},
        pages = {A21},
          doi = {10.1051/0004-6361/202449213},
archivePrefix = {arXiv},
       eprint = {2402.05171},
 primaryClass = {astro-ph.SR},
       adsurl = {https://ui.adsabs.harvard.edu/abs/2024A&A...685A..21M}
}

@ARTICLE{Moyano2023,
       author = {{Moyano}, F.~D. and {Eggenberger}, P. and {Salmon}, S.~J.~A.~J. and {Mombarg}, J.~S.~G. and {Ekstr{\"o}m}, S.},
        title = "{Angular momentum transport by magnetic fields in main-sequence stars with Gamma Doradus pulsators}",
      journal = {\aap},
         year = 2023,
        month = sep,
       volume = {677},
          eid = {A6},
        pages = {A6},
          doi = {10.1051/0004-6361/202346548},
archivePrefix = {arXiv},
       eprint = {2304.00674},
 primaryClass = {astro-ph.SR},
       adsurl = {https://ui.adsabs.harvard.edu/abs/2023A&A...677A...6M}
}

@ARTICLE{Maeder1997,
       author = {{Maeder}, A.},
        title = "{Stellar evolution with rotation. II. A new approach for shear mixing.}",
      journal = {\aap},
         year = 1997,
        month = may,
       volume = {321},
        pages = {134-144},
       adsurl = {https://ui.adsabs.harvard.edu/abs/1997A&A...321..134M}
}

@ARTICLE{Maeder1998,
       author = {{Maeder}, Andre and {Zahn}, Jean-Paul},
        title = "{Stellar evolution with rotation. III. Meridional circulation with MU -gradients and non-stationarity}",
      journal = {\aap},
         year = 1998,
        month = jun,
       volume = {334},
        pages = {1000-1006},
       adsurl = {https://ui.adsabs.harvard.edu/abs/1998A&A...334.1000M}
}

@ARTICLE{Takata2026_gdor_magnetic_fields,
       author = {{Takata}, Masao and {Murphy}, Simon J. and {Kurtz}, Donald W. and {Saio}, Hideyuki and {Shibahashi}, Hiromoto},
        title = "{Asteroseismic detection of a predominantly toroidal magnetic field in the deep interior of the main-sequence F star KIC 9244992}",
      journal = {\mnras},
         year = 2026,
        month = jan,
       volume = {545},
       number = {3},
          eid = {staf2153},
        pages = {staf2153},
          doi = {10.1093/mnras/staf2153},
archivePrefix = {arXiv},
       eprint = {2512.00786},
 primaryClass = {astro-ph.SR},
       adsurl = {https://ui.adsabs.harvard.edu/abs/2026MNRAS.545f2153T}
}

@ARTICLE{Mombarg2023calibrating_AM,
       author = {{Mombarg}, J.~S.~G.},
        title = "{Calibrating angular momentum transport in intermediate-mass stars from gravity-mode asteroseismology}",
      journal = {\aap},
         year = 2023,
        month = sep,
       volume = {677},
          eid = {A63},
        pages = {A63},
          doi = {10.1051/0004-6361/202345956},
archivePrefix = {arXiv},
       eprint = {2306.17211},
 primaryClass = {astro-ph.SR},
       adsurl = {https://ui.adsabs.harvard.edu/abs/2023A&A...677A..63M}
}

@ARTICLE{Rogers2013,
       author = {{Rogers}, T.~M. and {Lin}, D.~N.~C. and {McElwaine}, J.~N. and {Lau}, H.~H.~B.},
        title = "{Internal Gravity Waves in Massive Stars: Angular Momentum Transport}",
      journal = {\apj},
         year = 2013,
        month = jul,
       volume = {772},
       number = {1},
          eid = {21},
        pages = {21},
          doi = {10.1088/0004-637X/772/1/21},
archivePrefix = {arXiv},
       eprint = {1306.3262},
 primaryClass = {astro-ph.SR},
       adsurl = {https://ui.adsabs.harvard.edu/abs/2013ApJ...772...21R}
}

@ARTICLE{VanReeth2015ApJS,
       author = {{Van Reeth}, T. and {Tkachenko}, A. and {Aerts}, C. and {P{\'a}pics}, P.~I. and {Triana}, S.~A. and {Zwintz}, K. and {Degroote}, P. and {Debosscher}, J. and {Bloemen}, S. and {Schmid}, V.~S. and {De Smedt}, K. and {Fremat}, Y. and {Fuentes}, A.~S. and {Homan}, W. and {Hrudkova}, M. and {Karjalainen}, R. and {Lombaert}, R. and {Nemeth}, P. and {{\O}stensen}, R. and {Van De Steene}, G. and {Vos}, J. and {Raskin}, G. and {Van Winckel}, H.},
        title = "{Gravity-mode Period Spacings as a Seismic Diagnostic for a Sample of {\ensuremath{\gamma}} Doradus Stars from Kepler Space Photometry and High-resolution Ground-based Spectroscopy}",
      journal = {\apjs},
         year = 2015,
        month = jun,
       volume = {218},
       number = {2},
          eid = {27},
        pages = {27},
          doi = {10.1088/0067-0049/218/2/27},
archivePrefix = {arXiv},
       eprint = {1504.02119},
 primaryClass = {astro-ph.SR},
       adsurl = {https://ui.adsabs.harvard.edu/abs/2015ApJS..218...27V}
}

@ARTICLE{Ouazzani2020,
       author = {{Ouazzani}, R. -M. and {Ligni{\`e}res}, F. and {Dupret}, M. -A. and {Salmon}, S.~J.~A.~J. and {Ballot}, J. and {Christophe}, S. and {Takata}, M.},
        title = "{First evidence of inertial modes in {\ensuremath{\gamma}} Doradus stars: The core rotation revealed}",
      journal = {\aap},
         year = 2020,
        month = aug,
       volume = {640},
          eid = {A49},
        pages = {A49},
          doi = {10.1051/0004-6361/201936653},
archivePrefix = {arXiv},
       eprint = {2006.09404},
 primaryClass = {astro-ph.SR},
       adsurl = {https://ui.adsabs.harvard.edu/abs/2020A&A...640A..49O}
}

@ARTICLE{Saio2021,
       author = {{Saio}, Hideyuki and {Takata}, Masao and {Lee}, Umin and {Li}, Gang and {Van Reeth}, Timothy},
        title = "{Rotation of the convective core in {\ensuremath{\gamma}} Dor stars measured by dips in period spacings of g modes coupled with inertial modes}",
      journal = {\mnras},
         year = 2021,
        month = apr,
       volume = {502},
       number = {4},
        pages = {5856-5874},
          doi = {10.1093/mnras/stab482},
archivePrefix = {arXiv},
       eprint = {2102.08548},
 primaryClass = {astro-ph.SR},
       adsurl = {https://ui.adsabs.harvard.edu/abs/2021MNRAS.502.5856S}
}

@ARTICLE{Spruit2002,
       author = {{Spruit}, H.~C.},
        title = "{Dynamo action by differential rotation in a stably stratified stellar interior}",
      journal = {\aap},
         year = 2002,
        month = jan,
       volume = {381},
        pages = {923-932},
          doi = {10.1051/0004-6361:20011465},
archivePrefix = {arXiv},
       eprint = {astro-ph/0108207},
 primaryClass = {astro-ph},
       adsurl = {https://ui.adsabs.harvard.edu/abs/2002A&A...381..923S}
}

@ARTICLE{Nieva2012,
       author = {{Nieva}, M. -F. and {Przybilla}, N.},
        title = "{Present-day cosmic abundances. A comprehensive study of nearby early B-type stars and implications for stellar and Galactic evolution and interstellar dust models}",
      journal = {\aap},
         year = 2012,
        month = mar,
       volume = {539},
          eid = {A143},
        pages = {A143},
          doi = {10.1051/0004-6361/201118158},
archivePrefix = {arXiv},
       eprint = {1203.5787},
 primaryClass = {astro-ph.SR},
       adsurl = {https://ui.adsabs.harvard.edu/abs/2012A&A...539A.143N}
}

@ARTICLE{Verma2019,
       author = {{Verma}, Kuldeep and {Raodeo}, Keyuri and {Basu}, Sarbani and {Silva Aguirre}, V{\'\i}ctor and {Mazumdar}, Anwesh and {Mosumgaard}, Jakob R{\o}rsted and {Lund}, Mikkel N. and {Ranadive}, Pritesh},
        title = "{Helium abundance in a sample of cool stars: measurements from asteroseismology}",
      journal = {\mnras},
         year = 2019,
        month = mar,
       volume = {483},
       number = {4},
        pages = {4678-4694},
          doi = {10.1093/mnras/sty3374},
archivePrefix = {arXiv},
       eprint = {1812.02751},
 primaryClass = {astro-ph.SR},
       adsurl = {https://ui.adsabs.harvard.edu/abs/2019MNRAS.483.4678V}
}

@ARTICLE{Ricker2015,
       author = {{Ricker}, George R. and {Winn}, Joshua N. and {Vanderspek}, Roland and {Latham}, David W. and {Bakos}, G{\'a}sp{\'a}r {\'A}. and {Bean}, Jacob L. and {Berta-Thompson}, Zachory K. and {Brown}, Timothy M. and {Buchhave}, Lars and {Butler}, Nathaniel R. and {Butler}, R. Paul and {Chaplin}, William J. and {Charbonneau}, David and {Christensen-Dalsgaard}, J{\o}rgen and {Clampin}, Mark and {Deming}, Drake and {Doty}, John and {De Lee}, Nathan and {Dressing}, Courtney and {Dunham}, Edward W. and {Endl}, Michael and {Fressin}, Francois and {Ge}, Jian and {Henning}, Thomas and {Holman}, Matthew J. and {Howard}, Andrew W. and {Ida}, Shigeru and {Jenkins}, Jon M. and {Jernigan}, Garrett and {Johnson}, John Asher and {Kaltenegger}, Lisa and {Kawai}, Nobuyuki and {Kjeldsen}, Hans and {Laughlin}, Gregory and {Levine}, Alan M. and {Lin}, Douglas and {Lissauer}, Jack J. and {MacQueen}, Phillip and {Marcy}, Geoffrey and {McCullough}, Peter R. and {Morton}, Timothy D. and {Narita}, Norio and {Paegert}, Martin and {Palle}, Enric and {Pepe}, Francesco and {Pepper}, Joshua and {Quirrenbach}, Andreas and {Rinehart}, Stephen A. and {Sasselov}, Dimitar and {Sato}, Bun'ei and {Seager}, Sara and {Sozzetti}, Alessandro and {Stassun}, Keivan G. and {Sullivan}, Peter and {Szentgyorgyi}, Andrew and {Torres}, Guillermo and {Udry}, Stephane and {Villasenor}, Joel},
        title = "{Transiting Exoplanet Survey Satellite (TESS)}",
      journal = {Journal of Astronomical Telescopes, Instruments, and Systems},
         year = 2015,
        month = jan,
       volume = {1},
          eid = {014003},
        pages = {014003},
          doi = {10.1117/1.JATIS.1.1.014003},
       adsurl = {https://ui.adsabs.harvard.edu/abs/2015JATIS...1a4003R}
}

@ARTICLE{Meynet1997,
       author = {{Meynet}, G. and {Maeder}, A.},
        title = "{Stellar evolution with rotation. I. The computational method and the inhibiting effect of the {\ensuremath{\mu}}-gradient.}",
      journal = {\aap},
         year = 1997,
        month = may,
       volume = {321},
        pages = {465-476},
       adsurl = {https://ui.adsabs.harvard.edu/abs/1997A&A...321..465M}
}

@ARTICLE{Hekker2011,
       author = {{Hekker}, S. and {Basu}, S. and {Stello}, D. and {Kallinger}, T. and {Grundahl}, F. and {Mathur}, S. and {Garc{\'\i}a}, R.~A. and {Mosser}, B. and {Huber}, D. and {Bedding}, T.~R. and {Szab{\'o}}, R. and {De Ridder}, J. and {Chaplin}, W.~J. and {Elsworth}, Y. and {Hale}, S.~J. and {Christensen-Dalsgaard}, J. and {Gilliland}, R.~L. and {Still}, M. and {McCauliff}, S. and {Quintana}, E.~V.},
        title = "{Asteroseismic inferences on red giants in open clusters NGC 6791, NGC 6819, and NGC 6811 using Kepler}",
      journal = {\aap},
         year = 2011,
        month = jun,
       volume = {530},
          eid = {A100},
        pages = {A100},
          doi = {10.1051/0004-6361/201016303},
archivePrefix = {arXiv},
       eprint = {1104.4393},
 primaryClass = {astro-ph.SR},
       adsurl = {https://ui.adsabs.harvard.edu/abs/2011A&A...530A.100H}
}

@ARTICLE{Basu2011,
       author = {{Basu}, Sarbani and {Grundahl}, Frank and {Stello}, Dennis and {Kallinger}, Thomas and {Hekker}, Saskia and {Mosser}, Benoit and {Garc{\'\i}a}, Rafael A. and {Mathur}, Savita and {Brogaard}, Karsten and {Bruntt}, Hans and {Chaplin}, William J. and {Gai}, Ning and {Elsworth}, Yvonne and {Esch}, Lisa and {Ballot}, Jerome and {Bedding}, Timothy R. and {Gruberbauer}, Michael and {Huber}, Daniel and {Miglio}, Andrea and {Yildiz}, Mutlu and {Kjeldsen}, Hans and {Christensen-Dalsgaard}, J{\o}rgen and {Gilliland}, Ronald L. and {Fanelli}, Michael M. and {Ibrahim}, Khadeejah A. and {Smith}, Jeffrey C.},
        title = "{Sounding Open Clusters: Asteroseismic Constraints from Kepler on the Properties of NGC 6791 and NGC 6819}",
      journal = {\apjl},
         year = 2011,
        month = mar,
       volume = {729},
       number = {1},
          eid = {L10},
        pages = {L10},
          doi = {10.1088/2041-8205/729/1/L10},
archivePrefix = {arXiv},
       eprint = {1102.2231},
 primaryClass = {astro-ph.SR},
       adsurl = {https://ui.adsabs.harvard.edu/abs/2011ApJ...729L..10B}
}

@ARTICLE{Lund2016,
       author = {{Lund}, Mikkel N. and {Basu}, Sarbani and {Silva Aguirre}, V{\'\i}ctor and {Chaplin}, William J. and {Serenelli}, Aldo M. and {Garc{\'\i}a}, Rafael A. and {Latham}, David W. and {Casagrande}, Luca and {Bieryla}, Allyson and {Davies}, Guy R. and {Viani}, Lucas S. and {Buchhave}, Lars A. and {Miglio}, Andrea and {Soderblom}, David R. and {Valenti}, Jeff A. and {Stefanik}, Robert P. and {Handberg}, Rasmus},
        title = "{Asteroseismology of the Hyades with K2: first detection of main-sequence solar-like oscillations in an open cluster}",
      journal = {\mnras},
         year = 2016,
        month = dec,
       volume = {463},
       number = {3},
        pages = {2600-2611},
          doi = {10.1093/mnras/stw2160},
archivePrefix = {arXiv},
       eprint = {1608.07290},
 primaryClass = {astro-ph.SR},
       adsurl = {https://ui.adsabs.harvard.edu/abs/2016MNRAS.463.2600L}
}

@ARTICLE{Sandquist2020,
       author = {{Sandquist}, Eric L. and {Stello}, Dennis and {Arentoft}, Torben and {Brogaard}, Karsten and {Grundahl}, Frank and {Vanderburg}, Andrew and {Hedlund}, Anne and {DeWitt}, Ryan and {Ackerman}, Taylor R. and {Aguilar}, Miguel and {Buckner}, Andrew J. and {Juarez}, Christian and {Ortiz}, Arturo J. and {Richarte}, David and {Rivera}, Daniel I. and {Schlapfer}, Levi},
        title = "{Variability in the Massive Open Cluster NGC 1817 from K2: A Rich Population of Asteroseismic Red Clump, Eclipsing Binary, and Main-sequence Pulsating Stars}",
      journal = {\aj},
         year = 2020,
        month = mar,
       volume = {159},
       number = {3},
          eid = {96},
        pages = {96},
          doi = {10.3847/1538-3881/ab68df},
archivePrefix = {arXiv},
       eprint = {2001.01839},
 primaryClass = {astro-ph.SR},
       adsurl = {https://ui.adsabs.harvard.edu/abs/2020AJ....159...96S}
}

@ARTICLE{Dahm2015,
       author = {{Dahm}, S.~E.},
        title = "{Reexamining the Lithium Depletion Boundary in the Pleiades and the Inferred Age of the Cluster}",
      journal = {\apj},
         year = 2015,
        month = nov,
       volume = {813},
       number = {2},
          eid = {108},
        pages = {108},
          doi = {10.1088/0004-637X/813/2/108},
       adsurl = {https://ui.adsabs.harvard.edu/abs/2015ApJ...813..108D}
}

@ARTICLE{Murphy2022Pleiades,
       author = {{Murphy}, Simon J. and {Bedding}, Timothy R. and {White}, Timothy R. and {Li}, Yaguang and {Hey}, Daniel and {Reese}, Daniel and {Joyce}, Meridith},
        title = "{Five young {\ensuremath{\delta}} Scuti stars in the Pleiades seen with Kepler/K2}",
      journal = {\mnras},
         year = 2022,
        month = apr,
       volume = {511},
       number = {4},
        pages = {5718-5729},
          doi = {10.1093/mnras/stac240},
archivePrefix = {arXiv},
       eprint = {2111.04203},
 primaryClass = {astro-ph.SR},
       adsurl = {https://ui.adsabs.harvard.edu/abs/2022MNRAS.511.5718M}
}

@ARTICLE{LiGang_2024_NGC2516,
       author = {{Li}, Gang and {Aerts}, Conny and {Bedding}, Timothy R. and {Fritzewski}, Dario J. and {Murphy}, Simon J. and {Van Reeth}, Timothy and {Montet}, Benjamin T. and {Jian}, Mingjie and {Mombarg}, Joey S.~G. and {Gossage}, Seth and {Sreenivas}, Kalarickal R.},
        title = "{Asteroseismology of the young open cluster NGC 2516. I. Photometric and spectroscopic observations}",
      journal = {\aap},
         year = 2024,
        month = jun,
       volume = {686},
          eid = {A142},
        pages = {A142},
          doi = {10.1051/0004-6361/202348901},
archivePrefix = {arXiv},
       eprint = {2311.16991},
 primaryClass = {astro-ph.SR},
       adsurl = {https://ui.adsabs.harvard.edu/abs/2024A&A...686A.142L}
}

@ARTICLE{Reyes2025Natur,
       author = {{Reyes}, Claudia and {Stello}, Dennis and {Ong}, Joel and {Lindsay}, Christopher and {Hon}, Marc and {Bedding}, Timothy R.},
        title = "{Acoustic modes in M67 cluster stars trace deepening convective envelopes}",
      journal = {\nat},
         year = 2025,
        month = apr,
       volume = {640},
       number = {8058},
        pages = {338-342},
          doi = {10.1038/s41586-025-08760-2},
archivePrefix = {arXiv},
       eprint = {2504.01828},
 primaryClass = {astro-ph.SR},
       adsurl = {https://ui.adsabs.harvard.edu/abs/2025Natur.640..338R}
}

@ARTICLE{LiGang_2025_NGC2516_modelling,
       author = {{Li}, Gang and {Mombarg}, Joey S.~G. and {Guo}, Zhao and {Aerts}, Conny},
        title = "{Asteroseismology of the young open cluster NGC 2516: II. Constraining cluster age using gravity-mode pulsators}",
      journal = {\aap},
         year = 2025,
        month = nov,
       volume = {703},
          eid = {A116},
        pages = {A116},
          doi = {10.1051/0004-6361/202556409},
archivePrefix = {arXiv},
       eprint = {2509.05824},
 primaryClass = {astro-ph.SR},
       adsurl = {https://ui.adsabs.harvard.edu/abs/2025A&A...703A.116L}
}

@ARTICLE{Fritzewski2024UBC1,
       author = {{Fritzewski}, D.~J. and {Van Reeth}, T. and {Aerts}, C. and {Van Beeck}, J. and {Gossage}, S. and {Li}, G.},
        title = "{Age-dating the young open cluster UBC 1 with g-mode asteroseismology, gyrochronology, and isochrone fitting}",
      journal = {\aap},
         year = 2024,
        month = jan,
       volume = {681},
          eid = {A13},
        pages = {A13},
          doi = {10.1051/0004-6361/202347618},
archivePrefix = {arXiv},
       eprint = {2310.18426},
 primaryClass = {astro-ph.SR},
       adsurl = {https://ui.adsabs.harvard.edu/abs/2024A&A...681A..13F}
}

@ARTICLE{Mathis2025,
       author = {{Mathis}, S.},
        title = "{Non-linear internal waves breaking in stellar-radiation zones: Parametrisation for the transport of angular momentum: Bridging geophysical-to-stellar fluid dynamics}",
      journal = {\aap},
         year = 2025,
        month = feb,
       volume = {694},
          eid = {A173},
        pages = {A173},
          doi = {10.1051/0004-6361/202452066},
archivePrefix = {arXiv},
       eprint = {2411.13925},
 primaryClass = {astro-ph.SR},
       adsurl = {https://ui.adsabs.harvard.edu/abs/2025A&A...694A.173M}
}

@ARTICLE{Tayar2025,
       author = {{Tayar}, Jamie and {Joyce}, Meridith},
        title = "{Star-crossed Clusters: Asteroseismic Ages for Individual Stars are in Tension with the Ages of their Host Clusters}",
      journal = {arXiv e-prints},
         year = 2025,
        month = feb,
          eid = {arXiv:2502.09582},
        pages = {arXiv:2502.09582},
          doi = {10.48550/arXiv.2502.09582},
archivePrefix = {arXiv},
       eprint = {2502.09582},
 primaryClass = {astro-ph.SR},
       adsurl = {https://ui.adsabs.harvard.edu/abs/2025arXiv250209582T}
}

@ARTICLE{Paxton2019ApJS,
       author = {{Paxton}, Bill and {Smolec}, R. and {Schwab}, Josiah and {Gautschy}, A. and {Bildsten}, Lars and {Cantiello}, Matteo and {Dotter}, Aaron and {Farmer}, R. and {Goldberg}, Jared A. and {Jermyn}, Adam S. and {Kanbur}, S.~M. and {Marchant}, Pablo and {Thoul}, Anne and {Townsend}, Richard H.~D. and {Wolf}, William M. and {Zhang}, Michael and {Timmes}, F.~X.},
        title = "{Modules for Experiments in Stellar Astrophysics (MESA): Pulsating Variable Stars, Rotation, Convective Boundaries, and Energy Conservation}",
      journal = {\apjs},
         year = 2019,
        month = jul,
       volume = {243},
       number = {1},
          eid = {10},
        pages = {10},
          doi = {10.3847/1538-4365/ab2241},
archivePrefix = {arXiv},
       eprint = {1903.01426},
 primaryClass = {astro-ph.SR},
       adsurl = {https://ui.adsabs.harvard.edu/abs/2019ApJS..243...10P}
}

@ARTICLE{Heger2000,
       author = {{Heger}, A. and {Langer}, N. and {Woosley}, S.~E.},
        title = "{Presupernova Evolution of Rotating Massive Stars. I. Numerical Method and Evolution of the Internal Stellar Structure}",
      journal = {\apj},
         year = 2000,
        month = jan,
       volume = {528},
       number = {1},
        pages = {368-396},
          doi = {10.1086/308158},
archivePrefix = {arXiv},
       eprint = {astro-ph/9904132},
 primaryClass = {astro-ph},
       adsurl = {https://ui.adsabs.harvard.edu/abs/2000ApJ...528..368H}
}

@ARTICLE{Barrault2025_magnetic,
       author = {{Barrault}, L. and {Bugnet}, L. and {Mathis}, S. and {Mombarg}, J.~S.~G.},
        title = "{Exploring the probing power of {\ensuremath{\gamma}} Dor's inertial dip for core magnetism: The case of a toroidal field}",
      journal = {\aap},
         year = 2025,
        month = sep,
       volume = {701},
          eid = {A253},
        pages = {A253},
          doi = {10.1051/0004-6361/202555213},
archivePrefix = {arXiv},
       eprint = {2507.00308},
 primaryClass = {astro-ph.SR},
       adsurl = {https://ui.adsabs.harvard.edu/abs/2025A&A...701A.253B}
}

@ARTICLE{Barrault2025_dip,
       author = {{Barrault}, L. and {Mathis}, S. and {Bugnet}, L.},
        title = "{Constraining differential rotation in {\ensuremath{\gamma}} Doradus stars from the properties of inertial dips}",
      journal = {\aap},
         year = 2025,
        month = feb,
       volume = {694},
          eid = {A225},
        pages = {A225},
          doi = {10.1051/0004-6361/202451541},
archivePrefix = {arXiv},
       eprint = {2412.02849},
 primaryClass = {astro-ph.SR},
       adsurl = {https://ui.adsabs.harvard.edu/abs/2025A&A...694A.225B}
}

@ARTICLE{Espinosa_Lara2011,
       author = {{Espinosa Lara}, F. and {Rieutord}, M.},
        title = "{Gravity darkening in rotating stars}",
      journal = {\aap},
         year = 2011,
        month = sep,
       volume = {533},
          eid = {A43},
        pages = {A43},
          doi = {10.1051/0004-6361/201117252},
archivePrefix = {arXiv},
       eprint = {1109.3038},
 primaryClass = {astro-ph.SR},
       adsurl = {https://ui.adsabs.harvard.edu/abs/2011A&A...533A..43E}
}

@ARTICLE{BauerFasching2024,
       author = {{Bauer-Fasching}, B. and {Bernhard}, K. and {Br{\"a}ndli}, E. and {Burger}, H. and {Eisele}, B. and {H{\"u}mmerich}, S. and {Neuhold}, J. and {Paunzen}, E. and {Piecka}, M. and {Ratzenb{\"o}ck}, S. and {Pri{\v{s}}egen}, M.},
        title = "{New ACV variables discovered in the Zwicky Transient Facility survey}",
      journal = {\aap},
         year = 2024,
        month = jul,
       volume = {687},
          eid = {A211},
        pages = {A211},
          doi = {10.1051/0004-6361/202347476},
archivePrefix = {arXiv},
       eprint = {2406.06203},
 primaryClass = {astro-ph.SR},
       adsurl = {https://ui.adsabs.harvard.edu/abs/2024A&A...687A.211B}
}

@ARTICLE{Endal1976,
       author = {{Endal}, A.~S. and {Sofia}, S.},
        title = "{The evolution of rotating stars. I. Method and exploratory calculations for a 7 M sun star.}",
      journal = {\apj},
         year = 1976,
        month = nov,
       volume = {210},
        pages = {184-198},
          doi = {10.1086/154817},
       adsurl = {https://ui.adsabs.harvard.edu/abs/1976ApJ...210..184E}
}

@ARTICLE{Borucki2010,
       author = {{Borucki}, William J. and {Koch}, David and {Basri}, Gibor and {Batalha}, Natalie and {Brown}, Timothy and {Caldwell}, Douglas and {Caldwell}, John and {Christensen-Dalsgaard}, J{\o}rgen and {Cochran}, William D. and {DeVore}, Edna and {Dunham}, Edward W. and {Dupree}, Andrea K. and {Gautier}, Thomas N. and {Geary}, John C. and {Gilliland}, Ronald and {Gould}, Alan and {Howell}, Steve B. and {Jenkins}, Jon M. and {Kondo}, Yoji and {Latham}, David W. and {Marcy}, Geoffrey W. and {Meibom}, S{\o}ren and {Kjeldsen}, Hans and {Lissauer}, Jack J. and {Monet}, David G. and {Morrison}, David and {Sasselov}, Dimitar and {Tarter}, Jill and {Boss}, Alan and {Brownlee}, Don and {Owen}, Toby and {Buzasi}, Derek and {Charbonneau}, David and {Doyle}, Laurance and {Fortney}, Jonathan and {Ford}, Eric B. and {Holman}, Matthew J. and {Seager}, Sara and {Steffen}, Jason H. and {Welsh}, William F. and {Rowe}, Jason and {Anderson}, Howard and {Buchhave}, Lars and {Ciardi}, David and {Walkowicz}, Lucianne and {Sherry}, William and {Horch}, Elliott and {Isaacson}, Howard and {Everett}, Mark E. and {Fischer}, Debra and {Torres}, Guillermo and {Johnson}, John Asher and {Endl}, Michael and {MacQueen}, Phillip and {Bryson}, Stephen T. and {Dotson}, Jessie and {Haas}, Michael and {Kolodziejczak}, Jeffrey and {Van Cleve}, Jeffrey and {Chandrasekaran}, Hema and {Twicken}, Joseph D. and {Quintana}, Elisa V. and {Clarke}, Bruce D. and {Allen}, Christopher and {Li}, Jie and {Wu}, Haley and {Tenenbaum}, Peter and {Verner}, Ekaterina and {Bruhweiler}, Frederick and {Barnes}, Jason and {Prsa}, Andrej},
        title = "{Kepler Planet-Detection Mission: Introduction and First Results}",
      journal = {Science},
         year = 2010,
        month = feb,
       volume = {327},
       number = {5968},
        pages = {977},
          doi = {10.1126/science.1185402},
       adsurl = {https://ui.adsabs.harvard.edu/abs/2010Sci...327..977B}
}

@ARTICLE{Garcia2022,
       author = {{Garcia}, S. and {Van Reeth}, T. and {De Ridder}, J. and {Tkachenko}, A. and {IJspeert}, L. and {Aerts}, C.},
        title = "{Detection of period-spacing patterns due to the gravity modes of rotating dwarfs in the TESS southern continuous viewing zone}",
      journal = {\aap},
         year = 2022,
        month = jun,
       volume = {662},
          eid = {A82},
        pages = {A82},
          doi = {10.1051/0004-6361/202141926},
archivePrefix = {arXiv},
       eprint = {2202.10507},
 primaryClass = {astro-ph.SR},
       adsurl = {https://ui.adsabs.harvard.edu/abs/2022A&A...662A..82G}
}

@ARTICLE{Li2019_splitting_gdor,
       author = {{Li}, Gang and {Bedding}, Timothy R. and {Murphy}, Simon J. and {Van Reeth}, Timothy and {Antoci}, Victoria and {Ouazzani}, Rhita-Maria},
        title = "{Period spacings of {\ensuremath{\gamma}} Doradus pulsators in the Kepler field: detection methods and application to 22 slow rotators}",
      journal = {\mnras},
         year = 2019,
        month = jan,
       volume = {482},
       number = {2},
        pages = {1757-1785},
          doi = {10.1093/mnras/sty2743},
archivePrefix = {arXiv},
       eprint = {1810.03362},
 primaryClass = {astro-ph.SR},
       adsurl = {https://ui.adsabs.harvard.edu/abs/2019MNRAS.482.1757L}
}
  \bibliographystyle{aasjournalv7}


\end{CJK*}
\end{document}